\documentclass[%
reprint,prl,
amsmath,amssymb,
aps,
]{revtex4-1}

\usepackage{xparse}

\usepackage{mathtools}
\usepackage{appendix}
\usepackage{graphicx}
\usepackage{dcolumn}
\usepackage{bm}
\usepackage{hyperref}
\usepackage{xcolor}
\usepackage[normalem]{ulem}

\usepackage{float}
\usepackage{braket}
\usepackage{bbold}
\usepackage{mathtools}

\DeclareMathOperator{\sgn}{sgn}

\DeclareMathOperator{\Tr}{Tr}

\newcommand{\msf}[1]{\mathsf{#1}}

\newcommand{\D}{\mathcal{D}}

\newcommand{\intl}[1]{\int\limits_{#1}}
\newcommand{\suml}[1]{\sum\limits_{#1}}

\newcommand{\R}{\mathtt{R}}

\newcommand{\GG}{\mathcal{G}}

\newcommand{\ww}{\omega}

\newcommand{\e}{\varepsilon}

\newcommand{\tel}{\tau_{\msf{el}}}

\newcommand{\bpsi}{\bar{\psi}}

\newcommand{\re}{\operatorname{Re}}

\usepackage{bbold}

\begin{document}
	
	\title{Emergence of many-body quantum chaos via spontaneous breaking of unitarity}

	\author{Yunxiang Liao}
	\author{Victor Galitski}
	\affiliation{Joint Quantum Institute and Condensed Matter Theory Center, Department of Physics, University of Maryland, College Park, MD 20742, USA.}
	
	\date{\today}
	
	\begin{abstract}
		It is suggested that many-body quantum chaos appears as the spontaneous symmetry breaking of unitarity in interacting quantum many-body systems.   It has been shown that many-body level statistics, probed by the spectral form factor (SFF) defined as $K(\eta,t)=\langle|{\rm Tr}\, \exp(-\eta H  + itH)|^2\rangle$, is dominated by a diffuson-type mode in a field theory analysis. The key finding of this Letter is that the ``unitary'' $\eta=0$ case is different from the $\eta \to 0^{\pm}$ limit, with the latter leading to a finite mass of these modes due to interactions. This mass suppresses a rapid exponential  ramp in the SFF,   which is responsible for the fast emergence of Poisson statistics in the non-interacting case, and gives rise to a non-trivial random matrix structure of many-body levels. The interaction-induced mass in the SFF shares similarities with the dephasing rate in the theory of weak localization and the Lyapunov exponent of the out-of-time-ordered correlators.
		
	\end{abstract}	
	
	
	\maketitle
	
	Both our everyday experience and laboratory experiments indicate that physical systems, initially prepared in a non-equilibrium state,  time-evolve into thermal equilibrium. Sets of axioms, such as the eigenstate thermalization hypothesis (ETH)~\cite{srednicki1994chaos,srednicki1999,Deutsch,rigol2008,Deutsch_2018}, have been formulated to justify this generic behavior and the emergence of statistical mechanics.  Yet, there is no formal proof of ETH, nor a clear understanding of the fundamental principles underlying the universality of thermalization and ergodicity in a  variety of quantum many-body systems.  Perhaps the most perplexing is the disconnect between the reversibility of physical laws,  governing the unitary evolution of closed quantum systems, and the irreversibility of thermodynamic behavior, which they eventually exhibit.  A well-known example of this puzzle is the black hole information paradox~\cite{paradox}, but the question itself can be posed for a much wider class of quantum systems~\cite{Rigol-review,Gogolin-review}.  How does irreversible dynamics emerge in quantum systems?
	
	Closely related to this line of inquiry is research on many-body quantum chaos, which has attracted much interest recently~\cite{ Prosen-chaos,Prosen-chaos2, Prosen-chaos3,  Prosen-Fermion, Chalker, Chalker-2, Chalker-3, Chalker-4, Chalker-5, Chalker-6, SYK-Stanford,SYK-Cotler,SYK-Altland,SYK-Altland-2,SYK-Altland-3,Altland-MBL, SYK-garcia,SYK-garcia-2,SYK-Tezuka,SYK-You,Cotler-2020,Poilblanc,Poilblanc-2,abergchaos,Jacquod,Green,gornyimanybodychaos,Altland-2020,Altland-2021, Muller-s,Guhr-s,Richter-s,Urbina-s}.  It can be defined as the presence of Wigner-Dyson level statistics~\cite{Mehta,Dyson,Wigner,Wigner2012} of many-body energy levels in an interacting quantum system~\cite{BGS,Guhr,Haake}.  Chaoticity so defined, ETH,  and the thermal behavior  in non-equilibrium settings are often assumed to be nearly equivalent notions,  although no such equivalence has been proven.  Furthermore,  no generic derivation of many-body quantum chaos exists, and only a handful of  rather fine-tuned models allow a microscopic insight into the fine structure of many-body levels~\cite{Prosen-chaos,Prosen-chaos2, Prosen-chaos3, Prosen-Fermion, Chalker,Chalker-2,Chalker-3}.  However, given the ubiquity of thermal, ``chaotic''  behavior,  it is natural to ask whether there are generic underlying reasons for its emergence. 
	
	This Letter suggests that the spontaneous breaking of unitarity may play a role in the emergence of chaotic behavior of interacting many-body systems.  Specifically, we study  the distribution of many-body energy levels in the system of $N$ weakly-interacting particles, which populate Wigner-Dyson distributed single-particle levels.   The choice of the model of interacting 	fermions ``embedded'' in a single-particle chaotic background is motivated by the following considerations: First, most actual physical media (e.g.,  disordered metals) are chaotic from the single-particle perspective (see Ref.~\cite{Guhr} for a review).  Integrable environments require fine tuning and are not representative of typical physical systems.  Second,  single-particle quantum chaos gives rise to a universal long-wavelength description (e.g.,  diffusion in disordered metals) in contrast to a non-universal description of systems without intrinsic randomness where ultra-violet physics (e.g., ballistic motion in clean conductors) complicates matters.  Third, we notice that in a previous analytical derivation of the many-body level statistics of a Floquet-driven spin chain~\cite{Prosen-chaos2},  the many-body  random matrix structure  emerged after an ensemble averaging. Indeed, there had been arguments that level statistics is not self-averaging and hence introducing an ensemble averaging (or imposing bare single-particle quantum chaos) may be a necessary step to see many-body quantum chaos~\cite{Prange}.

	One quantity that carries useful statistical information about the spectrum  is the two-level correlation function, $R_2(E - E')$.  Its Fourier transform can be shown to give rise to the spectral form factor (SFF)~\cite{Guhr,Haake,BerrySFF,Cotler-2017,Liu} $ K(t)=	\left\langle \Tr e^{ -iHt } \Tr e^{ +iHt  }\right\rangle$.  This relates the statistical properties of the static spectrum to the SFF, which explicitly involves unitary time-evolution operators.  More concrete links between the SFF and actual thermalization dynamics have been considered~\cite{Reimann}.  
	
	This Letter considers the generalized SFF defined as
	\begin{equation}\label{eq:K}
	K(\eta, t)	= 	\braket{Z(it+\eta)Z(-it+\eta)}		=
	\left\langle \left| \Tr e^{ -i H (t-i\eta) } \right|^2 \right\rangle,
	\end{equation}
	where $\eta \rightarrow 0^{\pm}$ is an infinitesimal.
	The key observation of the Letter is that 
	\begin{align}
	\begin{aligned}
	\label{SBU}
		\lim\limits_{\eta \to 0} \lim\limits_{N \to \infty}  K(\eta, t) \ne 
		\lim\limits_{N \to \infty} \lim\limits_{\eta \to 0}  K(\eta, t),
	\end{aligned}
	\end{align}
and random matrix theory (RMT) statistics appears if the $\eta \to 0$ limit is taken after the $N \to \infty$ limit, suggesting a possible connection between the emergence of quantum chaos and the spontaneous breaking of unitarity.
We note that although the calculation is done with $\eta\rightarrow 0^{+}$ being a positive infinitesimal, it can be easily generalized to the case of $\eta \rightarrow 0^{-}$ and the main conclusion still holds.
	
We consider an interacting random matrix model of fermions populating Wigner-Dyson single-particle levels and generic nonrandom two-body interactions. For simplicity, we restrict ourselves to the case of broken time-reversal symmetry.
	The Hamiltonian assumes the form
	\begin{align}\label{eq:H}
	\begin{aligned}
	H=\sum_{i,j=1}^{N} \psi^{\dagger}_i h_{ij}  \psi_j 
	+ 
	\frac{1}{2} \sum_{i,j,k,l=1}^{N}  \psi^{\dagger}_i\psi^{\dagger}_j V_{ij;kl}  \psi_k\psi_l,
	\end{aligned}
	\end{align}
 	where $h$ is a
	$N \times N$ random Hermitian matrix drawn from a Gaussian unitary ensemble (GUE)~\cite{Mehta} with the distribution function
	$
	P(h) \propto \exp \left(-\frac{N}{2J^2} \Tr h^2   \right).
	$
	The interaction matrix $V$ is antisymmetric: $V_{ij;kl}=-V_{ji;kl}=-V_{ij;lk}=V_{kl;ij}^*$ and not random. We consider an arbitrary fixed realization of $V$.

\begin{figure}[t!]
		\centering
		\includegraphics[width=1\linewidth]{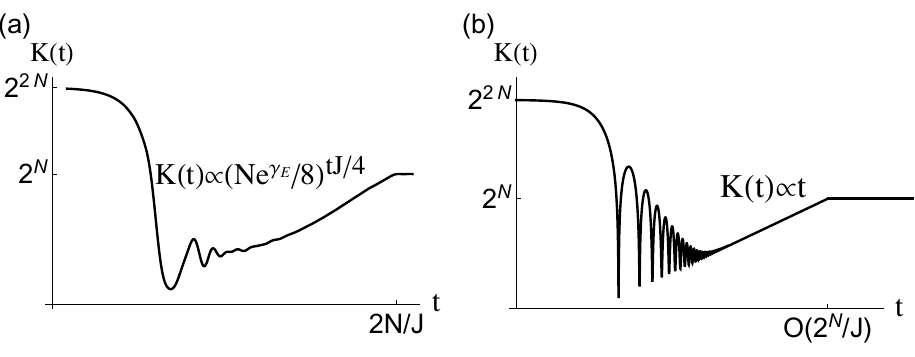}
		\caption{
				Schematic log-log plots of the SFF $K(t)$ of the random matrix model given by Eq.~\ref{eq:H}  for the (a) noninteracting and (b) interacting case.  (a) is obtained using the numerical data from our previous study~\cite{PRL}, while  (b) is the RMT prediction applying the analytical expression for the GUE SFF in Ref.~\cite{Liu}.
				(a) In the non-interacting case, the SFF  consists of an initial slope, an exponential-in-$t$ ramp, and a plateau that starts at $t=2N/J$. 
				(b) In the presence of interactions, the many-body spectrum exhibits RMT statistics and the exponential ramp is expected to be replaced with a linear one which approaches the plateau at a much larger time $t\sim 2^N/J$. See also Ref.~\cite{SYK2-2020} where this SFF transition has been observed numerically in a similar model of Majorana fermions.
		}
		\label{fig:1}
	\end{figure}

	In the absence of interactions, the model is trivially many-body integrable. However, the SFF is still non-trivial showing an initial slope falling off the value of $K(t=0)=2^{2N}$ and then rapidly increasing via an exponential ramp to a plateau $K(t) = 2^N$ at $t \geq 2N/J$, see Ref.~\cite{PRL} and Fig.~\ref{fig:1}(a). This implies residual correlations in the many-particle spectrum on single-particle energy scales, $\Delta E \sim N^{-1}J$. However, there are negligible correlations on smaller many-body energy scales and the statistics there becomes Poisson -- the plateau. Interactions are expected to break integrability and give rise to Wigner-Dyson many-body level statistics, whereas the exponential ramp is replaced with a much slower linear one leading to a plateau at $\Delta E \sim 2^{-N}J$ (see Fig.~\ref{fig:1}(b)). The question is how does this transition occur? It is argued below that this involves two ``steps'' -- spontaneous breaking of unitarity, which selects a unique saddle point out of a manifold of ``unitary'' saddle points, and gapping out soft modes, which are responsible for the exponential ramp in the non-interacting theory~\cite{PRL,Winer}. 
	
	The SFF defined in Eq.~\ref{eq:K} for this model can be expressed as the following functional integral, 
	\begin{subequations}\label{eq:K-0}
	\begin{align}
	&K(\eta, t)
	= 
	\int \D h P(h)
	\int \D (\bpsi, \psi) 
	e^{-S[\bpsi,\psi]} ,
	\\
	&\begin{aligned}\label{eq:S-0}
	S[\bpsi,\psi]
	= -&
	i\sum_{a=\pm}
	\int_0^{z_{a}} d t'
	\left[ 
	\bpsi_i^{a} (t') (i \partial_{t'} \delta_{ij}-\zeta_a h_{ij}) \psi_j^{a}(t')
	\right. 
	\\
	&\left. 
	-\frac{i}{2} 
	\zeta_a
	\bpsi_i^{a} (t') \bpsi_j^{a} (t') V_{ij;kl}
	\psi_k^{a} (t')  \psi_l^{a}	(t')
	\right] 
	.
	\end{aligned}	
	\end{align}
	\end{subequations}
	The Grassmann field $\psi^{a}_{i}$ is labeled by a flavor index $i=1,...,N$ along with a replica index $a=\pm$, and is subject to the antiperiodic boundary condition
	$
	\psi^{a}(z_a)=-\psi^{a}(0),
	$
	with $z_a=t-i\zeta_a \eta$ and $\zeta_a=\pm 1$ for $a=\pm$.
	The integration over $\psi^{a}$ yields the partition function
	 $Z(i\zeta_a z_a)$. 
	
	Starting from Eq.~\ref{eq:K-0} and following the standard $\sigma$-model derivation procedure~\cite{Efetov,Wegner-1979,Kamenev-GUE,Kamenev-Keldysh}, we obtain
	\begin{subequations}\label{eq:K-1}
	\begin{align}
	& K(\eta,t) =\frac{1}{Z_{\phi}Z_Q}
	\int \D \phi \int \D Q \exp \left( -S[Q,\phi] \right) ,
	\\
	&
	\begin{aligned}\label{eq:S-1}
	&S[Q,\phi]
	=\,		
	-\frac{i}{2} \sum_{a} 
	\zeta_a z_a \phi_{il}^a(-\ww_m^a)  V^{-1}_{ij;kl} \phi_{jk}^a(\ww_m^a)
	\\
	&+
	\frac{N}{2J^2}
	\Tr  Q^2
	-
	\Tr
	\ln
	\left[
	\left( 
	\mathcal{E}\sigma^3
	+
	iQ
	\right) 
	\otimes I_f
	+
	\Phi 
	\right].
	\end{aligned}
	\end{align}
	\end{subequations}
	Here 
	$\mathcal{E}$ and $\Phi$ are matrices with elements:
	\begin{align}\label{eq:E}
	\begin{aligned}
	\mathcal{E}^{ab}_{nn'}=\delta_{ab}\delta_{nn'}\e_n^ae^{-i\e_n^a \delta z_a},
	\,
	\Phi^{ab}_{ij;nn'}=\delta_{ab}\phi_{ij}^a(\ww_{n-n'}^a),
	\end{aligned}
	\end{align} 
	where $\e_n^a=2\pi (n+1/2)/z_a$ and $\ww_m^a=2\pi m/z_a$ denote the fermionic and bosonic Matsubara frequencies, respectively.  
	$\sigma^3$ indicates the direct product of the third Pauli matrix in the replica space (labeled by $a$) and the identity matrix in the frequency space (indexed by $n$), while $I_f$ represents the $N \times N$ identity matrix in the flavor space (labeled by $i$).
	The phase factor $e^{-i\e_n^a \delta z_a}$ in Eq.~\ref{eq:E}, with $\delta z_a$ being the time discretization interval for path $a$, ensures convergence of the integral.
	We have reduced the problem to a theory of the matrix field $Q$ and bosonic field $\phi$.
	$Q$ is a Hermitian matrix acting in the replica and Matsubara frequency spaces. It decouples the ensemble-average-generated four-fermion term. The bosonic field $\phi$ is introduced to decouple the interactions~\cite{FNa}, and carries replica, frequency and flavor indices.
	See Sec.~I.A. of the Supplemental Material~\cite{Sup} for the detailed derivation and the expressions for normalization constants
	$Z_{\phi}$ and $Z_Q$
	
	In the large $N\rightarrow \infty$ limit, the functional integral Eq.~\ref{eq:K-1} can be evaluated by considering the saddle points and small fluctuations around them. We assume that the decoupling field $\phi$ does not influence the stationary configuration of matrix field $Q$, and obtain from the non-interacting action $S[Q,\phi=0]$ (Eq.~\ref{eq:S-1}) the saddle point equation:
	\begin{align}
	\begin{aligned}
	Q_{sp}
	=
	J^2\left( 
	-i\mathcal{E} \sigma^3
	+
	Q_{sp}
	\right) ^{-1}.
	\end{aligned}
	\end{align}
	This is solved by diagonal matrices $\Lambda$ with
	\begin{align}\label{eq:sp}
	\begin{aligned}
	\Lambda^{ab}_{nm}=\pm J \delta_{ab}\delta_{nm}, 
	\qquad
	|\e_n^a| \ll J.
	\end{aligned}
	\end{align}
	Each diagonal element can take two possible values, resulting in $2^{2N_{\e}}$ distinct diagonal saddle points $\Lambda$, with $N_{\e}$ being the total number of Matsubara frequencies considered.
	We investigate  the long-time behavior of the SFF $K(t \gg J^{-1})$,  which allows us to focus on the low energy sector of the theory ($|\e_n^a|\ll J$) and ignore the correction from nonuniversal single-particle density of states.

	In the zero $\eta$ and $\delta z_a$ limit, the non-interacting action remains invariant under unitary transformations $Q\rightarrow U^{\dagger}Q U$ satisfying $U\mathcal{E}\sigma^3U^{\dagger}=\mathcal{E}\sigma^3$. These unitary transformations $U$ are of the form $U=\prod_{n=1}^{N_{\e}} U_n$, where $U_n \in U(2)$ is a rotation acting on the subblock 
	$\begin{bmatrix} Q^{++}_{nn} & Q^{+-}_{n,-n-1}\\Q^{-+}_{-n-1,n} & Q^{--}_{-n-1,-n-1} \end{bmatrix}$. 
	More saddle points can be found by
	applying the symmetry transformations $U$ to the diagonal ones $Q_{sp}=U^{\dagger} \Lambda U$.

\begin{figure}[t!]
		\centering
		\includegraphics[width=0.8\linewidth]{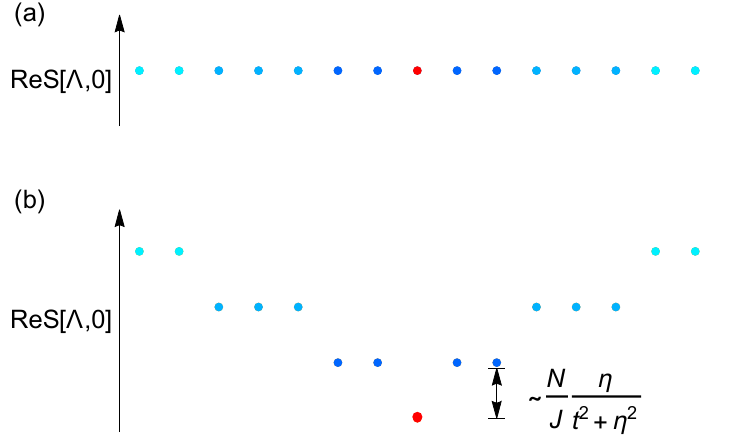}
		\caption{
		Schematic plot of the real part of the non-interacting action $\re S[\Lambda,0]$ (vertical displacement)  for various diagonal saddle points (solid circles).
		(a) For $\eta=0$, $\re S[\Lambda,0]$  of all diagonal saddle points are equivalent, and the corresponding contributions to the SFF differ only by phase factors. It is essential to take into account fluctuations around all saddle points for the computation of SFF for zero $\eta$. (b) For nonzero $\eta$, the differences in $\re S[\Lambda,0]$ between various saddle points are non-negligible as long as $\eta/(t^2+\eta^2) \gtrsim J/N$. The standard saddle point $\Lambda^{(0)}$ (solid red circle)  acquires the minimum $\re S[\Lambda,0]$  and dominates over all remaining saddle points. In this case, the SFF is dominated by fluctuations around $\Lambda^{(0)}$. }
		\label{fig:2}
	\end{figure}

	The presence of nonzero $\eta$ breaks this $\prod_{n=1}^{N_{\e}} U(2)$ symmetry of the non-interacting theory and allows us to select one dominant saddle point. The non-interacting action of a diagonal saddle point acquires the form
	\begin{align}\label{eq:S0}
	\begin{aligned}
	S[\Lambda,0]
	=\,	
	i\frac{N}{J^2}	
	\sum_{a,n}
	\zeta_a 
	\Lambda^{aa}_{nn}
	\e_n^ae^{-i\e_n^a\delta z_a}
	+\text{const.}.
	\end{aligned}
	\end{align}
The phase factor $e^{-i\e_n^a \delta z_a}$ is only needed for convergence and can be omitted here. 
	In the case of 
	$\eta/(t^2+\eta^2) \gtrsim J/N$, among various saddle points,
	 $(\Lambda^{(0)})^{ab}_{nm}=J \sgn(n+1/2)\delta_{ab}\delta_{nm}$ yields the minimum $\re S[\Lambda,0]$ and consequently the dominant contribution.
	 By contrast, for $\eta=0$, the contributions from various saddle points to the SFF differ only by phase factors (see Fig.~\ref{fig:2}).
	We call $\Lambda^{(0)}$ the standard saddle point and all remaining diagonal saddle points the nonstandard ones~\cite{AndreevAltshuer,Kamenev-GUE}.

	Let us now consider the fluctuations of $Q$ around the diagonal saddle points $\Lambda$, which fall into two categories~\cite{PRL,Kamenev}: 
	(i) soft modes (or Goldstone modes) generated by unitary rotations of saddle points $Q=R^{\dagger}\Lambda R$ and associated with the explicitly breakdown of the $U(2N_{\e})$ symmetry of the non-interacting action by the $\mathcal{E}\sigma^3$ term;
	(ii) massive modes that can not be obtained by rotation.
	In the zero $\eta$ case, there is a subset of unitary transformations $U\in \prod_{n=1}^{N_{\e}} U(2)$ which leaves the non-interacting action invariant. Applying these symmetry transformations to diagonal saddle points $Q=U^{\dagger}\Lambda U$ generates a special type of soft modes called zero mode with $\delta S[Q,0]=0$~\cite{FNb}.

	In the non-interacting case, the soft modes are responsible for the exponential-in-$t$ ramp, whereas the massive modes only give rise to a nonessential constant~\cite{PRL}. 
    Therefore, we focus on the interaction effects on the soft modes.
	If $\eta=0$, to evaluate the SFF, one must sum over contributions of fluctuations around various saddle points. 
	By contrast, for $\eta \rightarrow 0^+$  (or more precisely $\eta/(t^2+\eta^2) \gtrsim J/N$), there is one dominant saddle point $\Lambda^{(0)}$ determined by the $\eta$-induced symmetry breaking, and it is sufficient to consider only soft mode fluctuations around it. 

	In the following, we consider the latter case and explore the influence of interactions on soft mode fluctuations around $\Lambda^{(0)}$  and their contribution to the SFF.
	In this case, the SFF can be approximated by (see Sec. I.D. of the Supplemental Material \cite{Sup}),
	\begin{align}\label{eq:K-2}
	\begin{aligned}
	&K(\eta,t)=
	\frac{Z_{m} e^{-S[\Lambda^{(0)},0]}}{ Z_{\phi} Z_Q}
	\int \!\!\D \phi
	\int_{ \mathcal{M} }\D Q e^{ -\delta S[Q,\phi;\Lambda^{(0)}]},
	\\
	&\delta S[Q,\phi; \Lambda^{(0)}]
	=
	i\frac{N}{J^2}
	\Tr
	\left[ \sigma^3 \mathcal{E} \left( Q-\Lambda^{(0)} \right)   \right]
	\\
	&
	+
	\frac{i}{J^2} \!\!\! \suml{a,n,n',i} \!\!\! Q^{aa}_{nn'}\phi^a_{ii}(\ww_{n'-n}^a)
	\\
	&	- \sum_{a,m,i,j} \frac{i\zeta_a z_a}{2}
	 \phi_{il}^a(-\ww_m^a)  \left( V^{-1}_{ij;kl}+i\frac{\zeta_a}{z_a} \tilde{M}^a_m\delta_{ik}\delta_{lj} \right) \phi_{jk}^a(\ww_m^a),
	\end{aligned}
	\end{align}
	where $	
	\tilde{M}^a_m
	=
	-\sum_{n} (\Lambda^{(0)})_{nn}^{aa} (\Lambda^{(0)})_{n+m,n+m}^{aa}/J^4$, and $Z_{m}$ denotes the nonessential contribution from massive modes.
		The integration $\int \D Q$ is over the manifold $\mathcal{M}$ of matrices $Q$ which are generated by unitary rotations of the standard saddle point $\Lambda^{(0)}$: 
	\begin{align}\label{eq:Q}
		Q=R^{-1}\Lambda^{(0)}R,
		\quad
		R \in \frac{U(2N_{\e})}{U(N_{\e})\times U(N_{\e})}.
	\end{align}
		These matrices $Q$ represent soft modes around $\Lambda^{(0)}$ and obey the nonlinear constraints:
		$\Tr Q=0$ and $Q^2/J^2=I$. The aforementioned zero modes are the special soft modes corresponding to
		\begin{align}\label{eq:zeroQ}
			Q=U^{-1}\Lambda^{(0)}U,
			\quad
			U \in \prod_{n=1}^{N_{\e}}\frac{U(2)}{U(1)\times U(1)}.
	\end{align}	

For the matrix field $Q$ governed by the action in Eq.~\ref{eq:K-2},
the non-interacting propagator is given by
\begin{align}\label{eq:GX0}
\begin{aligned}
(\GG_{X}^{(0)})^{ab;ba}_{nm;mn}
=
\frac{N}{2J^3}
\braket{Q^{ab}_{nm}Q^{ba}_{mn}}_0
=
\frac{i}{\zeta_a\e_n^a-\zeta_b\e_m^b},
\end{aligned}
\end{align}
for $n\geq0>m$.
Interactions between $Q$ and $\phi$ result in a self-energy correction, which gives rise to a mass term $\lambda^{ab}_{nm}$ in the interaction-dressed inter-replica $Q$ propagator (see Sec. I.D.3 of the Supplemental Material~\cite{Sup}): 
	\begin{align}\label{eq:GX}
	\begin{aligned}
	(\GG_{X})^{ab;ba}_{nm;mn}
	=
	\frac{i}{(\zeta_a\e_n^a-\zeta_b\e_m^b)(1+\delta z)+i\lambda^{ab}_{nm}},
	\end{aligned}
	\end{align}
	for $a\neq b$ and $n\geq 0 >m$.
	Note that the intra-replica propagator $\GG_{X}^{aa;aa}$ contributes to the disconnected SFF $K^{\msf{dis}}(\eta,t)=\braket{Z(iz_+)}\braket{Z(-iz_-)}$ which dominates the early time slope regime and will not be considered here.
	Assuming that the main interaction effect on the SFF comes from the mass $\lambda^{ab}_{nm}$, we ignore the renormalization effect and set $\delta z=0$.
	We obtain a self-consistent equation for the mass $\lambda^{ab}_{nm}$ (Eq.~S59 in the Supplemental Material~\cite{Sup}), whose explicit form depends on the interaction matrix $V$ and is not important for our purpose.
	 Eq.~\ref{eq:GX} shows that the fluctuations of the bosonic field $\phi$ lead to decoherence and introduce a cutoff $\lambda$ in the soft modes. 

	Carrying out the integration over $Q$ and $\phi$ in Eq.~\ref{eq:K-2}, we find that, up to an irrelevant constant, $K(\eta,t)$ is approximately given by
	\begin{align}
	\begin{aligned}\label{eq:K-3}
	&K(\eta,t)
	\propto
	\exp
	\left\lbrace 
	-\sum_{a b} \suml{n\geq 0>m}  \ln  
	\left[ (\GG_{X}^{-1})^{ab;ba}_{nm;mn} \right] 
	\right. 
	\\
	&\left. 
	+\frac{1}{2} \suml{a,m} \left(  \Tr \ln \tilde{V}^a(\ww_m^a) -\Tr \ln V \right) 
	-S[\Lambda^{(0)},0]
	\right\rbrace.
	\end{aligned}
	\end{align}
	Here $\tilde{V}^a_{ij;kl}(\ww_m^a)=-i\zeta_az_a \braket{\phi^a_{il}(\ww_m^a)\phi^a_{jk}(-\ww_m^a)}$ stands for the effective interaction matrix and is proportional to the interaction dressed propagator for the bosonic field $\phi$ 
	(see Eq.~S49 in the Supplemental Material~\cite{Sup}).
	
	In the exponent in Eq.~\ref{eq:K-3}, only the first term with $a\neq b$  contributes to the connected SFF $K^{\msf{con}}=K-K^{\msf{dis}}$. Substituting the explicit form of the interaction-dressed inter-replica propagator $\GG_{X}^{ab;ba}$ into this term and evaluating the Matsubara frequency summation by analytical continuation technique, one obtains
	\begin{align}\label{eq:lnK-1}
	\begin{aligned}
	&\ln \!K^{\msf{con}}(\eta,t)
	\!=\!\!
	\sum_{a\neq b}
	\frac{1}{4\pi^2} \! \intl{\e,\e'}\!
	\ln \left( 1+e^{-i \zeta_a z_a\e} \right) 
	\ln \left( 1+e^{-i \zeta_b z_b \e'} \right) 
	\\
	& \times
	\left( \dfrac{1}{-i\left( \e-\e' \right) +\lambda^{ab}(\e+\e')} \right)^2
	\left( \frac{\partial \lambda^{ab}}{\partial \e}-i \right) 
	\left( \frac{\partial \lambda^{ab}}{\partial \e'}+i\right).
	\end{aligned}
	\end{align}
	Here $\lambda^{ab}(\e+\e')$ now represents the analytic continuation of the inter-replica mass $\lambda^{ab}_{nm}$ from $\zeta_a\e_n^a\rightarrow \e$ and $\zeta_b\e_m^b\rightarrow \e'$, and is assumed to be a function of $\e+\e'$ (independent of $\e-\e'$).
	The non-interacting connected SFF can also be obtained from the equation above by setting  $\lambda^{ab}=0$. 
	
	
	Focusing on the difference between the interacting and non-interacting cases, we then replace the factor $\ln \left( 1+e^{-i \zeta_a z_a\e} \right) $ in Eq.~\ref{eq:lnK-1} with $e^{-i \zeta_a z_a\e} $. We assume that the neglected contribution cancels partially with higher order fluctuation correction (see Sec. I.F. of the Supplemental Material~\cite{Sup} and also Ref.~\cite{PRL}) and find
	\begin{align}\label{eq:lnK-2}
	\begin{aligned}
	\ln K^{\msf{con}}(\eta, t)
	=&
	t\suml{a\neq b}
	\int_{-E_{\msf{UV}}}^{E_{\msf{UV}}} \frac{dE}{2\pi}
	e^{-2 \eta E}
	e^{-\zeta_a\lambda^{ab}(E) t} \Theta(\zeta_a\re\lambda^{ab})
	\\
	&\times 
	\left[ 	\left( \frac{1}{2}\frac{\partial \lambda^{ab}}{\partial E}\right)^2+1\right],
	\end{aligned}
	\end{align}
	where $E_{\msf{UV}}\sim J$ denotes the ultraviolet cutoff.
	
	Let us now take the limit $\eta\rightarrow 0^+$.
	In the non-interacting case, after setting $\lambda^{ab}=0$, it is straightforward to see from the equation above that 
	$\ln K^{\msf{con}}(\eta\rightarrow 0^+, t)\propto t$, 
	and an infrared divergence originating from the zero modes occurs.
	This infrared divergence can be resolved by including higher order fluctuation correction~\cite{Sup,PRL} (which is also needed to recover the correct overall coefficient), and is cut off by the mass $\lambda^{ab}$ in the interacting theory at the quadratic order. 
	
    With interactions, fluctuations of the decoupling field $\phi$ lead to dephasing effect, reflected by the appearance of  a mass $\lambda$ to the soft modes.
	 The mass $\lambda^{ab}$ results in an exponential factor $e^{-\zeta_a\lambda^{ab}(E) t}\Theta(\zeta_a\re\lambda^{ab})$ in $\ln K^{\msf{con}}(\eta\rightarrow 0^+, t)$, and thus suppresses the exponential-in-$t$ growth of the connected SFF. We note that what matters is not the explicit form of the inter-replica mass $\lambda^{ab}(E)$, but its existence. 
	
	For the case where $\eta\rightarrow 0^{-}$ is a negative infinitesimal, the dominant saddle point is instead $-\Lambda^{(0)}$ and the soft mode fluctuations around this saddle point are described by a nonlinear $\sigma$-model which can be obtained by replacing $\Lambda^{(0)}$ in Eqs.~\ref{eq:K-2} and~\ref{eq:Q} with $-\Lambda^{(0)}$.
    Through a calculation similar to that of $\eta\rightarrow 0^{+}$ case, one can show that the inter-replica propagator of the fluctuations around $-\Lambda^{(0)}$ also acquire a mass, which arises from interaction-induced dephasing processes and is responsible for the suppression of the exponential ramp.
	 
	The suppression of the exponential ramp is a necessary prerequisite for the emergence of RMT statistics in a many-body spectrum.  In particular, for the interacting model (with broken time-reversal symmetry) whose many-body energy levels follow Wigner-Dyson statistics, the connected SFF should grow linearly in $t$ instead of exponentially.
	However, the derivation of the explicit expression for the SFF requires a consideration of the fluctuation corrections beyond the quadratic order. For the non-interacting case, the many-body SFF can be expressed in terms of the connected $n$-point single-particle level correlation function $\R_n^{\msf{con}}(\e_1,...,\e_n)$ as~\cite{Sup, PRL} 
	\begin{align}\label{eq:CumSum}
	\begin{aligned}
	\ln K(\eta,t)  
	=\,&
	\suml{n}
	\frac{N^n}{n!} 
	\int
	\prod_{k=1}^{n}
	d \e_k
	\left[ \suml{a_k}
	\ln \left( 1+e^{-i \zeta_{a_k} z_{a_k} \e_k }\right) \right] 
	\\
	&
	\times 
	\R_n^{\msf{con}}(\e_1,...,\e_n)
	.
	\end{aligned}
	\end{align}
	The saddle point action $S[\Lambda^{(0)},0]$ (Eq.~\ref{eq:S-1}) yields the $n=1$ term which results in the initial slope, whereas the quadratic fluctuation correction leads to the $n=2$ term which contributes to the exponential ramp.  The contributions from $n>2$ terms are as important and are necessary to obtain the correct overall coefficient in the exponent of the ramp.
	In the presence of interactions, Eq.~\ref{eq:CumSum} is no longer valid. However, we find that the quadratic fluctuation correction Eq.~\ref{eq:lnK-1} is analogous to the $n=2$ term.
If the interaction effects on all higher order fluctuations are similar, their contributions will be suppressed in a similar way.

In Sec II of the Supplemental Material~\cite{Sup}, we also perform an analogous calculation of the SFF for a two-dimensional (2D) disordered system of fermions interacting via density-density interactions. 
In the ergodic regime $t \gg L^2/D$,  with $D$ being the diffusion constant and $L$ the system size, the statistics of non-interacting single-particle energy levels can be universally described by RMT~\cite{Eliashberg,Efetov,Efetov2,Altshuler,AndreevAltshuer,Zirnbauer,Kravtsov,Kamenev}, and the corresponding field theory becomes effectively the same as the current theory of the random matrix model.  In the diffusive regime with $ L^2/D  \gg t \gg \tel$, where $\tel$ denotes the elastic scattering time,  the connected SFF is governed by inter-replica diffusons, which acquire a mass from the dephasing processes due to interactions (see Eq.~S132 in the Supplemental Material~\cite{Sup}). The inter-replica mass gives rise to an exponential decay factor in the exponent of the connected SFF,
and is crucial for the emergence of RMT statistics. This mass is similar to
the dephasing rate~\cite{AAK,AAG} that cuts off the quantum interference correction to conductance and the Lyapunov exponent of the out-of-time-ordered correlator (OTOC)~\cite{Patel,OTOC} in interacting disordered metals.
All these three quantities are given by the mass of diffusons, and the complex time $z_{\pm}=t \mp i\eta$ in the SFF now plays the role of inverse temperature in the dephasing of conductance correction and the Lyapunov exponent of OTOC.
 If the fluctuations of the interaction decoupling field become ineffective in destroying the coherence, these three types of diffuson mass vanish, Poisson statistics would appear, the quantum conductance correction would diverge~\cite{dephasing,Davis}, and the OTOC would no longer grow exponentially in time, suggesting a connection between dephasing failure and many-body localization. However, it is unclear under what condition dephasing failure will occur for these three separate cases, which requires a more detailed analysis of these masses and is a possible direction for future work.

	In conclusion, we have shown that the spectral form factor behaves in a drastically different way in the $\eta \to 0^{\pm}$ limit compared to the $\eta=0$ case. In the former case, the exponential ramp of the non-interacting theory is suppressed, which leads to the appearance of a plateau at a later time (or equivalently to the emergence of correlations on smaller many-body energy scales). Note that the mathematical problem of calculating level statistics studies properties of the spectrum, has no notion of temperature, and the SFF is described by the correlations of two unitary time-evolution operators. While the $\eta=0$ case is formally correct, it does not account for the possible external perturbations and noise that the physical system may experience, and which would lead to non-unitary time-evolution. Our proposed explanation of the result is that the $\eta \to 0^{\pm}$ limit (or any other way of imposing a non-unitary structure in the SFF) in effect encodes such tiny non-unitary perturbations, which get magnified in the thermodynamic limit. Mathematically this occurs via a spontaneous symmetry breaking in the $\sigma$-model saddle-point manifold and the appearance of  a ``dephasing'' mass of the Goldstone modes, while physically this may imply spontaneous breaking of unitarity triggered by any external perturbations. 
	
	
	This  work  was supported by the U.S. Department of Energy,  Office of Science,  Basic  Energy  Sciences  under  Award  No.   DE-SC0001911. Y. L. acknowledges a postdoctoral fellowship from  the  Simons  Foundation  ``Ultra-Quantum  Matter'' Research Collaboration.
	
\bibliography{main}

\begin{thebibliography}{76}
\bibitem [{FNa()}]{FNa}%
  \BibitemOpen
  \href@noop {} {}\bibinfo {howpublished} {{We assume that interaction matrix's
  inverse $V^{-1}$, defined by {$ \sum_{jk} V^{-1}_{i'j;kl'} V_{ji;lk}
  =\delta_{ii'}\delta_{ll'} $}, exists. Otherwise, one can proceed in an
  analogous manner by decoupling the interactions in a different
  channel.}}\BibitemShut {Stop}%
\bibitem [{FNb()}]{FNb}%
  \BibitemOpen
  \href@noop {} {}\bibinfo {howpublished} {Note that there are no soft modes
  around saddle point $\Lambda=\pm JI$, and no zero modes around saddle points
  satisfying $\Lambda^{++}_{nn}=\Lambda^{--}_{-n-1,-n-1}$ for any frequency
  index $n$.}\BibitemShut {Stop}%
\end{thebibliography}
		
\end{document}


\title{
		Emergence of many-body quantum chaos via spontaneous breaking of unitarity
		\\
		Supplemental Material
	}
	\author{Yunxiang Liao}
	\author{Victor Galitski}
	\affiliation{Joint Quantum Institute and Condensed Matter Theory Center, Department of Physics, University of Maryland, College Park, MD 20742, USA.}
	
	\date{\today}

	\maketitle
	
	\tableofcontents
	
	\bigskip
	
	In this supplemental material, we present the detailed derivation for the spectral form factors of (1) a random matrix model with additional nonrandom two-body interactions (Sec.~\ref{sec:RMT}); (2) a two-dimensional  disordered system of fermions interacting via short-range density-density interactions (Sec.~\ref{sec:disorder}).
	
	\section{Random matrix model with interactions}~\label{sec:RMT}
	
	In this section, we consider a system of $N \gg 1$ interacting fermions populating the single-particle energy levels of a Gaussian unitary ensemble, and is governed by the following Hamiltonian
	\begin{align}\label{eq:H}
	\begin{aligned}
	H=\sum_{i,j=1}^{N} \psi^{\dagger}_i h_{ij}  \psi_j 
	+ 
	\frac{1}{2} \sum_{i,j,k,l=1}^{N}  \psi^{\dagger}_i\psi^{\dagger}_j V_{ij;kl}  \psi_k\psi_l,
	\end{aligned}
	\end{align}
	where $h$ is a
	$N \times N$ random Hermitian matrix drawn from a Gaussian unitary ensemble (GUE) with the distribution function
	\begin{align}\label{eq:Ph}
	\begin{aligned}
	P(h) =\frac{1}{Z_h} \exp \left(-\frac{N}{2J^2} \Tr h^2   \right),
	\qquad
	Z_h=\int \D h \exp \exp \left(-\frac{N}{2J^2}\Tr  h^2   \right).
	\end{aligned}
	\end{align}
	The coupling $V$ of the two-body interactions is not random. We consider an arbitrary but fixed configuration of $V$ which is antisymmetrized:
	\begin{align}\label{eq:V}
	\begin{aligned}
	V_{ij;kl}=-V_{ji;kl}=-V_{ij;lk}=V_{kl;ij}^*.
	\end{aligned}
	\end{align}
	
	To investigate the spectral statistics of this model, we calculate the spectral form factor (SFF) defined as
	\begin{align}\label{eq:K}
	\begin{aligned}
	K(\eta, t)
	= \,&
	\braket{Z(\eta+it)Z(\eta-it)}.
	\end{aligned}
	\end{align}
	Here the angular bracket represents the ensemble averaging over random matrix $h$ with the weight $P(h)$ (Eq.~\ref{eq:Ph}), and $Z(\eta \pm it)=\Tr \left( e^ { -(\eta \pm it) H } \right)$ is the analytically-continued partition function.

	Without the interactions, the current model reduces to the $q=2$ complex Sachdev–Ye–Kitaev (SYK) model whose many-body level statistics has been investigated in Ref.~\cite{PRL} (see also Ref.~\cite{Winer} for a similar study of $q=2$ Majorana SYK model). Despite being integrable, the $q=2$ SYK model exhibits surprising rich structure which combines the single-particle chaotic and many-body integrable features, reflected by the exponential-in-$t$  ramp in its SFF. In the following, we show that, in the presence of interactions, $K(\eta\rightarrow 0^+,t)\neq K(\eta=0,t)$. Furthermore, in the case of $\eta\rightarrow 0^+$, the soft modes responsible for the exponential ramp acquire a mass, leading to the suppression of the ramp.
	
	
	\subsection{Preliminaries}
	
	We start with a path integral formula for the spectral form factor:
	\begin{align}\label{eq:ZZ}
	\begin{aligned}
	K(\eta,t)
	=&\left\langle Z(iz_+)Z(-iz_-) \right\rangle 
	\\
	=&
	\left\langle 
	\int \D (\bpsi, \psi)  
	\exp 
	\left\lbrace 
	i\sum_{a=\pm}
	\int_0^{z_{a}} d t'
	\left[ 
	\bpsi_i^{a} (t') (i \partial_{t'} \delta_{ij}-\zeta_a h_{ij}) \psi_j^{a}(t')
	-\frac{i}{2} 
	\zeta_a
	\bpsi_i^{a} (t') \bpsi_j^{a} (t') V_{ij;kl}
	\psi_k^{a} (t')  \psi_l^{a}	(t')
	\right] 
	\right\rbrace
	\right\rangle ,
	\end{aligned}	
	\end{align}
	where for simplicity we denote $t \mp i\eta$ by $z_{\pm}$. Besides the flavor index $i$, the Grassmann fields $\psi$ and $\bar{\psi}$ carry an additional replica index $a=\pm$. The path integral over $\psi^{a}$ and $\bar{\psi}^a$ is subject to the boundary condition:
	\begin{align}\label{eq:bc}
	\begin{aligned}
	\psi^{a}(z_a)=-\psi^{a}(0),
	\qquad
	\bar{\psi}^{a}(z_a)=-\bar{\psi}^{a}(0),
	\end{aligned}
	\end{align}
	and gives rise to $Z(i\zeta_az_a)$. The sign factor $\zeta_a$ takes the value of $\pm 1$ for $a=\pm$
	
	In the Fourier basis
	\begin{align}\label{eq:FT}
	\begin{aligned}
	\psi^a(\e_n^a)
	=
	\frac{1}{z_a}
	\int_0^{z_a}
	dt' \,
	\psi^a(t') e^{i \e_n^a t'},
	\qquad
	\e_n^a=\frac{2\pi}{z_a} (n+\frac{1}{2}),
	\end{aligned}
	\end{align}
	Eq.~\ref{eq:ZZ} can be rewritten as
	\begin{align}\label{eq:ZZn}
	\begin{aligned}
	K(\eta,t)
	=&
	\left\langle 
	\int \D (\bpsi, \psi)  
	\exp 
	\left\lbrace 
	\begin{aligned}
	&i\sum_{a,n} z_a
	\bpsi_i^{a}(\e_n^a)  (\e_n^a e^{-i\e_n^a \delta z_a} \delta_{ij}-\zeta_a h_{ij}) \psi_j^{a}(\e_n^a)
	\\
	&-\frac{i}{2} \sum_{a}
	\zeta_a z_a
	\bpsi_i^{a} (\e_n^a)\bpsi_j^{a} (\e_{n'}^a) V_{ij;kl}
	\psi_k^{a} (\e_{n'-m}^a)  \psi_l^{a}	(\e_{n+m}^a)
	\end{aligned}	
	\right\rbrace 
	\right\rangle .
	\end{aligned}	
	\end{align}
	We note that the phase factor $e^{-i\e_n^a \delta z_a}$ comes from the time discretization of the path integral, with $\delta z_a \rightarrow 0$ being the time discretization interval of the forward (backward) path for $a=+$ ($a=-$).
	
	We then introduce a bosonic field $\phi$ to decouple the four-fermion interaction term by Hubbard–Stratonovich (H.S.) transformation:
	\begin{align}\label{eq:Zt}
	\begin{aligned}
	&
	\exp 
	\left\lbrace 
	-\frac{i}{2} \sum_{a=\pm}
	\zeta_a z_a
	\bpsi_i^{a} (\e_n^a)\bpsi_j^{a} (\e_{n'}^a) V_{ij;kl}
	\psi_k^{a} (\e_{n'-m}^a)  \psi_l^{a}	(\e_{n+m}^a)
	\right\rbrace 
	\\
	=&
	\frac{1}{Z_{\phi}}
	\int D\phi
	\exp 
	\left\lbrace 
	\begin{aligned}
	\frac{i}{2} \sum_{a=\pm} 
	\zeta_a z_a  \phi_{il}^a(-\ww_m^a)  V^{-1}_{ij;kl} \phi_{jk}^a(\ww_m^a)
	+
	i\sum_{a=\pm} \zeta_a z_a
	\phi_{il}^a(-\ww_m^a)
	\bpsi_i^{a} (\e_n^a) \psi_l^{a}	(\e_{n+m}^a)
	\end{aligned}
	\right\rbrace,
	\end{aligned}	
	\end{align}
	where $\ww_m^a=2\pi m/z_a$ is the bosonic Matsubara frequency and $Z_{\phi}$ is the normalization constant given by
	\begin{align}
	Z_{\phi}
	=
	\int D\phi
	\exp 
	\left( 
	\frac{i}{2} \sum_{a} 
	\zeta_a z_a
	\phi_{il}^a(-\ww_m^a) 
	V^{-1}_{ij;kl} 
	\phi_{jk}^a(\ww_m^a)
	\right) .
	\end{align}
	$V^{-1}$ is defined as
	\begin{align}
	\sum_{jk} V^{-1}_{i'j;kl'} V_{ji;lk} =\delta_{ii'}\delta_{ll'},
	\end{align}
	and is assumed to exist. Otherwise, one can decouple the interactions in a different channel and proceed in an analogous manner. 
	
	Ensemble averaging the $h-$dependent term in Eq.~\ref{eq:ZZn} over the random matrix $h$, we arrive at an effective four-fermion term which can be  decoupled by introducing a Hermitian matrix field $Q$:
	\begin{align}
	\begin{aligned}
	&\left\langle 
	\exp 
	\left( 
	-
	i \suml{a}
	\zeta_a z_a
	\bpsi_{i}^{a} (\e_n^a)
	h_{ij} 
	\psi_{j}^{a} (\e_n^a)
	\right) 
	\right\rangle
	=\,
	\frac{1}{Z_h} \int \D h \exp \left(-\frac{N}{2J^2} \Tr h^2   
	-
	i \suml{a}
	\zeta_a z_a
	\bpsi_{i}^{a} (\e_n^a)
	h_{ij} 
	\psi_{j}^{a} (\e_n^a)
	\right) 
	\\ 
	=\,&
	\exp 
	\left(-\suml{a,b} \frac{J^2}{2N} \zeta_a \zeta_b z_a z_b
	\bpsi_{i}^{a} (\e_n^a) \psi_{j}^{a} (\e_n^a) 
	\bpsi_{j}^{b} (\e_{n'}^b) \psi_{i}^{b} (\e_{n'}^b)  
	\right) 
	\\
	=&
	\frac{1}{Z_Q}
	\int \D Q
	\exp
	\left\lbrace 
	-
	\frac{N}{2J^2}
	Q^{ab}_{nn'}
	Q^{ba}_{n'n}
	+
	\suml{ab}
	Q^{ab}_{nn'}
	\left( 
	\suml{i}
	z_a \zeta_b\psi_{i}^{b} (\e_{n'}^b) 
	\bpsi_{i}^{a} (\e_{n}^a)
	\right) 
	\right\rbrace. 
	\end{aligned}	
	\end{align}
	Here
	$Z_Q=\int \D Q \exp \left(-\frac{N}{2J^2} \Tr	Q^2 \right)  $ is the normalization constant.
	$Q^{ab}_{nn'}$ is a Hermitian matrix in the Matsubara frequency space (labeled by $n$) and the replica space (labeled by $a$).
	
	Combining everything, the SFF defined in Eq.~\ref{eq:K} can now be expressed as
	\begin{align}\label{eq:Ft-0}
	\begin{aligned}
	K(\eta,t)
	=&
	\frac{1}{Z_{\phi}Z_Q}
	\int \D Q
	\exp
	\left( 
	-
	\frac{N}{2J^2}
	Q^{ab}_{nn'}
	Q^{ba}_{n'n}
	\right) 
	\int D\phi
	\exp 
	\left( 
	\frac{i}{2} \sum_{a=\pm} 
	\zeta_a z_a \phi_{il}^a(-\ww_m^a)  V^{-1}_{ij;kl} \phi_{jk}^a(\ww_m^a)
	\right) 
	\\
	\times&
	\int \D (\bpsi, \psi)  
	\exp 
	\left[ 
	i\sum_{a=\pm} z_a
	\bpsi_i^{a}(\e_n^a) 
	\left(  
	\e_n^a e^{-i\e_n^a \delta z_a} \delta_{nn'}\delta_{ab}\delta_{ij}
	+iQ^{ab}_{nn'}\zeta_b\delta_{ij}
	+\zeta_a\phi_{ij}^a(\e_n^a-\e_{n'}^a)\delta_{ab}
	\right) 
	\psi_j^{b}(\e_{n'}^b)
	\right] .
	\end{aligned}	
	\end{align}
	Integrating out the fermionic fields $\psi$ and $\bpsi$, we are left with a theory of the matrix field $Q$ and the interaction decoupling field $\phi$:
	\begin{subequations}	\label{eq:K-1}
		\begin{align}
		&
		K(\eta,t)=\,
		\frac{1}{Z_{\phi}Z_Q}
		\int \D \phi \int \D Q \exp \left( -S_{\phi 0} [\phi]- S[Q,\phi]\right), 
		\\
		\label{eq:Sphi0}
		&
		S_{\phi 0} [\phi]
		=\,	
		-\frac{i}{2} \sum_{a=\pm} 
		\zeta_a z_a \phi_{il}^a(-\ww_m^a)  V^{-1}_{ij;kl} \phi_{jk}^a(\ww_m^a),
		\\
		\label{eq:SQP}
		&
		S[Q,\phi]
		=\,
		\frac{N}{2J^2}
		\Tr  Q^2
		-
		\Tr
		\ln
		\left[
		\left( 
		\mathcal{E}
		+
		iQ\sigma^3
		\right) 
		\otimes I_f
		+
		\Phi (\sigma^3\otimes I_f)
		\right]	
		+
		const..
		\end{align}
	\end{subequations}
	Here the first $\Tr$ in Eq.~\ref{eq:SQP} denotes the trace over the Matsubara frequency space (indexed by $n$) and the replica space  (indexed by $a$), while the second $\Tr$ acts on the flavor space (indexed by $i$) in addition to the frequency and replica spaces.
	$I_{f}$ denotes the $N \times N$ identity matrix in the flavor space, while
	$\sigma^3$ represents a direct product of the third Pauli matrix in the replica space and an identity matrix in the Matsubara frequency space.
	Matrices $\mathcal{E}$ and $\Phi$ are defined as
	\begin{align}
	\begin{aligned}
	\mathcal{E}^{ab}_{nn'}=\delta_{ab}\delta_{nn'}\e_n^ae^{-i\e_n^a \delta z_a},
	\qquad
	\Phi^{ab}_{ij;nn'}=\delta_{ab}\phi_{ij}^a(\e_n^a-\e_{n'}^a).
	\end{aligned}
	\end{align} 
	
	
	\subsection{The saddle points}
	
	To proceed, we assume that the interaction strength is weak enough so that
	the decoupling field $\phi$ do not disturb the matrix field's saddle point $Q_{sp}$. In other words, we assume the saddle point $Q_{sp}$ of the interacting theory is the same to that of the non-interacting case  whose action takes the form:
	\begin{align}\label{eq:S0}
	\begin{aligned}
	&S[Q,0]
	=\,	
	\frac{N}{2J^2}
	\Tr Q^2
	-
	N
	\Tr 
	\ln
	\left( 
	-i\mathcal{E} \sigma^3
	+
	Q
	\right)
	+\text{const.}.
	\end{aligned}
	\end{align}
	Taking the variation of this action with respect to $Q$, we arrive at the saddle point equation
	\begin{align}
	\begin{aligned}
	\frac{1}{J^2} Q_{sp}
	=
	\left( 
	-i\mathcal{E} \sigma^3
	+
	Q_{sp}
	\right) ^{-1} .
	\end{aligned}
	\end{align}
	It is straightforward to see that diagonal matrix $\Lambda$ with element 
	\begin{align}\label{eq:Lambda}
	\begin{aligned}
	&
	\Lambda_{nn'}^{ab}
	=
	\frac{1}{2}
	\left( 
	i\zeta_a\e_n^a e^{-i\e_n^a\delta z_a}
	+
	s_n^a
	\sqrt{4J^2 -(\e_n^ae^{-i\e_n^a\delta z_a})^2}
	\right)\delta_{ab}\delta_{nn'} ,
	\qquad
	s_n^a=\pm 1.
	\end{aligned}
	\end{align}
	is a solution to the saddle point equation. Here $s_n^a$ can takes the value of $+1$ or $-1$ (when $|\e_n^a|<2J)$~\cite{PRL}, and different choices of $\left\lbrace s_n^a\right\rbrace $ give rise to different diagonal saddle points.

	In the limit $\eta ,\delta z_a \rightarrow 0$, the non-interacting action in Eq.~\ref{eq:S0} is invariant under the transformation $Q\rightarrow U^{\dagger} Q U$ for any unitary matrix $U$ that satisfies $U\mathcal{E}\sigma^3 U^{\dagger}=\mathcal{E}\sigma^3$. The symmetry transformation $U$ is given by a direct product of multiple rotations $U=\prod_{n} U_n$, where $U_n$ is any $U(2)$ rotation that acts only on the following matrix subblock
	\begin{align}\label{eq:Un}
	\begin{aligned}
	\begin{bmatrix}
	Q^{++}_{n,n} & Q^{+-}_{n,-n-1}
	\\
	Q^{-+}_{-n-1,n} & Q^{--}_{-n-1,-n-1}
	\end{bmatrix}
	\rightarrow
	U_n^{\dagger}
	\begin{bmatrix}
	Q^{++}_{n,n} & Q^{+-}_{n,-n-1}
	\\
	Q^{-+}_{-n-1,n} & Q^{--}_{-n-1,-n-1}
	\end{bmatrix}
	U_n.
	\end{aligned}
	\end{align}
	More specifically, the matrix element $U^{ab}_{nm}$ takes nonzero value only when $n=m$ for $a=b$, and $n=-m-1$ for $a=-b$.
	Applying the symmetry transformations $U\in \prod_{n} U(2)$ to the diagonal saddle point $\Lambda$ can generate new saddle points $Q_{sp}=U^{\dagger} \Lambda U$.
	For finite $\eta $, the argument above no longer applies.	
	The presence of nonzero $\eta$ in the matrix $\mathcal{E}$ breaks the symmetry, and allows us to select one dominate saddle point.

	We now evaluate the non-interacting action at a diagonal saddle point $\Lambda$:
	\begin{align}\label{eq:S0-1}
	\begin{aligned}
	S[ \Lambda,0]
	=\,	
	\sum_{a,n}
	S_0(\e_n^ae^{-i\e_n^a\delta z_a})
	+\text{const.},
	\qquad
	S_0(\e_n^a e^{-i\e_n^a\delta z_a})=
	\frac{N}{2J^2}\left( \Lambda_{nn}^{aa}\right)^2
	-
	N
	\ln
	\left( 
	\e_n^a e^{-i\e_n^a\delta z_a}
	+
	i\Lambda_{nn}^{aa} \zeta_a
	\right)
	.
	\end{aligned}
	\end{align}
	Taking derivative of  $S_0(\e_n^ae^{-i\e_n^a\delta z_a})$ with respect to $\e_n^ae^{-i\e_n^a\delta z_a}$, we have
	\begin{align}
	\begin{aligned}
	&\frac{dS_0(\e_n^ae^{-i\e_n^a\delta z_a})}{d\e_n^a e^{-i\e_n^a\delta z_a}}
	=
	\frac{\partial S_0}{\partial \e_n^a e^{-i\e_n^a\delta z_a}}
	+
	\frac{\partial S_0}{\partial \Lambda_{nn}^{aa}}\frac{d \Lambda_{nn}^{aa}}{d\e_n^a e^{-i\e_n^a\delta z_a}}
	=
	\frac{\partial S_0}{\partial \e_n^a e^{-i\e_n^a\delta z_a}}
	=
	-N\dfrac{1}{
		\e_n^a e^{-i\e_n^a\delta z_a}
		+
		i\Lambda_{nn}^{aa} \zeta_a}
	=
	i N\zeta_a  \frac{\Lambda_{nn}^{aa}}{J^2}.
	\end{aligned}
	\end{align}
	In the last equality, the fact that $ \Lambda$ solves the saddle point equation has been used. 
	With the help of the equation above, we find that, to the leading order in an expansion in $\e_n^a/J$, 
	\begin{align}\label{eq:S0-2}
	\begin{aligned}
	S_0(\e_n^ae^{-i\e_n^a\delta z_a})
	=
	\frac{N}{J}	
	i\zeta_a 
	s_n^a
	\e_n^ae^{-i\e_n^a\delta z_a}
	+
	S_0(0),
	\qquad
	|\e_n^a|\ll J.
	\end{aligned}
	\end{align}
	We have used that, in the low energy limit $|\e_n^a| \ll J$, the saddle point $\Lambda_{nn}^{aa}$ can be approximated as $Js_n^a$. 
	In the following, we will focus on the low energy sector of the theory, sufficient for the investigation of SFF at $t\gg J^{-1}$, to ignore the correction from nonuniversal density of states. In particular, by considering only the low-energy sector, the bare average single-particle density of states $\braket{\nu(\e)}=\sqrt{4J^2-\e^2} /(2\pi J^2)$ is now approximated by a constant $\braket{\nu(\e)}\approx 1/\pi J$ at low energy $\e$.
	
	From Eqs.~\ref{eq:S0-1} and ~\ref{eq:S0-2}, one can see that the dominate saddle point can be determined by minimizing $S_0(\e_n^ae^{-i\e_n^a\delta z_a})$. In this problem, we assume $|\delta z_a|/t^2 \ll J/N$, and use the phase factor $e^{-i\e_n^a\delta z_a}$ only as a convergence-generating factor. We then discuss separately two cases: $\eta/(t^2+\eta^2) \gtrsim  J/N$ and $\eta/(t^2+\eta^2) \ll  J/N$ (including $\eta=0$). In the case of $\eta/(t^2+\eta^2) \gtrsim  J/N$, using $\im \e_n^a= \zeta_a 2\pi(n+1/2) \eta/(t^2+\eta^2)$, we find that the  diagonal saddle point $\Lambda^{(0)}$ with matrix element \begin{align}\label{eq:Lambda0}
	(\Lambda^{(0)})^{aa'}_{nn'}=J\sgn (n+1/2)\delta_{aa'}\delta_{nn'},
	\end{align} 
	dominates. By contrast, for $\eta/(t^2+\eta^2) \ll  J/N$, the actions of various saddle points differ by phase factors, and their contributions are equally important. 
	In the following, we will call $\Lambda^{(0)}$ the standard saddle point, and the remaining diagonal saddle points the nonstandard ones. 
	We refer to Refs.~\cite{AndreevAltshuer,Kamenev-GUE,Kamenev-Keldysh} for a detailed discussion of the role played by the standard  and nonstandard saddle points in the context of single-particle level statistics. It has been shown there that the soft mode fluctuations around the standard saddle point give rise to the smoothed part of the single-particle level correlation, whereas those around nonstandard saddle points lead to the nonperturbative-oscillatory part (see also the discussion in Sec.~\ref{sec:high}).
	

	\subsection{Quadratic fluctuations and the non-interacting SFF}\label{sec:quad}
	
	We will now examine the fluctuations around the diagonal saddle points.
	Expanding the action $S[Q,\phi]$ up to quadratic order in decoupling field $\phi$ and $\delta Q=Q-\Lambda$, which represents the fluctuation around the diagonal saddle point $\Lambda$ (Eq.~\ref{eq:Lambda}), we find
	\begin{align}\label{eq:S2}
	\begin{aligned}
	&
	\delta S[\delta Q,\phi;\Lambda]
	\equiv
	S[\Lambda+\delta Q,\phi]-S[\Lambda,0]
	=
	\suml{a,b,m,n}
	M_{nm}^{ab} \delta Q_{nm}^{ab} \delta Q_{mn}^{ba}
	+
	\suml{a,m,n}
	\bar{M}^{a}_{nm} 
	\delta Q_{nm}^{aa}
	\suml{i}\phi^{a}_{ii}(\e_m^a-\e_n^a)
	\\
	&
	+
	\suml{a}  M'^a \suml{i} \phi^{a}_{ii}(0)
	+
	\frac{1}{2}
	\suml{a,m} \tilde{M}^a_m \sum_{ij}
	\phi^{a}_{ij}(-\ww_m^a) \phi^{a}_{ij}(\ww_m^a),
	\end{aligned}	
	\end{align}
	where
	\begin{align}\label{eq:Mab}
	\begin{aligned}
	&M^{ab}_{nm}
	=\, 
	\frac{N}{2J^2} 
	\left( 1-J^2
	G^{a}_n
	G^{b}_m \right) 
	\approx 
	\frac{N}{2J^2} 
	\left[ 
	1+s_n^a s_m^b
	+\frac{i}{2J}  \left( s_n^a \zeta_b \e_m^be^{-i\e_m^b\delta z_b}  +s_m^b  \zeta_a \e_n^a e^{-i\e_n^a\delta z_a} \right)
	\right] ,
	\\
	&\bar{M}^{a}_{nm}
	=\, 
	iG^{a}_n
	G^{a}_m
	\approx 
	-\frac{i}{J^2} s_n^a s_m^a,
	\qquad
	M'^a=-\suml{n} G^{a}_n
	\approx
	\frac{i}{J} \suml{n} s_n^a,
	\qquad
	\tilde{M}^a_m
	=\, 
	\sum_{n}
	G^{a}_n
	G^{a}_{n+m}
	\approx
	-\frac{1}{J^2} 
	\sum_{n} s_n^a s_{n+m}^a.
	\end{aligned}
	\end{align}
	$G$ is defined as
	\begin{align}
	G^{a}_n
	=
	\frac{1}{\zeta_a \e_n^a e^{-i\e_n^a\delta z_a} +i (\Lambda^{(s)})_{nn}^{aa} },
	\end{align}
	and from the saddle point equation satisfies $G^{a}_n=-i\Lambda_{nn}^{aa}/J^2$. 
	$M$, $\bar{M}$, $M'$ and $\tilde{M}$ depend on the choice of the diagonal saddle point $\Lambda$, and
	in Eq.~\ref{eq:Mab} they are approximated by the leading order terms in the expansion in energy $\e_n^a/J$.

	Let us first consider the case of  $\eta/(t^2+\eta^2) \ll  J/N$. Depending on the value of the matrix kernel $M$, the fluctuations $\delta Q$ can be divided into two types. 
	\begin{enumerate}
		\item 
		If $s_n^a=s_m^b$, $M^{ab}_{nm}$ is given by, to the leading order in energy, 
		\begin{align}\label{eq:M_m}
		M^{ab}_{nm} \approx N/J^2,
		\end{align}
		and the corresponding modes are the massive modes. 
		
		\item 
		If $s_n^a=-s_m^b$, we have instead
		\begin{align}\label{eq:M_s}
		M^{ab}_{nm} 
		\approx
		\frac{N}{4J^3} i s_n^a\left(  \zeta_b \e_m^be^{-i\e_m^b\delta z_b}  -  \zeta_a \e_n^a e^{-i\e_n^a\delta z_a} \right),
		\end{align}
		and the associated modes are soft modes which can be generated by unitary transformation 
		\begin{align}\label{eq:soft}
		Q=R^{-1} \Lambda R,
		\end{align}
		with $R$ being an unitary matrix in the replica and Matsubara frequency spaces.
		We note that, without the $\mathcal{E} \sigma^3$ term, the non-interacting action $S[Q,0]$ in Eq.~\ref{eq:S0} is invariant under any unitary transformation $Q \rightarrow R^{-1} Q R$. The soft modes are the Goldstone modes associated with the unitary transformation symmetry explicitly broken by the $\mathcal{E} \sigma^3$ term. 
		In the special case where $n=-m-1$ and $a=-b$, Eq.~\ref{eq:M_s} reduces to $M^{a,-a}_{n,-n-1}=0$.  
		This corresponds to the soft mode generated by symmetry transformation $Q=U^{-1} \Lambda U$ with $U$ satisfying $U^{-1}\mathcal{E}\sigma^3 U=\mathcal{E} \sigma^3$. 
		We note that, for this particular type of soft mode, called zero mode in the following, fluctuation correction to the action vanishes $\delta S=0$ not just at the quadratic order but at all higher orders as well. 
	\end{enumerate}
	
	For the case of $\eta/(t^2+\eta^2)\gtrsim  J/N$, the situation is similar. There also exist two types of fluctuations -- the massive mode and the soft mode with the kernels given by Eq.~\ref{eq:M_m} and Eq.~\ref{eq:M_s}, respectively. However, the presence of nonvanishing $\eta$ breaks the symmetry, and $M^{ab}_{nm}$ is no longer vanishing for the ``zero modes". (For simplicity, we still call fluctuation $\delta Q^{ab}_{nm}$ with $n=-m-1$, $a=-b$ and $s_n^a=-s_m^b$ the zero mode, although $\delta S$ is no longer zero for nonzero $\eta$.) In particular,
	we have  
	\begin{align}\label{eq:M_z}
	M^{a,-a}_{n,-n-1} 
	\approx
	\frac{N}{4J^3} i s_n^a\zeta_a\left( \e_{n}^{-a} e^{i\e_{n}^{-a}\delta z_{-a} } -  \e_n^a e^{-i\e_n^a\delta z_a} \right)
	\approx
	\frac{N}{J^3} s_n^a\frac{\eta}{t^2+\eta^2} \pi(n+1/2),
	\end{align}
	which is nonvanishing ($M^{a,-a}_{n,-n-1} \gtrsim 1$) but much smaller compared with the matrix kernel element $M^{ab}_{nm}$ associated with nonzero mode in the limit of $\eta\rightarrow 0^+$.
	
	Ignoring fluctuation correction beyond quadratic order,  the contribution to the SFF $K(\eta,t)$ from saddle point $\Lambda^{(s)}$ and the fluctuations $\delta Q^{(s)}$ around it can be expressed as
	\begin{align}\label{eq:K-2}
	\begin{aligned}
	K'(\eta,t;\Lambda^{(s)})
	=\,
	\frac{e^{-S[\Lambda^{(s)},0]}}{Z_{\phi}Z_Q}
	\int \D \phi
	e^{ -S_{\phi 0}[\phi]}
	\int \D \delta Q^{(s)} \exp \left( -\delta S[\delta Q^{(s)},\phi;\Lambda^{(s)}]\right).
	\end{aligned}
	\end{align}
	$S[\Lambda^{(s)},0]$ (Eq.~\ref{eq:S0}) and $\delta S[\delta Q^{(s)},\phi;\Lambda^{(s)}]$ (Eq.~\ref{eq:S2}) represent, respectively, the non-interacting  action of the saddle point $\Lambda^{(s)}$ and the associated fluctuation correction in the presence of $\phi$.
	
	In the non-interacting case, the equation above simply becomes
	\begin{align}\label{eq:K0-1}
	\begin{aligned}
	&K_0'(\eta,t;\Lambda^{(s)})
	=\,
	e^{-S[\Lambda^{(s)},0]}
	\dfrac{\int \D \delta Q^{(s)} \exp 
		\left( -	\suml{a,b,m,n}
		M_{nm}^{ab} (\delta Q^{(s)})_{nm}^{ab} (\delta Q^{(s)})_{mn}^{ba}
		\right)}
	{\int \D \delta Q^{(s)} \exp \left(-\frac{N}{2J^2} \suml{a,b,m,n} (\delta Q^{(s)})_{nm}^{ab} (\delta Q^{(s)})_{mn}^{ba} \right)  }.
	\end{aligned}
	\end{align} 
	Integration over $\delta Q^{(s)}$ leads to
	\begin{align}\label{eq:K0-2}
	\begin{aligned}
	K_0'(\eta,t;\Lambda^{(s)})
	=
	e^{-S[\Lambda^{(s)},0]}
	\exp 
	\left[  -	
	\sum_{a,b,m,n}'
	\ln \left( \frac{2J^2}{N} M_{nm}^{ab}\right) 
	\right] 
	\mathcal{Z}_0^{(s)}.
	\end{aligned}
	\end{align} 
	Here the summation $\sum_{a,b,m,n}'$ excludes the zero modes, whose contribution is denoted by $\mathcal{Z}_0^{(s)}$. 
	In the limit of $\eta \rightarrow 0$, at the quadratic level, it might seem that zero modes' contribution is divergent. However, taking into account the fact that $\delta Q^{(s)}$ associated with zero modes stay on the corresponding saddle point manifold generated by symmetry rotation $U$: $\delta Q^{(s)}=U^{-1}\Lambda^{(s)} U-\Lambda^{(s)}$,
	one finds
	\begin{align}\label{eq:K0-2a}
	\begin{aligned}
	&\mathcal{Z}_0^{(s)}
	=\,
	\mathcal{V}_0^{(s)}
	\exp 
	\left[ 
	-	\frac{1}{2}
	\mathcal{N}_0^{(s)}
	\ln \left( \frac{2J^2}{N} \pi \right) 
	\right] .
	\end{aligned}
	\end{align} 
	$\mathcal{V}_0^{(s)}$ represents the volume of the associated saddle point manifold, while $\mathcal{N}_0^{(s)}$ denotes the  number of the corresponding zero modes. See Refs.~\cite{Kamenev,Winer,PRL} for more details about zero modes's contribution. 
	
	
	From Eq.~\ref{eq:K0-2}, we can see that the massive modes yield a nonessential constant to the non-interacting SFF $K_0(\eta,t)$ at the leading order.
	We believe that the influence of interactions on the massive modes is not significant and therefore will focus on the soft modes, and in particular the soft modes associated with the standard saddle point $\Lambda^{(0)}$. 
	In the case of $\eta/(t^2+\eta^2) \gtrsim  J/N$, the standard saddle point dominates over all remaining nonstandard ones, and we may approximate $K(\eta,t)$ by $K'(\eta,t;\Lambda^{(0)})$, i.e., the contribution to the SFF from  $\Lambda^{(0)}$ and the fluctuations around it. 
	By contrast, for $\eta/(t^2+\eta^2) \ll J/N$, the contributions associated with nonstandard saddle points must also be included.  
	In the following, we will consider the former case, and evaluate $K'(\eta,t;\Lambda^{(0)})$ for both the interacting and non-interacting theories.

	\subsection{Nonlinear $\sigma$-model for the standard saddle point}
	
	In this subsection, we investigate the interactions effect on the soft mode fluctuations around the standard saddle point $\Lambda^{(0)}$.
	We parameterize
	the soft mode around $\Lambda^{(0)}$ as $Q=R^{-1} \Lambda^{(0)} R$, where $R$ is a unitary rotation belonging to the coset space $U(2N_{\e})/U(N_{\e}) \times U(N_{\e})$, with $N_{\e}$ being the total number of Matsubara frequencies.
	Inserting $Q=R^{-1} \Lambda^{(0)} R$ into the action $S[Q,\phi]$ (Eq.~\ref{eq:SQP}) and expanding in terms of gradient of rotation matrix $R$ and decoupling field $\phi$, we arrive at the low energy effective action for the soft modes:
	\begin{align}\label{eq:NLSM}
	\begin{aligned}
	&\delta S[Q,\phi; \Lambda^{(0)}]
	\equiv
	S[Q,\phi]-S[\Lambda^{(0)},0]
	\\
	=&
	i\frac{N}{J^2}
	\Tr
	\left[ \sigma^3 \mathcal{E} \left( Q-\Lambda^{(0)} \right)   \right]
	+
	i\frac{1}{J^2} \suml{a,n,n'} Q^{aa}_{nn'}\suml{i}\phi^a_{ii}(\e_{n'}^a-\e_n^a)
	+
	\frac{1}{2}  
	\suml{a,m}
	\tilde{M}^a_m
	\suml{ij}\phi^{a}_{ij}(-\ww_m^a)  \phi^{a}_{ji}(\ww_m^a)
	.
	\end{aligned}
	\end{align}
	The low energy saddle point $\Lambda^{(0)}$ is given by approximately
	Eq.~\ref{eq:Lambda0}
	which leads to the following constraints for matrix field $Q$
	\begin{align}\label{eq:cons}
	\Tr Q=0, 
	\qquad  
	\frac{1}{J^2}Q^2=I.
	\qquad
	Q^{\dagger}=Q.
	\end{align}
		The coefficient $\tilde{M}^a_m$ for the $\phi^2$ term is defined in Eq.~\ref{eq:Mab}, and from the low energy approximation is given by
		\begin{align}
		\begin{aligned}
		\tilde{M}^a_m=-\frac{1}{J^4}\sum_{n}(\Lambda^{(0)})_{nn}^{aa}(\Lambda^{(0)})_{n+m,n+m}^{aa}
		\approx 
		-\frac{z_a E_{\msf{UV}}}{\pi J^2} +O(\e_m^a),
		\end{aligned}
		\end{align}
		with $E_{\msf{UV}} \sim J$ being the ultraviolet energy cutoff ($|\e_n^{\pm}|\leq E_{\msf{UV}}$).
	
	The contribution to the SFF from the standard saddle point $\Lambda^{(0)}$ as well as the soft modes around it can then be obtained by substituting Eq.~\ref{eq:NLSM} into Eq.~\ref{eq:K-2} (ignoring the unessential contribution from massive modes). The integration $\int \D Q$ is now over the soft mode manifold characterized by the constraints in Eq.~\ref{eq:cons}.
	
	To proceed, we parametrize the Hermitian matrix $Q$ as:
	\begin{align}\label{eq:para}
	\begin{aligned}
	&
	\begin{array}{cc}
	\quad m \geq 0 &  \qquad\qquad m<0
	\end{array}
	\\
	Q_{nm}
	=&
	J
	\begin{bmatrix}
	\sqrt{I- X X^\dagger } &  X
	\smallskip
	\\
	X^\dagger   & -\sqrt{I- X^\dagger X } 
	\end{bmatrix}
	\begin{array}{c}
	n\geq 0\smallskip
	\\
	n<0
	\end{array},
	\end{aligned}
	\end{align}		
	where $X^{ab}_{nm}$ ($X^{\dagger}\,^{ab}_{nm}$) is an unconstrained $N_{\e} \times N_{\e}$ complex matrix labeled by replica indices $a,b$, nonnegative (negative) row frequency index $n$ and negative (nonnegative) column frequency index $m$. It is straightforward to prove that the constraints in Eq.~\ref{eq:cons} are automatically satisfied using this parametrization.
	
	We then substitute Eq.~\ref{eq:para} into the effective action Eq.~\ref{eq:NLSM} and expand in terms of $X$ and $X^{\dagger}$.
	To simplify the power counting of small perturbation parameter $J/N$, we also rescale $X$ and $X^{\dagger}$ by
	\begin{align}
	\begin{aligned}\label{eq:rescale}
	X\rightarrow \sqrt{\frac{2J}{N}} X,
	\qquad
	X^{\dagger} \rightarrow \sqrt{\frac{2J}{N}}  X^{\dagger}.
	\end{aligned}
	\end{align}
	After the substitution, expansion and rescaling, we find that the action in Eq.~\ref{eq:NLSM} becomes
	\begin{align}\label{eq:SX24}
	\begin{aligned}
	&\delta S[Q,\phi; \Lambda^{(0)}]=
	S_X^{(2)}+S_X^{(4)}+S_{\phi},
	\\
	&S_{\phi}
	=
	\frac{1}{2}  
	\suml{a,m}
	\tilde{M}^a_m
	\suml{ij}\phi^{a}_{ij}(-\ww_m^a)  \phi^{a}_{ji}(\ww_m^a),
	\\
	&S_X^{(2)} 
	=\, 
	\suml{abcd}
	\suml{m,u\geq 0>n,v}
	X^{\dagger}\,^{ab}_{nm}
	\mathcal{K}^{ba;dc}_{mn;vu}
	X^{cd}_{uv}
	+
	\suml{ab}\suml{m\geq 0>n}
	\left( 
	\bar{J}\,^{ab}_{nm}
	X\,^{ba}_{mn}
	+
	X^{\dagger}\,^{ab}_{nm}
	J\,^{ba}_{mn}
	\right) ,
	\\
	& S_X^{(4)}
	= 
	i\frac{1}{4}\frac{J}{N}\suml{abcd}\suml{m,v \geq 0>n,u}
	X^{\dagger}\,^{ab}_{nm} X^{bc}_{mu}X^{\dagger}\,^{cd}_{uv} X^{da}_{vn}
	\left( \zeta_a \e_n^ae^{-i\e_n^a \delta z_a}-\zeta_b \e_m^be^{-i\e_m^b \delta z_b}+
	\zeta_c \e_u^ce^{-i \e_u^c \delta t_c}-\zeta_d \e_v^de^{-i\e_v^d \delta t_d}
	\right) ,
	\end{aligned}
	\end{align}	
	where
	\begin{align}\label{eq:MJ}
	\begin{aligned}
	\mathcal{K}^{ba;dc}_{mn;vu}
	=\, &
	\delta_{ad}\delta_{bc}
	i
	\left[ 
	(\zeta_a \e_n^ae^{-i \e_n^a \delta z_a}-\zeta_b \e_m^be^{-i\e_m^b \delta z_b}) 
	\delta_{nv} \delta_{mu} 
	-
	\delta_{nv} \frac{1}{N}\suml{i}\phi^{b}_{ii}(\e_m^b-\e_u^b)
	+
	\delta_{mu} \frac{1}{N}\suml{i} \phi^{a}_{ii}(\e_v^a-\e_n^a)
	\right] ,
	\\
	\bar{J}\,^{ab}_{nm}
	=\,&
	\delta_{ab}
	i  \sqrt{\frac{2N}{J}} \frac{1}{N} \suml{i}
	\phi^{a}_{ii}(\e_n^a-\e_m^a),
	\qquad
	J\,^{ba}_{mn}
	=\,
	\delta_{ab}
	i  \sqrt{\frac{2N}{J}} \frac{1}{N} 
	\suml{i}
	\phi^{a}_{ii}(\e_m^a-\e_n^a).
	\end{aligned}
	\end{align}
	All remaining terms of higher orders in $X$ and $X^{\dagger}$ are also of higher order in small parameter $J/N$. These higher order terms might give rise to non-negligible contribution to the SFF. However, they are ignored in the current study which focuses on the difference between the leading order fluctuation corrections in the interacting and non-interacting theories.
	We note that the first few leading terms in Eq.~\ref{eq:SX24} are consistent with the action in Eq.~\ref{eq:S2} for quadratic fluctuations (around $\Lambda^{(0)}$)
	if we equate $\delta Q_{nm}^{ab}$ with
	\begin{align}\label{eq:dQ}
	\delta Q_{nm}^{ab}=
	\begin{cases}
	\sqrt{(2J^3/N)} X^{ab}_{nm}, & n \geq 0 >m,\\
	\sqrt{(2J^3/N)} X^{\dagger}\,^{ab}_{nm}, &m \geq 0 >n.
	\end{cases}
	\end{align}
	
	
	\subsubsection{Feynman rules}
	
	Ignoring the interactions between matrix field $X$ and bosonic field $\phi$, the bare $X$ propagator is given by the inverse of matrix kernel $\mathcal{K}$ (Eq.~\ref{eq:MJ}) with $\phi=0$:
	\begin{align}\label{eq:GX0}
	\begin{aligned}
	(\GG^{(0)}_{X}\,)^{ab;b'a'}_{nm;m'n'}
	\equiv 
	\braket{X^{ab}_{nm}X^{\dagger}\,^{b'a'}_{m'n'}}_0
	=
	(\mathcal{K}^{-1})^{ab;b'a'}_{nm;m'n'}|_{\phi=0}
	=
	\frac{i}{\zeta_a\e_n^ae^{-i \e_n^a \delta z_a}-\zeta_b\e_m^be^{-i\e_m^b \delta z_b}}\delta_{nn'}\delta_{mm'}\delta_{aa'}\delta_{bb'}.
	\end{aligned}
	\end{align}
	The bare action for the bosonic field $\phi$ is given by the sum of $S_{\phi0}[\phi]$ (Eq.~\ref{eq:SQP}) originating from the decoupling of the four-fermion interactions 
	and  $S_{\phi}[\phi]$ (Eq.~\ref{eq:SX24}) arising from the $\Tr \ln$ expansion.
	It leads to the following bare propagator for $\phi$:
	\begin{align}\label{eq:Gphi0}
	\begin{aligned}
	&i\left[ \GG_{\phi}^{(0)}(\ww_m^a)\right]^{ab}_{ij;kl} 
	\equiv 
	\braket{\phi^{a}_{il}(\ww_m^a)\phi^{b}_{jk}(-\ww_m^b)}_0
	=
	i \zeta_a \frac{1}{z_a}(V'^a)_{ij;kl} \delta_{ab},
	\end{aligned}
	\end{align}
	where $V'\,^{a}$ is defined as
	\begin{align}\label{eq:Va}
	\begin{aligned}
	&(V'^{a})^{-1}_{ij;kl}
	=
	V^{-1}_{ij;kl}+i \zeta_a\frac{1}{z_a} M^a_m \delta_{ik}\delta_{lj}.
	\end{aligned}
	\end{align}
	
	\begin{figure}[h!]
		\centering
		\includegraphics[width=0.5\linewidth]{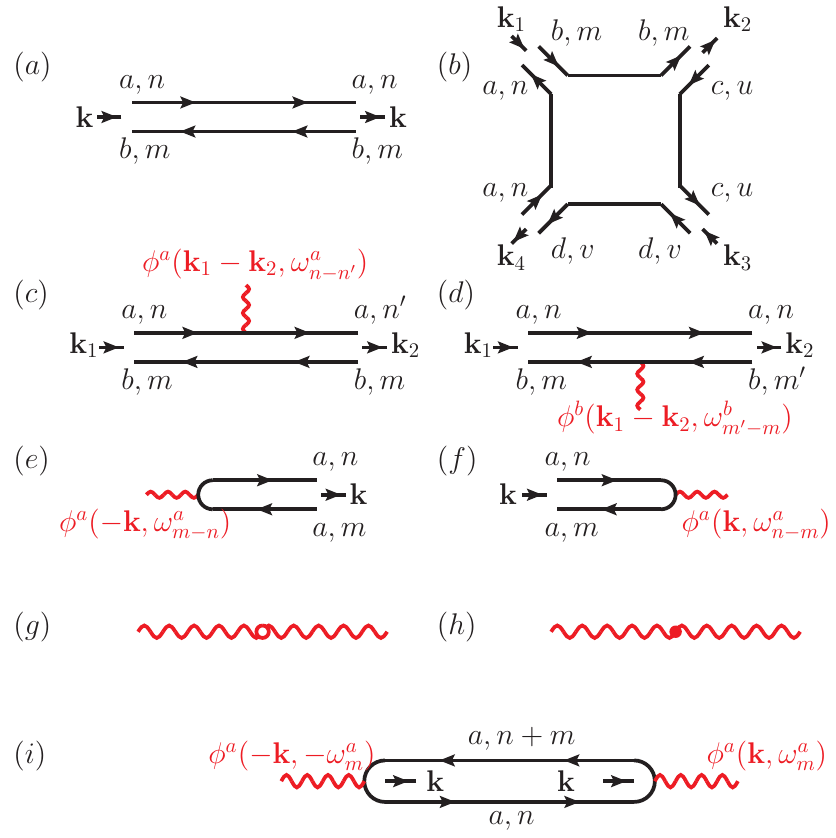}
		\caption{Feynman rules for both interacting RMT model and 2D disordered system: (a) shows a diagrammatic representation of the bare $X$ propagator; (b) illustrates the Hikami box that couples four $X$ matrices; (c)-(f) depict the interaction vertices coupling between the matrix field $X$ and the bosonic field $\phi$. (g) and (h) represent, respectively, the bare and interaction dressed $\phi$ propagators. (i) shows the leading order self-energy diagram for the bosonic field $\phi$. Note that the momentum labels are used for the 2D disordered system only.}
		\label{fig:p1}
	\end{figure}
	
	In Figs.~\ref{fig:p1}(a) and~(g), we show the diagrammatic representations of the bare $X$ propagator $(\GG^{(0)}_{X})^{ab;ba}_{nm;mn}$ and the bare $\phi$ propagator $\GG_{\phi}^{(0)}$.	 We use two black solid lines with arrows pointing in the opposite directions to indicate the matrix fields $X$ and $X^{\dagger}$. The short arrows in between these two lines are introduced to distinguish $X$ (incoming arrow) and $X^{\dagger}$ (outgoing arrow). The red wavy line represents diagrammatically the bosonic field $\phi$, and the one with a open dot in the middle indicates the bare $\phi$ propagator $\GG_{\phi}^{(0)}$ (see Fig.~\ref{fig:p1}(g)). Fig.~\ref{fig:p1}(b) depicts the four-point vertex which comes from the quartic action $S_X^{(4)}$, while Figs.~\ref{fig:p1}(c)-(f) show the interaction vertices arising from the $\phi$-dependent part of the action $S_X^{(2)}$. The corresponding amplitudes are, in respective order,
	\begin{align}
	\begin{aligned}
	&
	(b)=-i\frac{1}{2}\frac{J}{N}
	X^{\dagger}\,^{ab}_{nm} X^{bc}_{mu}X^{\dagger}\,^{cd}_{uv} X^{da}_{vn}
	\left( \zeta_a \e_n^ae^{-i\e_n^a \delta z_a}-\zeta_b \e_m^be^{-i\e_m^b \delta z_b}+
	\zeta_c \e_u^ce^{-i \e_u^c \delta t_c}-\zeta_d \e_v^de^{-i\e_v^d \delta t_d}
	\right),
	\\
	&
	(c)=i\frac{1}{N}\suml{i}\phi^{a}_{ii}(\e_n^a-\e_{n'}^a)
	X^{\dagger}\,^{ba}_{mn}
	X^{ab}_{n'm},
	\qquad
	(d)=-i
	\frac{1}{N}\suml{i} \phi^{b}_{ii}(\e_{m'}^b-\e_m^b)
	X^{\dagger}\,^{ba}_{mn}
	X^{ab}_{nm'},
	\\
	&
	(e)
	=\,
	-i  \sqrt{\frac{2N}{J}} \frac{1}{N} 
	\suml{i}
	\phi^{a}_{ii}(\e_m^a-\e_n^a)
	X^{aa}_{nm},
	\quad\,\,\,
	(f)=
	-i  \sqrt{\frac{2N}{J}} \frac{1}{N} \suml{i}
	\phi^{a}_{ii}(\e_n^a-\e_m^a) X^{\dagger}\,^{aa}_{mn}.
	\end{aligned}
	\end{align}

	\subsubsection{Interaction dressed $\phi$ propagator}
	
	Taking into account the interactions between $\phi$ and $X$, we find that, to the leading order in small parameter $J/N$, the self-energy of the bosonic field $\phi$ is given by the diagram depicted in Fig.~\ref{fig:p1}(i) with the analytical expression:
	\begin{align}
	\begin{aligned}
	-i (\Sigma_{\phi})^{ab}_{ij;kl}(\ww_m^a)
	=&
	-i \frac{1}{\pi}\frac{1}{JN} \zeta_a z_a (1-\delta_{m,0})\delta_{ab}\delta_{il}\delta_{jk}.
	\end{aligned}
	\end{align}
	The full propagator for  $\phi$ can be obtained from the Dyson equation,  and takes the form
	\begin{align}\label{eq:Gphi}
	\begin{aligned}
	&i(\GG_{\phi})^{ab}_{ij;kl}(\ww_m^a)
	=
	\left[ \left( 
	-i(\GG^{(0)}_{\phi})^{-1} +i\Sigma_{\phi}
	\right)^{-1}\right]^{ab}_{ij;kl} (\ww_m^a)
	=
	i \zeta_a \frac{1}{z_a}\tilde{V}^a_{ij;kl}(\ww_m^a)\delta_{ab},
	\end{aligned}
	\end{align} 
	where we have defined $\tilde{V}^{a}$ as
	\begin{align}\label{eq:tV}
	\begin{aligned}
	&(\tilde{V}^{a})^{-1}_{ij;kl}(\ww_m^a)
	=
	V^{-1}_{ij;kl}
	+
	i \zeta_a\frac{1}{z_a} M^a_m \delta_{ik}\delta_{lj} 
	-
	\frac{1}{\pi}\frac{1}{JN}(1-\delta_{m,0})\delta_{il}\delta_{jk}
	.
	\end{aligned}
	\end{align} 
	We use a red wavy line with a solid dot in the middle to denote the full  $\phi$ propagator $\GG_{\phi}$ (to distinguish it from the bare propagator $\GG^{(0)}_{\phi}$ indicated by a open dot), see Fig.~\ref{fig:p1} (h).
	
	\subsubsection{Interaction dressed $X$ propagator}
	
	\begin{figure}
		\centering
		\includegraphics[width=0.5\linewidth]{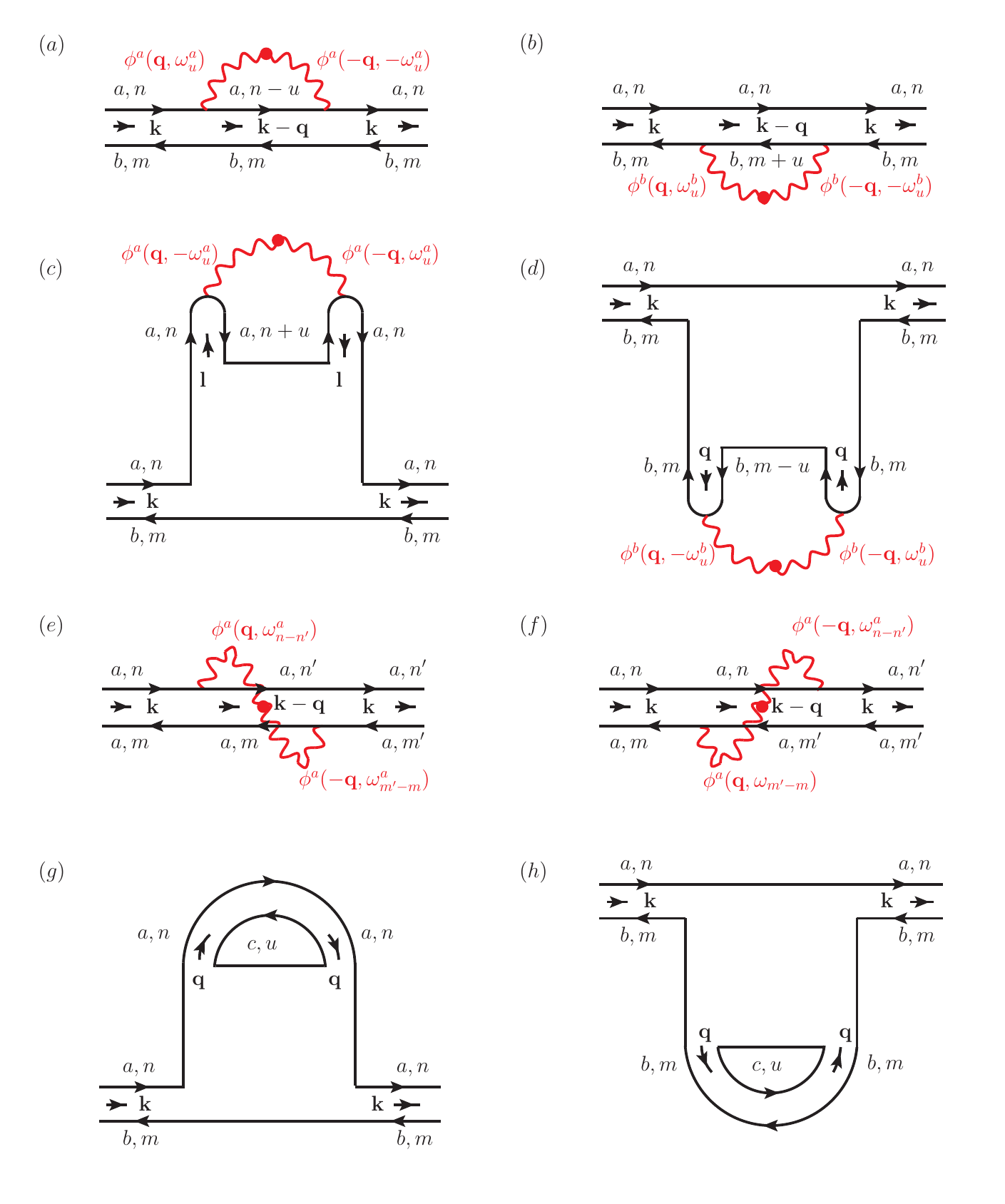}
		\caption{Leading order $X$ self energy diagrams for both the interacting RMT model and 2D disordered system. The momentum labels in all self-energy diagrams are used for the latter case.}
		\label{fig:p2}
	\end{figure}

	In Fig.~\ref{fig:p2}, we show the leading order self-energy diagrams for the matrix field $X$. In respective order, diagrams (a)-(f) give rise to the following contributions to the $X$ self-energy $\Sigma_{X}$:
	\begin{align}\label{eq:Sigmaa-d}
	\begin{aligned}
	(\Sigma^{(a)}_{X})^{ab;ba}_{nm;mn}
	=&
	-\frac{i}{N^2}\sum_{ij}
	\suml{u \leq n}
	\dfrac{1}{\zeta_a (\e_n^a-\ww_u^a) e^{-i (\e_n^a-\ww_u^a) \delta z_a}-\zeta_b \e_m^b e^{-i\e_m^b \delta z_b} } 
	i \frac{\zeta_a }{z_a}\tilde{V}^a_{ij;ji}(\ww_u^a),
	\\
	(\Sigma^{(b)}_{X})^{ab;ba}_{nm;mn}
	=&
	-\frac{i}{N^2}\sum_{ij}
	\suml{u<-m}
	\dfrac{1}{\zeta_a \e_n^ae^{-i \e_n^a \delta z_a}-\zeta_b (\e_m^b +\ww_u^b) e^{-i(\e_m^b +\ww_u^b) \delta z_b} } 
	i \frac{\zeta_b}{z_b}\tilde{V}^b_{ij;ji}(\ww_u^b),
	\\
	(\Sigma^{(c)}_{X})^{ab;ba}_{nm;mn}
	=&
	+i \frac{1}{N^2}\suml{ij}
	\suml{u<-n}
	\dfrac{	
		\left[ 
		\zeta_a \e_n^ae^{-i \e_n^a \delta z_a}-\zeta_b \e_m^b  e^{-i\e_m^b \delta z_b}
		+
		\zeta_a \e_n^ae^{-i \e_n^a \delta z_a}-\zeta_a (\e_n^a +\ww_u^a) e^{-i(\e_n^a +\ww_u^a) \delta z_a}
		\right] }
	{
		\left[ \zeta_a \e_n^ae^{-i \e_n^a \delta z_a}-\zeta_a (\e_n^a +\ww_u^a) e^{-i(\e_n^a +\ww_u^a) \delta z_a}\right]^2 
	} 
	i \frac{\zeta_a }{z_a}\tilde{V}^a_{ij;ji}(\ww_u^a),
	\\
	(\Sigma^{(d)}_{X})^{ab;ba}_{nm;mn}
	=&
	+i \frac{1}{N^2}\suml{ij}
	\suml{u\leq m}
	\dfrac{
		\left[ 
		\zeta_a \e_n^ae^{-i \e_n^a \delta z_a}-\zeta_b \e_m^b  e^{-i\e_m^b \delta z_b}
		+
		\zeta_b (\e_m^b-\ww_u^b)e^{-i (\e_m^b-\ww_u^b) \delta z_b}-\zeta_b \e_m^b e^{-i\e_m^b \delta z_b}
		\right] 
	}
	{
		\left[ \zeta_b (\e_m^b-\ww_u^b)e^{-i (\e_m^b-\ww_u^b) \delta z_b}-\zeta_b \e_m^b e^{-i\e_m^b \delta z_b} \right] ^2 
	}
	i \frac{\zeta_b }{z_b}\tilde{V}^b_{ij;ji}(\ww_u^b),
	\\
	(\Sigma^{(e)}_{X})\,^{aa;aa}_{nm;m'n'}
	=&
	+ \frac{i}{N^2}\sum_{ij}
	\dfrac{1}{\zeta_a \e_{n'}^a e^{-i \e_{n'}^a \delta z_a}-\zeta_a \e_m^a e^{-i\e_m^a \delta z_a} } 
	i \frac{\zeta_a }{z_a}\tilde{V}^a_{ij;ji}(\e_{n'}^a-\e_n^a)
	\delta_{n-n',m-m'},
	\\
	(\Sigma^{(f)}_{X})\,^{aa;aa}_{nm;m'n'}
	=&
	+ \frac{i}{N^2}\sum_{ij}
	\dfrac{1}{\zeta_a \e_{n}^a e^{-i \e_{n}^a \delta z_a}-\zeta_a \e_{m'}^a e^{-i\e_{m'}^a \delta z_a} } 
	i \frac{\zeta_a }{z_a}\tilde{V}^a_{ij;ji}(\e_{n'}^a-\e_n^a)
	\delta_{n-n',m-m'},
	\end{aligned}
	\end{align}
	where $n,n' \geq 0 >m,m'$.
	We note that diagrams (g) and (h) contain a free replica summation $(\suml{c})$. Diagrams of this kind usually cancel with the Jacobian for parametrization Eq.~\ref{eq:para} in NL$\sigma$M calculation, and as a result are ignored here.
	
	The total contribution from all diagrams in Fig.~\ref{fig:p2}(a)-(f) is composed of two parts:
	\begin{align}\label{eq:SigmaX}
	(\Sigma_{X})\,^{ab;b'a'}_{nm;m'n'} 
	=
	\left[ 
	(\Sigma_{\msf{diag}})\,^{ab;ba}_{nm;mn}\delta_{nn'}\delta_{mm'} 
	+
	(\Sigma_{\msf{off}})\,^{aa;aa}_{nm;m'n'}\delta_{n-m,n'-m'}
	\delta_{ab}
	\right]
	\delta_{aa'}\delta_{bb'},
	\qquad
	n,n' \geq 0>m,m'.
	\end{align} 
	Here $\Sigma_{\msf{diag}}$ represents the contribution from diagrams (a)-(d), while $\Sigma_{\msf{off}}$ denotes the remaining contribution from diagrams (e)-(f) in Fig.~\ref{fig:p2}. 
	$\Sigma_{\msf{diag}}$ is diagonal in the Matsubara frequency space, meaning that the matrix element $(\Sigma_{\msf{diag}})\,^{ab;ba}_{nm;m'n'} $ is nonvanishing only when $n=n'$ and $m=m'$. By contrast, $\Sigma_{\msf{off}}$ has non-zero off-diagonal components in the frequency space.
	In addition, $\Sigma_{\msf{off}}^{ab;ba}$ vanishes when $a \neq b$, while $\Sigma_{\msf{diag}}^{ab;ba}$ takes non-zero value for any pair of $a$ and $b$. 
	
	The interaction dressed $X$ propagator $\GG_{X}$ can be deduced from the Dyson equation for $X$
	\begin{align}\label{eq:GX1}
	\begin{aligned}
	\suml{a',b'}\suml{n'\geq 0>m'}
	\left( 
	(\GG^{(0)}_{X} )^{-1}-\Sigma_{X}
	\right)^{ab;b'a'}_{nm;m'n'}
	(\GG_{X})^{a'b';b''a''}_{n'm';m''n''}
	=
	\delta_{aa''}\delta_{bb''}\delta_{nn''}\delta_{mm''},
	\end{aligned}
	\end{align}
	which reduces to the following equation for $a \neq b$ with the help of Eqs.~\ref{eq:GX0} and~\ref{eq:SigmaX},
	\begin{align}
	&\begin{aligned}\label{eq:GX2a}
	(\GG_{X})^{ab;b''a''}_{nm;m''n''}
	=
	\dfrac{1}{	  
		(\GG^{(0)}_{X})^{-1}\,^{ab;ba}_{nm;mn}-\left( \Sigma_{\msf{dia}}\right)^{ab;ba}_{nm;mn}
	}
	\delta_{aa''}\delta_{bb''}\delta_{nn''}\delta_{mm''},
	\qquad \qquad\qquad \qquad\quad\,\,
	a \neq b.
	\end{aligned}
	\end{align}
	In the following, we will call the $X$ propagator $\GG_{X}^{ab;ba}$ ($X$ self-energy $\Sigma_{X}^{ab;ba}$) with $a\neq b$ the inter-replica propagator (inter-replica self-energy) and the one with $a=b$ the intra-replica propagator (intra-replica self-energy).

	We emphasize that the intra-replica $X$ propagator $\GG_{X}^{aa;aa}$ contribute to the disconnected part of the SFF 
	\begin{align} \label{eq:disSFF}
	K^{\msf{dis}}(t)=\braket{Z(\eta+it)}\braket{Z(\eta-it)},
	\end{align}
	while its inter-replica counterpart $\GG_{X}^{a,-a;-a,a}$ is associated with the connected part of the SFF
	\begin{align} \label{eq:conSFF}
	K^{\msf{con}}(\eta,t)=\braket{Z(\eta+it)Z(\eta-it)}- \braket{Z(\eta+it)}\braket{Z(\eta-it)}.
	\end{align} 
	In this paper, we're interested in the connected SFF $K_{\msf{con}}(t)$ which governs the ramp regime, and therefore will focus on the inter-replica propagator $\GG_{X}^{a,-a;-a,a}$. The inter-replica $X$ propagator can be directly obtained from Eq.~\ref{eq:GX2a} after substituting the explicit forms of the bare propagator $(\GG_X^{(0)})^{a,-a;-a,a}$ (Eq.~\ref{eq:GX0}) and the self-energy component $\Sigma_{\msf{dia}}^{a,-a;-a,a}$.
	
	The self-energy component diagonal in the Matsubara frequency space $\Sigma_{\msf{diag}}$ is given by the summation of $\Sigma_X^{(a)}$ to $\Sigma_X^{(d)}$ in Eq.~\ref{eq:Sigmaa-d} and acquires the form of
	\begin{align}
	(\Sigma_{\msf{diag}})\,^{ab;ba}_{nm;mn}
	=
	i\delta z (\zeta_a \e_n^ae^{-i \e_n^a \delta z_a}-\zeta_b \e_m^b  e^{-i\e_m^b \delta z_b})
	-
	\lambda^{ab}_{nm}.
	\end{align} 
	Here $\delta z$ contributes to the renormalization effect and will be neglected in the following discussion. 
	We believe the main interaction effect is from the mass term $\lambda^{ab}$ given by
	\begin{align}
	\begin{aligned}
	\lambda^{ab}_{nm}
	=&\,
	\suml{-n \leq u \leq n}
	\dfrac{1}{z_a \ww_u^a} 
	\sum_{ij}\frac{1}{N^2}\tilde{V}^a_{ij;ji}(\ww_u^a)
	+
	\suml{m< u <-m}
	\dfrac{1}{z_b \ww_u^b } 
	\sum_{ij}\frac{1}{N^2}\tilde{V}^b_{ij;ji}(\ww_u^b),
	\end{aligned}
	\end{align}
	which is divergent at $u=0$.
	This infrared divergence can be curved by inclusion of higher order self-energy diagrams.
	
	We use the self-consistent Born approximation (SCBA) to solve the infared divergence. In particular, we ignore all self-energy diagrams which contribute to the off-diagonal component in the Matsubara frequency space $\Sigma_{\msf{off}}$,  and substitute the bare $X$ propagator $\GG_{X}^{(0)}$ in the self-energy diagrams (a)-(d) with 
	\begin{align}\label{eq:GXab}
	\begin{aligned}
	(\GG_{X}')^{ab;ba}_{nm;mn}
	=
	\frac{1}{-i\left( \zeta_a\e_n^ae^{-i \e_n^a \delta z_a}-\zeta_b\e_m^be^{-i\e_m^b \delta z_b}\right) +\lambda^{ab}_{nm}}.
	\end{aligned}
	\end{align}
	We note that for $a\neq b$, this is the full interaction dressed inter-replica $X$ propagator (without considering the renormalization effect). For $a=b$, this propagator takes into account only the self-energy component $\Sigma_{\msf{dia}}^{aa;aa}$ but ignores $\Sigma_{\msf{off}}^{aa;aa}$.
	Using SCBA, we obtain the following self-consistent equation for the mass $\lambda^{ab}$:   
	\begin{align}\label{eq:sceq}
	\begin{aligned}
	\lambda^{ab}_{nm}
	=&
	+
	\suml{-n \leq u \leq n}
	\dfrac{1}{z_a \ww_u^a-i \zeta_a z_a \lambda^{ab}_{nm}} 
	\sum_{ij}\frac{1}{N^2}\tilde{V}^a_{ij;ji}(\ww_u^a)
	+
	\suml{m<u<-m}
	\dfrac{1}{z_b \ww_u^b -i \zeta_b z_b \lambda^{ab}_{nm} } 
	\sum_{ij}\frac{1}{N^2}\tilde{V}^b_{ij;ji}(\ww_u^b)
	\\
	&
	+
	\suml{u<-n}
	\left( 
	-\frac{z_a\ww_u^a }{\left( z_a\ww_u^a- i\zeta_a z_a\lambda^{aa}_{nn}\right)^2 } 
	+\frac{1 }{z_a\ww_u^a -i\zeta_az_a\lambda^{ab}_{nm}} 
	\right) 
	\suml{ij}\frac{1}{N^2}\tilde{V}^a_{ij;ji}(\ww_u^a)
	\\
	&
	+ 
	\suml{u\leq m}
	\left( 
	-\frac{ z_b\ww_u^b}{\left( z_b \ww_u^b -i\zeta_bz_b\lambda^{bb}_{mm}\right)^2 } 
	+\frac{1}{ z_b \ww_u^b -i\zeta_bz_b\lambda^{ab}_{nm} } 
	\right) 
	\suml{ij}\frac{1}{N^2}\tilde{V}^b_{ij;ji}(\ww_u^b).
	\end{aligned}
	\end{align}
	After summing over the Matsubara frequencies, this equation reduces to
		\begin{align}\label{eq:sceq-1}
		\begin{aligned}
		&\lambda^{ab}_{nm} 
		=
		\dfrac{\bar{V}_0^a}{2\pi}\dfrac{1}{-i \zeta_a z_a \lambda^{ab}_{nm}} 
		+
		\dfrac{\bar{V}^a}{2\pi}
		\left[ 
		\begin{aligned}
		&-\psi (1-i\frac{\zeta_a z_a}{2\pi} \lambda^{ab}_{nm})
		+
		\psi (1-i\frac{\zeta_a z_a}{2\pi} \lambda^{ab}_{nm}+n)
		+
		\psi (1+i\frac{\zeta_a z_a}{2\pi} \lambda^{ab}_{nm})
		-
		\psi (1+i\frac{\zeta_a z_a}{2\pi} \lambda^{ab}_{nm}+n)
		\\
		&+\psi (1+i\frac{\zeta_a z_a}{2\pi} \lambda^{ab}_{nm}+n)
		-
		\psi (1+i\frac{\zeta_a z_a}{2\pi} \lambda^{aa}_{nn}+n)
		-i
		\frac{\zeta_a z_a}{2\pi} \lambda^{aa}_{nn}
		\psi' (1+i\frac{\zeta_a z_a}{2\pi} \lambda^{aa}_{nn}+n)
		\end{aligned}
		\right] 
		\\
		&+
		\dfrac{\bar{V}_0^b}{2\pi}\dfrac{1}{-i \zeta_b z_b \lambda^{ab}_{nm}} 
		+
		\dfrac{\bar{V}^b}{2\pi}
		\left[ 
		\begin{aligned}
		&-\psi (1-i\frac{\zeta_b z_b}{2\pi} \lambda^{ab}_{nm})
		+
		\psi (1-i\frac{\zeta_b z_b}{2\pi} \lambda^{ab}_{nm}-m-1)
		+
		\psi (1+i\frac{\zeta_b z_b}{2\pi} \lambda^{ab}_{nm})
		-
		\psi (1+i\frac{\zeta_b z_b}{2\pi} \lambda^{ab}_{nm}-m-1)
		\\
		&+\psi (1+i\frac{\zeta_b z_b}{2\pi} \lambda^{ab}_{nm}-m-1)
		-
		\psi (1+i\frac{\zeta_b z_b}{2\pi} \lambda^{bb}_{mm}-m-1)
		-i
		\frac{\zeta_b z_b}{2\pi} \lambda^{bb}_{mm}
		\psi' (1+i\frac{\zeta_b z_b}{2\pi} \lambda^{bb}_{mm}-m-1)
		\end{aligned}
		\right] ,
		\end{aligned}
		\end{align}
		where $\bar{V}^a$ and $\bar{V}^a_0$ are defined as $\sum_{ij}\frac{1}{N^2}\tilde{V}^a_{ij;ji}(\ww_u^a)$ for $u\neq 0$ and $u= 0$, respectively.
		The problem then reduces to solving three coupled self-consistent equations with three variables $\lambda^{+-}$, $\lambda^{++}$, $\lambda^{--}$. We have checked numerically that the solution to these self-consistent equations exist and it depends on the external frequency indices $n,m$ in addition to interaction matrix $V$ and complex time $z_a$.

	Now let us consider what happen to the ``zero modes" of the non-interacting theory after introducing the interactions.
	From Eq.~\ref{eq:dQ} and earlier discussion about quadratic fluctuations, we can see that ``zero modes" are encoded by $X^{ab}_{nm}$ and $X^{\dagger}\,^{ba}_{mn}$ for $a=-b$ and $m=-n-1$. It is straightforward to check that in the non-interacting theory ($\phi=0$), by setting $X^{ab}_{nm}=X^{\dagger}\,^{ab}_{nm}=0$ when $a\neq -b$ or $m\neq-n-1$, $S_X^{(2)}$ and $S_X^{(4)}$ (Eq.~\ref{eq:SX24}) as well as all higher order terms in the non-interacting action vanish.
	In the presence of interactions, we find from the caluclation above that that the soft modes encoded by $X^{a,-a}_{n,-n-1}$ and $X^{\dagger}\,^{-a,a}_{-n-1,n}$ acquire a mass $\lambda^{a,-a}$ (the solution to Eq.~\ref{eq:sceq-1} with $b=-a$ and $m=-n-1$). 
	
	\subsection{Contribution to the SFF from the soft modes around the standard saddle point}
	
	In terms of the interaction dressed propagators $\GG_{X}$ and $\GG_{\phi}$, we can express the contribution to the SFF from the standard saddle point $\Lambda^{(0)}$ and the corresponding fluctuations around it as
	\begin{align}\label{eq:K2-1}
	\begin{aligned}
	&K'(\eta, t;\Lambda^{(0)})
	\propto
	e^{-S[\Lambda^{(0)},0]}
	\frac{
		\int \D \phi
		\exp \left[ \frac{i}{2} \phi\GG_{\phi}^{-1} \phi \right] 
		\int \D (X^{\dagger},X) 
		\exp \left( 
		-\frac{N}{2J} X^{\dagger} \GG_X^{-1} X
		\right) 
	}
	{
		\int \D \phi
		\exp \left( -S_{\phi 0} [\phi]
		\right) 
		\int \D (X^{\dagger},X) 
		\exp \left( 
		-\frac{N}{2J} X^{\dagger} X	
		\right) 
	}.
	\end{aligned}
	\end{align}
	Here we have ignored the contribution from the massive modes, and kept only the leading order soft mode fluctuation correction. The two integrals in the denominators arise from the normalization constants $Z_{\phi}$ and $Z_Q$.
	As mentioned earlier, in the case of $\eta/(t^2+\eta^2) \gtrsim {J}/{N}$, contributions from all remaining saddle points can be ignored, and the expression above gives approximately the SFF, i.e., $K(\eta,t)\approx K'(\eta,t;\Lambda^{(0)})$.
	By contrast, for $\eta/(t^2+\eta^2) \ll {J}/{N}$ (including $\eta=0$), all saddle points are equally important, and $K'(\eta,t;\Lambda^{(0)}) $ represents only the contribution associated with $\Lambda^{(0)}$. 
	
	For non-interacting case, we instead have
	\begin{align}\label{eq:K2-10}
	\begin{aligned}
	&K_0'(\eta, t;\Lambda^{(0)})
	\propto
	e^{-S[\Lambda^{(0)},0]}
	\frac{
		\int \D (X^{\dagger},X) 
		\exp 
		\left( 
		-\frac{N}{2J} X^{\dagger} (\GG_X^{(0)})^{-1} X
		\right) 
	}
	{
		\int \D (X^{\dagger},X) 
		\exp \left( 
		-\frac{N}{2J} X^{\dagger} X	
		\right) 
	},
	\end{aligned}
	\end{align}
	which is consistent with Eq.~\ref{eq:K0-1}. In the following, we will analyze the difference between $K'(\eta,t;\Lambda^{(0)})$ (Eq.~\ref{eq:K2-1}) and $K_0'(\eta,t;\Lambda^{(0)})$ (Eq.~\ref{eq:K2-10}), and in particular how the interaction effect influence the leading order fluctuation correction to the SFF. 
	
	Carrying out the integrations in Eq.~\ref{eq:K2-1} and Eq.~\ref{eq:K2-10} and using Eq.~\ref{eq:Gphi}, we find 
	\begin{subequations}
		\begin{align}
		&\begin{aligned}\label{eq:K2-2a}
		&K'(\eta, t;\Lambda^{(0)})
		\propto
		e^{-S[\Lambda^{(0)},0]}
		\exp
		\left\lbrace 
		-\sum_{a b} \suml{n\geq 0>m}  \ln  \left( 
		(\GG_{X}^{-1})^{ab;ba}_{nm;mn} \right) 
		+\frac{1}{2} \suml{a,m} \left(  \Tr \ln \tilde{V}^a(\ww_m^a) -\Tr \ln V \right) 
		\right\rbrace,
		\end{aligned}
		\\
		&\begin{aligned}\label{eq:K2-2b}
		&K'_0(\eta, t;\Lambda^{(0)})
		\propto
		e^{-S[\Lambda^{(0)},0]}
		\exp
		\left\lbrace 
		-\sum_{a b} \suml{n\geq 0>m}  \ln  \left( 
		(\GG_{X}^{(0)})^{-1}\,^{ab;ba}_{nm;mn} \right) 
		\right\rbrace,
		\end{aligned}
		\end{align}
	\end{subequations}
	where $\Tr$ indicates a trace over the flavor space only.
	In Eq.\ref{eq:K2-2a} (Eq.\ref{eq:K2-2b}), only the terms involving the inter-replica $X$ propagator $\GG_{X}^{a,-a;-a,a}$ ($(\GG_{X}^{(0)})^{a,-a;-a,a}$) contribute to the connected SFF $K^{\msf{con}}(\eta,t)$ ($K_{0}^{\msf{con}}(\eta,t)$). All remaining terms, including the ones associated with intra-replica $X$ propagator $\GG_{X}^{aa;aa}$ ($(\GG_{X}^{(0)})^{aa;aa}$) 
	and $\phi$ propagator $\GG_{\phi}$ contribute to the disconnected SFF $K^{\msf{dis}}(\eta,t)$ ($K_{0}^{\msf{dis}}(\eta,t)$). 
	As mentioned earlier, we are interested in the connected SFF, and therefore will consider only the terms involving the inter-replica $X$ propagators in Eqs.~\ref{eq:K2-2a} and~\ref{eq:K2-2b}.
	
	For $V \neq 0$, ignoring the renormalization effect and using Eq.~\ref{eq:GXab}, we find the contribution to $\ln K'^{\msf{con}}$ from the inter-replica $X$ propagator  $\GG_{X}^{ab;ba}$:
	\begin{align}\label{eq:lnK-1}
	\begin{aligned}
	&\ln K'^{\msf{con}}(\eta, t;\Lambda^{(0)})
	=-\sum_{a\neq b} \suml{n \geq 0>m} \ln \left( (\GG_{X}^{-1})\,^{ab;ba}_{nm;mn}\right) 
	=	
	-\sum_{a\neq b} \suml{n \geq 0>m}
	\ln \left[  -i\left( \zeta_a \e_n^ae^{-i \e_n^a \delta z_a}-\zeta_b \e_m^b  e^{-i\e_m^b \delta z_b}\right) +\lambda^{ab}_{nm}\right] .
	\end{aligned}
	\end{align}
	We then use the anlytical continuation technique, and convert the double Matsubara summation in the equation above into a two-variables continous integral:
	\begin{align}\label{eq:lnK-2}
	\begin{aligned}
	\ln K'^{\msf{con}}(\eta, t)
	=&
	\sum_{a\neq b}
	\int_{-\infty}^{\infty} \frac{d\e}{2\pi}
	\int_{-\infty}^{\infty} \frac{d\e'}{2\pi}
	\ln \left( 1+e^{-i \zeta_a z_a\e} \right) 
	\ln \left( 1+e^{-i \zeta_b z_b \e'} \right) 
	\\
	& \times
	\left( \dfrac{1}{-i\left( \e-\e' \right) +\lambda^{ab}(\zeta_a\e,\zeta_b\e')} \right)^2
		\left( \frac{\partial \lambda^{ab}}{\partial \e}-i \right) 
		\left( \frac{\partial \lambda^{ab}}{\partial \e'}+i\right).
	\end{aligned}
	\end{align}
	Here $\lambda^{ab}(\zeta_a\e,\zeta_b\e')$ stands for $\lambda^{ab}_{nm}$ after the analytical continuation $\e_n^a\rightarrow\zeta_a\e$ and $\e_m^b\rightarrow\zeta_b\e'$.
	
	In the non-interacting case, the corresponding contribution can be found by replacing the interaction dressed propagator $\GG_{X}^{ab;ba}$  with the bare one $(\GG_{X}^{(0)})^{ab;ba}$, and setting the mass $\lambda^{ab}$ to $0$ in Eqs.~\ref{eq:lnK-1} and~\ref{eq:lnK-2}, which leads to
	\begin{align}\label{eq:lnK-0}
	\begin{aligned}
	\ln K_{0}'^{\msf{con}}(\eta,t)
	=&-\sum_{a\neq b} \suml{n>0>m} \Tr \ln  (\GG_{X}^{(0)}\,^{-1})^{ab;ba}_{nm;mn}
	\\
	=&
	\sum_{a\neq b}
	\int_{-\infty}^{\infty} \frac{d\e}{2\pi}
	\int_{-\infty}^{\infty} \frac{d\e'}{2\pi}
	\ln \left( 1+e^{-i \zeta_a z_a\e} \right) 
	\ln \left( 1+e^{-i \zeta_b z_b \e'} \right) 
	\left( \dfrac{1}{-i\left( \e-\e' \right) }\right)^2.
	\end{aligned}
	\end{align}
	
	In Ref.~\cite{PRL}, using a cumulant expansion, we showed that, to study the quadratic order fluctuation correction to the non-interacting SFF $K_0(\eta,t)$,  
		one can replace 
		$	\ln \left( 1+e^{-i \zeta_a z_a\e} \right) \ln \left( 1+e^{-i \zeta_b z_a \e'} \right) $ with $e^{-i \zeta_a z_a \e-i \zeta_b z_b\e'}$. The neglected term from this replacement cancels partially with higher order contributions. See also the discussion in Sec.~\ref{sec:high}. We apply the same substitution here for both the interacting and non-interacting theories, with the focus placed on the difference between the two cases, and find that
	Eq.~\ref{eq:lnK-2} and Eq.~\ref{eq:lnK-0} become, respectively, 			
	\begin{align}\label{eq:K2-3}
	\begin{aligned}
	\ln K''^{\msf{con}}(\eta,t)=&
	\suml{a}
	\int_{-\infty}^{\infty} \frac{d\e}{2\pi}
	\int_{-\infty}^{\infty} \frac{d\e'}{2\pi}
	e^{-i \zeta_a t (\e-\e' )} e^{- \eta (\e+\e' )} 
	\left( \dfrac{1}{-i\left( \e-\e' \right) +\lambda^{a,-a}(\zeta_a\e,-\zeta_a\e')} \right)^2
	\\
	&\times 
		\left( \frac{\partial \lambda^{a,-a}}{\partial \e}-i \right) 
		\left( \frac{\partial \lambda^{a,-a}}{\partial \e'}+i\right)
	\\
	=&
	\suml{a}
	\int_{-E_{\msf{UV}}}^{E_{\msf{UV}}} \frac{dE}{2\pi}
	\int_{-\infty}^{\infty} \frac{d\ww}{2\pi}
	e^{-2\eta E} 
	e^{-i \zeta_a t \ww} 
	\left( \dfrac{1}{-i \ww+\lambda^{a,-a}(E)} \right)^2
		\left[ 	\left(\frac{1}{2} \frac{  \partial \lambda^{a,-a}(E)}{\partial E}\right)^2+1\right] 
	\\
	=&
	\suml{a}
	\int_{-E_{\msf{UV}}}^{E_{\msf{UV}}} \frac{dE}{2\pi} e^{-2\eta E} 
	t e^{-\zeta_a\lambda^{a,-a}(E) t} \Theta(\zeta_a\re\lambda^{a,-a}(E))
		\left[ 	\left( \frac{1}{2}\frac{\partial \lambda^{a,-a}(E)}{\partial E}\right)^2+1\right] ,
	\end{aligned}
	\end{align}
	and
	\begin{align}\label{eq:K2-30}
	\begin{aligned}
	\ln K_{0}^{\msf{con}}\,''(\eta, t)=&
	\suml{a}
	\int_{-\infty}^{\infty} \frac{d\e}{2\pi}
	\int_{-\infty}^{\infty} \frac{d\e'}{2\pi}
	e^{-i \zeta_a t (\e-\e' )} e^{-\eta (\e+\e' )} 
	\left( \dfrac{1}{-i\left( \e-\e' \right)} \right)^2
	\\
	=&
	-\suml{a}
	\int_{-E_{\msf{UV}}}^{E_{\msf{UV}}} \frac{dE}{2\pi}e^{-2 \eta E} 
	\int_{-\infty}^{\infty} \frac{d\ww}{2\pi}
	e^{-i \zeta_a t \ww} 
	\dfrac{1}{\ww^2} .
	\end{aligned}
	\end{align}
	Here in the second equalities of both equations, we have made the transformation
	\begin{align}\label{eq:cov}
	\begin{aligned}
	\ww=\e-\e',\qquad
	E=\frac{1}{2}(\e+\e').
	\end{aligned}
	\end{align}		
	We have also employed the box approximation which ignores the energy dependence of average bare single-particle density of states and imposes the ultraviolet cutoff $E_{\msf{UV}}$ (see Refs.~\cite{PRL,Cotler-2017,Liu}). Furthermore, in Eq.~\ref{eq:K2-3} we have assumed that $\lambda^{a,-a}$ depends on $E$ but not $\ww$ (the $\ww$-dependent part contributes to the renormalization and can be neglected). 
	
	For $\eta/(t^2+\eta^2) \gtrsim {J}/{N}$, Eqs.~\ref{eq:K2-3} and~\ref{eq:K2-30} represent approximately the leading order fluctuation correction to  $\ln K^{\msf{con}}(\eta, t)$ in the interacting and non-interacting cases, respectively.
	Taking the $\eta\rightarrow 0^+$ limit and  comparing these two equations,
	we can see that the presence of inter-replica mass $\lambda^{\pm,\mp}$ results in the suppression of exponential growth of the connected SFF, which is necessary for the emergence of RMT statistics in many-body energy spectrum.
	We emphasize that Eqs.~\ref{eq:K2-3} and~\ref{eq:K2-30}  only takes into account the contribution from leading order fluctuations. Superficially, it might seems that corrections from higher order fluctuations are of higher order in $J/N$, but they are in fact non-negligible and essential to extract the explicit expression for the SFF.
	
	
	\subsection{Cumulant expansion and higher order fluctuations}\label{sec:high}

	To see that the contribution from higher order fluctuations is crucial to the derivation of the explicit form of the SFF, let us first consider the non-interacting case. 
	Setting the decoupling field $\phi$ to $0$ and integrating out the fermionic field $\psi$ in Eq.~\ref{eq:ZZn}, we find that the non-interacting SFF can be written as
	\begin{align}\label{eq:Z1}
	\begin{aligned}
	K_0(\eta,t)
	=&
	\left\langle 
	\exp \left\lbrace  
	\suml{a,n} 
	\Tr \ln \left[ 
	- i z_a  \left(  \e_n^a e^{-i\e_n^a \delta z_a }I_f- \zeta_a h\right) 
	\right] 
	\right\rbrace 
	\right\rangle ,
	\end{aligned}
	\end{align}
	Converting the summation over Matsubara frequency $\e_n^a=\frac{2\pi}{z_a} (n+\frac{1}{2})$ into a continuous integral, the above equation becomes
	\begin{align}\label{eq:Kave}
	\begin{aligned}
	K_0(\eta,t)
	=
	\left\langle
	\exp
	\left\lbrace 
	N
	\suml{a}
	\int_{-\infty}^{\infty} d\e
	\ln \left[ 1+e^{-i \zeta_a z_a \e } \right] 
	\nu(\e)
	\right\rbrace 
	\right\rangle ,
	\end{aligned}
	\end{align}
	where $\nu(\e)$ is the single-particle density of states:
	\begin{align}\label{eq:nu}
	\nu(\e)=\frac{1}{N} \Tr \delta (\e - h).
	\end{align}
	After a cumulant expansion of Eq.~\eqref{eq:Kave}~\cite{PRL}, we obtain
	\begin{align}\label{eq:CumSum}
	\begin{aligned}
	\ln K_0(\eta,t)  
	=\,&
	N	
	\int_{-\infty}^{\infty} d \e
	\suml{a}\ln \left( 1+e^{-i \zeta_a z_a \e } \right) 
	\R_1(\e)
	\\
	&
	+
	\frac{N^2}{2} 
	\int_{-\infty}^{\infty} d \e
	\int_{-\infty}^{\infty} d \e'
	\left[ \suml{a}\ln \left(  1+e^{-i \zeta_a z_a \e } \right) \right] 
	\left[ \suml{b}\ln \left( 1+e^{-i \zeta_{b} z_b \e' } \right) \right] 
	\R_2^{\msf{con}}(\e,\e')
	\\
	&+...+
	\frac{N^n}{n!} 
	\int_{-\infty}^{\infty} ...\int_{-\infty}^{\infty} 
	\left\lbrace 
	\prod_{k=1}^{n}
	d \e_k
	\left[ \suml{a_k}
	\ln \left( 1+e^{-i \zeta_{a_k} z_{a_k} \e_k }\right) \right] 
	\right\rbrace 
	\R_n^{\msf{con}}(\e_1,...,\e_n)
	+... \,\,\,.
	\end{aligned}
	\end{align}
	Here
	$\R_n^{\msf{con}}(\e_1,...,\e_n)$ is the connected part of the bare $n$-point single-particle energy level correlation function $\R_n(\e_1,...,\e_n)$, which is defined as
	\begin{align}\label{eq:Rn}
	\begin{aligned}
	\R_n(\e_1,...,\e_n)
	=
	\left\langle 
	\nu(\e_1)...\nu(\e_n)
	\right\rangle.
	\end{aligned}
	\end{align}
	
	It is known that, for a GUE ensemble following the probability $P(h)$ in Eq.~\ref{eq:Ph}, the two-level connected correlation function assumes the form~\cite{Mehta,Guhr}
	\begin{align}\label{eq:R2}
	\begin{aligned}
	N^2\R_2^{\msf{con}}(\e,\e')=-\frac{1-\cos\left(2 N (\e-\e')/J \right) }{2\pi^2(\e-\e')^2}+\frac{N}{\pi J}\delta(\e-\e'),
	\end{aligned}
	\end{align}
	which at $|\e-\e'|\gg J/N$ can be approximated as
	\begin{align}
	\begin{aligned}
	N^2 \R_2^{\msf{con}}(\e,\e')=-\frac{1 }{2\pi^2(\e-\e')^2}.
	\end{aligned}
	\end{align}
	Substituting this expression into Eq.~\ref{eq:CumSum}, we find that the second term on the right-hand side of Eq.~\ref{eq:CumSum} is equivalent to Eq.~\ref{eq:lnK-0} after restoring the contribution to the disconnected SFF from the intra-replica $X$ propagators (restoring $a=b$ terms).
	In other words,  the quadratic order fluctuation correction obtained earlier yields the second order term in the cumulant expansion:
	\begin{align}\label{eq:CUM-2}
	\begin{aligned}
	&-\sum_{a, b} \suml{n>0>m} \Tr \ln  (\GG_{X}^{(0)}\,^{-1})^{ab;ba}_{nm;mn}
	=
	\frac{N^2}{2}  \suml{a,b}
	\int_{-\infty}^{\infty} d \e
	\int_{-\infty}^{\infty} d \e'
	\ln \left(  1+e^{-i \zeta_a z_a \e } \right) 
	\ln \left( 1+e^{-i \zeta_{b} z_b \e' } \right)
	\R_2^{\msf{con}}(\e,\e')
	\end{aligned}
	\end{align}
	To recover the contribution from the oscillatory part of $\R_2^{\msf{con}}(\e,\e')$ (the $\cos$ term in Eq.~\ref{eq:R2}), one needs to consider fluctuations around the nonstandard saddle points~\cite{AndreevAltshuer}.
	
	Using Eqs.~\ref{eq:S0} and~\ref{eq:Lambda}, we find that the action at the standard saddle point $\Lambda^{(0)}$ is equivalent to the first term in the cumulant expansion in Eq.~\ref{eq:CumSum}:
	\begin{align}\label{eq:CUM-1}
	\begin{aligned}
	S[\Lambda^{(0)},0]
	=
	-N	
	\int_{-\infty}^{\infty} d \e
	\suml{a}\ln \left( 1+e^{-i \zeta_a z_a \e } \right) 
	\R_1(\e),
	\end{aligned}
	\end{align}
	where the average single-particle level density is given by
	\begin{align}
	\R_1(\e)=\frac{1}{2\pi J^2} \sqrt{4J^2-\e^2}.
	\end{align}
	Combining Eqs.~\ref{eq:CUM-2} and~\ref{eq:CUM-1}, one can see that Eq.~\ref{eq:K2-2a} obtained from considering only the quadratic order soft mode fluctuations yields only the leading two terms in the cumulant expansions. 
	It has been found in Ref.~\cite{PRL} that all higher order terms in the cumulant expansion are equally important in the ramp region, and they are essential to extract the correct overall coefficient in the exponent of ramp expression.
	For this reason, we believe the higher order correction is indispensable for the derivation the SFF. 
	
	Note that the level correlation function is sometimes defined differently in some literatures~\cite{Mehta}, and $\R_n$ here contains several $\delta$ functions for $n \geq 2$ (see for example Eq.~\ref{eq:R2}).
	In Ref.~\cite{PRL}, we show that if one throws away all $\delta$-functions in $\R_n^{\msf{con}}$ and at the same time replaces $\suml{a} \ln(1+e^{-i\zeta_a z_a\e})$ with $\frac{1}{2}\prod_{a}\left( 1+e^{-i\zeta_a z_a\e}\right) -1$,  the left hand side of Eq.~\ref{eq:CumSum} remains the same up to some unessential constant.
	This justifies the replacement 
		$	\ln \left( 1+e^{-i \zeta_a z_a\e} \right) \ln \left( 1+e^{-i \zeta_b z_a \e'} \right) \rightarrow e^{-i \zeta_a z_a \e-i \zeta_b z_b\e'}$ performed in Eqs.~\ref{eq:K2-3} and~\ref{eq:K2-30}.

	In the presence of interactions, as shown in the earlier section, the contribution from the quadratic order fluctuations to $K^{\msf{con}}(\eta,t)$ is also given by the second term in the cumulant expansion Eq.~\ref{eq:CumSum}  (with $a\neq b$) after making the replacement
	\begin{align}
	\begin{aligned}
	2\pi^2 N^2 \R_2^{\msf{con}}(\e,\e')= -\frac{1 }{(\e-\e')^2} \rightarrow \left( \frac{1 }{-i(\e-\e')+\lambda}\right)^2	
	\left( \frac{\partial \lambda^{ab}}{\partial \e}-i \right) 
	\left( \frac{\partial \lambda^{ab}}{\partial \e'}+i\right),
	\qquad
	|\e-\e'| \gg J/N
	\end{aligned}
	\end{align} 
	We expect that, for higher order fluctuations, the interaction effect will be similar and the corresponding contribution to the SFF is also be suppressed because of the inter-replica masses $\lambda^{\pm,\mp}$. However, this is beyond the scope of the current study.

	\section{Disordered interacting systems in 2D}\label{sec:disorder}
	
	In this section, we will perform an analogous calculation of the SFF for a two-dimensional disordered interacting system of spinless fermions with broken time reversal symmetry, which falls into the unitary Wigner-Dyson class and is governed by the following Hamiltonian
	\begin{align}\label{eqd:H}
	\begin{aligned}
	H=\int d^2\rb \,\psi^{\dagger} (\rb) \left( -\frac{\nabla^2}{2m} -\mu +U_{\msf{imp}}(\rb) \right) \psi (\rb)  
	+\frac{1}{2}\int d^2\rb \int d^2\rb' \psi^{\dagger} (\rb) \psi^{\dagger}  (\rb') V(|\rb-\rb'|) \psi (\rb') \psi (\rb).  
	\end{aligned}
	\end{align}
	Here $U_{\msf{imp}}(\rb)$ represents the random impurity potential, and follows the Gaussian distribution function
	\begin{align}\label{eqd:PU}
	P(U_{\msf{imp}})
	=
	\frac{1}{Z_U}
	\exp{\left[-\pi \nu_0 \tel \intl{\rb} U_{\msf{imp}}^2(\rb)\right]},
	\qquad
	Z_U=\int \D U_{\msf{imp}} \exp \left[ -\pi \nu_0 \tel \intl{\rb} U_{\msf{imp}}^2(\rb) \right],
	\end{align}
	where $\nu_0$ denotes the bare average single-particle density of states at Fermi energy $\mu$, and $\tel$ indicates the elastic scattering time.
	For simplicity, we consider short-range density-density interactions with $V(r)=V\delta(r)$. The current calculation can be generalized to other types of interaction.
	As in the previous section, we calculate the SFF $K(\eta,t)=\braket{Z(\eta+it)Z(\eta-it)}$, where the angular bracket now indicates averaging over the ensemble of random impurity potential $U_{\msf{imp}}(\rb)$ according to the distribution function $P(U_{\msf{imp}})$ in Eq.~\ref{eqd:PU}.
	We will focus on the time regime $ \Delta^{-1}\gg t\gg \tel$, where
	$\Delta \equiv 1/\nu_0 L^2$ is the average bare single-particle level spacing, with $L \rightarrow \infty$ being system size.
	We also consider weak enough disorder so that a perturbation in terms of disorder strength can be performed.
	
	\subsection{Preliminaries}
	
	The SFF for this disordered interacting fermion system
	can be expressed as the following path integral:
	\begin{align}\label{eqd:KT-0}
	\begin{aligned}
	K(\eta,t)
	=&
	\frac{1}{Z_U}\int \D U_{\msf{imp}}
	\exp\left( -\pi \nu_0 \tel \intl{\rb} U_{\msf{imp}}^2(\rb) \right) 
	\\
	\times &
	\int \D (\bpsi, \psi)  
	\exp 
	\left\lbrace 
	i \suml{a=\pm}
	\int_0^{z_a} d t' \intl{\rb} 
	\bpsi^{a} (\rb, t') \left[ i \partial_{t'} -\zeta_a \left( -\frac{\nabla^2}{2m} -\mu +U_{\msf{imp}}(\rb) \right)  \right]  \psi^{a} (\rb, t')
	\right\rbrace 
	\\
	\times&
	\exp 
	\left\lbrace 
	- \frac{i}{2}
	\suml{a=\pm}	\zeta_a
	\int_0^{z_a} d t' 
	\intl{\rb} \bpsi^{a} (\rb,t') \bpsi^{a} (\rb,t') V \psi^{a} (\rb,t') \psi^{a} (\rb,t')  
	\right\rbrace .
	\end{aligned}	
	\end{align}
	The Grassmann fields $\bpsi^{a}$ and $\psi^{a}$ 
	carry a replica index $a=\pm$ which corresponds to the forward/backward path for $Z(\eta\pm it)$, and they are subject to the antiperiodic boundary condition Eq.~\ref{eq:bc}. Complex time $z_a$ is defined as $z_a=t-i\zeta_a \eta$ with $\zeta_a=\pm 1$ for $a=\pm$.

	Introducing an auxiliary bosonic field $\phi$ to decouple the interactions, we obtain
	\begin{align}\label{eqd:KT-1}
	\begin{aligned}
	&K(\eta,t)
	=\,
	\frac{1}{Z_U Z_{\phi}}
	\int \D U_{\msf{imp}}
	\exp \left[ -\pi \nu_0\tel \intl{\rb} U_{\msf{imp}}^2(\rb)  \right]	
	\int \D \phi
	\exp 
	\left( \suml{a,n}   \frac{ i }{2} \zeta_a  z_a
	\intl{\rb}
	\phi^{a}_{-n}(\rb) V^{-1} \phi^{a}_n(\rb)\right) 
	\\
	\times &
	\int \D (\bpsi, \psi)  
	\exp 
	\left\lbrace 
	\begin{aligned}
	\suml{a,n,m}
	i z_a 
	\intl{\rb} 
	\bpsi^{a}_n (\rb)
	\left\lbrace  
	\e_n^a e^{-i \e_n^a\delta z_a }\delta_{nm} 
	-\zeta_a \left( -\frac{\nabla^2}{2m} -\mu +U_{\msf{imp}}(\rb) \right) \delta_{nm}+\zeta_a \phi^{a}_{n-m}(\rb)  \right\rbrace  
	\psi^{a}_m (\rb)  
	\end{aligned}
	\right\rbrace.
	\end{aligned}	
	\end{align}
	Here we have performed a temporal Fourier transformation of the fermionic field $\psi$ as in Eq.~\ref{eq:FT}, and similarly for the bosonic field $\phi$.
	The normalization constant $Z_{\phi}$ is given by
	$Z_{\phi} =\int \D \phi
	\exp \left( \suml{a,n}  \frac{i}{2}  \zeta_a z_a \int_{\rb} 
	\phi_{-n}^a (\rb) V^{-1} \phi_n^a (\rb)\right)$.

	To proceed, we perform the ensemble averaging in Eq.~\ref{eqd:KT-1}, and decouple the generated quartic term by a Hermitian matrix field $Q(\rb)$: 
	\begin{align}\label{eqd:FT-6}
	\begin{aligned}
	&K_{\eta}(\eta,t)
	=\,
	\frac{1}{Z_{\phi}Z_Q}
	\int \D \phi
	\exp 
	\left( 
	\frac{i}{2}\suml{a,n} \zeta_a z_a
	\intl{\rb}
	\phi_{-n}^a(\rb)V^{-1}  \phi_n^a (\rb)
	\right) 
	\int \D Q
	\exp
	\left( 
	-
	\frac{\pi \nu_0}{4\tel}
	\intl{\rb}
	\Tr Q^2(\rb)
	\right) 
	\\
	\times &
	\int \D (\bpsi, \psi)  
	\exp 
	\left\lbrace 
	\begin{aligned}
	\suml{a,b,n,m}
	i z_a
	\intl{\rb} 
	\bpsi_n^a (\rb) 
	\left[ 
	\left(
	\e_n^a e^{-i\e_n^a \delta z_a}\zeta_a
	+\frac{\nabla^2}{2m} +\mu \right)
	\delta_{nm}\delta_{ab}
	+\phi_{n-m}^a(\rb)\delta_{ab}
	+\frac{i}{2\tel} Q_{nm}^{ab}(\rb)
	\right]  
	\zeta_b
	\psi_m^{b} (\rb)  
	\end{aligned}
	\right\rbrace,
	\end{aligned}	
	\end{align}
	where the normalization constant is
	$Z_Q
	=
	\int \D Q
	\exp
	\left(
	-
	\frac{\pi \nu_0}{4\tel}
	\intl{\rb} \Tr Q^2(\rb)
	\right)$.
	After integrating out the fermionic field $\psi$ and performing the spatial Fourier transformation, 
	\begin{align}
	\begin{aligned}
	\phi(\qb)=\frac{1}{L^2}\int d^2\rb \phi(\rb) e^{-i\qb\cdot \rb},
	\qquad
	Q(\qb)=\frac{1}{L^2}\int d^2\rb Q(\rb) e^{-i\qb\cdot \rb},
	\end{aligned}
	\end{align}
	we arrive at the following $\sigma-$model 
	\begin{subequations}
		\begin{align}
		\label{eqd:K}
		&K(\eta,t)
		=\,
		\frac{1}{Z_{\phi}Z_Q}
		\int \D \phi
		\exp 
		\left( -S_{\phi 0}[\phi]\right) 
		\int \D Q \exp \left( -S_{\phi 0}[\phi]- S[Q,\phi]\right) ,
		\\
		\label{eqd:Sphi0}
		&S_{\phi 0}[\phi]
		=- \frac{i}{2}\suml{a}
		\zeta_a z_a L^2
		\suml{\qb,m}
		\phi_{-m}^a(-\qb) V^{-1} \phi_m^a(\qb),
		\\
		\label{eqd:SQ}
		&S[Q,\phi]=\,	
		\frac{\pi \nu_0}{4\tel} L^2
		\suml{\qb} \Tr Q(\qb) Q(-\qb)
		-
		\Tr \ln  
		\left[ 
		\left(\mathcal{E}\sigma^3 -\xi_{\kb}\right) \delta_{\qb,0}
		+\frac{i}{2\tel} Q(\qb)
		+ \Phi(\qb)
		\right]
		+
		\text{const}.,
		\end{align}
	\end{subequations}
	with $\xi_{\kb}=k^2/2m-\mu$. 
	The second $\Tr$ in Eq.~\ref{eqd:SQ} represents a trace over momentum space in addition to the Matsubara frequency and replica spaces.
	$\mathcal{E}$, $\Phi(\qb)$, and $\sigma^3$ are defined such that their matrices elements are given by $\mathcal{E}^{ab}_{nm}=\delta_{nm}\delta_{ab} \e_n^a e^{-i\e_n^a \delta z_a}$ ,
	$\Phi^{ab}_{nm}(\qb)=\delta_{ab}\phi^{a}_{n-m} (\qb)$, and $(\sigma^3)^{ab}_{nm}=\zeta_a \delta_{ab}\delta_{nm}$, respectively. 
	
	\subsection{The saddle points}
	
	Starting from the action $S[Q,\phi]$ in Eq.~\ref{eqd:SQ} and ignoring  
	the influence of the decoupling field $\phi$, we now look for saddle point $Q_{sp}$ of the matrix field.
	Taking variation of the non-interacting action $S[Q,0]$ over $Q$, we find that the saddle point  $Q_{sp}$ is determined by the equation
	\begin{align}
	\begin{aligned}
	Q_{sp}(\qb)
	=
	\frac{i}{\pi } 
	\frac{1}{\nu_0 L^2}
	\sum_{\kb}
	\left[ 
	\left( \mathcal{E}\sigma^3 -\xi_{\kb} \right) \delta_{\qb,0}
	+\frac{i}{2\tel} Q_{sp}(-\qb)
	\right]^{-1}.
	\end{aligned}
	\end{align}
	We first look for spatially uniform solution $Q_{sp}(\qb)=\Lambda \delta_{\qb,0}$ diagonal in Matsubara frequency and replica spaces.
	Assuming weak enough disorder such that the chemical potential $\mu \gg \tel^{-1}$, and neglecting the variation of single-particle density states near the Fermi surface, we find 
	\begin{align}
	\Lambda_{nm}^{ab}=s_n^a \delta_{nm}\delta_{ab},
	\qquad
	s_n^a=\pm 1.
	\end{align} 
	We therefore obtain a series of spatially uniform saddle points diagonal in both Matsubara frequency and replica spaces, corresponding to different choices of $\left\lbrace s_n^a \right\rbrace $.
	
	The non-interacting action $S[Q,\phi=0]$ (see Eq.~\ref{eqd:SQ}) is invariant under any spatially uniform unitary rotation $Q \rightarrow R^{-1} \Lambda R$ without the symmetry breaking term $\mathcal{E} \sigma^3$. In the limits $|\delta z_a|,\eta \rightarrow 0$, one can find that the symmetry transformation $U$, under which the non-interacting action $S[Q,\phi=0]$ remains invariant, is given by direct product of multiple spatially uniform $U(2)$ rotations $U=\prod_{n} U_n$. Here $U_n$ rotates only the matrix subblock
	$\begin{bmatrix}
	Q^{++}_{nn} & Q^{+-}_{n,-n-1}
	\\
	Q^{-+}_{-n-1,n} &Q^{--}_{-n-1,-n-1}
	\end{bmatrix}$
	as in Eq.~\ref{eq:Un}. New saddle points can be generated by applying the symmetry transformation $U$ to the diagonal saddle points $Q \rightarrow U^{-1} \Lambda U$. 
	
	At a diagonal saddle point $\Lambda$, the non-interacting action $S[Q,\phi=0]$ acquires the form
	\begin{align}\label{eq:Ssp}
	\begin{aligned}
	S[\Lambda,0]
	=
	\suml{n}\suml{a}
	i \frac{\pi}{\Delta} \zeta_a \e_n^a s_{n}^{a} e^{-i\e_n^a \delta z_a} 
	+\text{const}.
	\end{aligned}
	\end{align}
	As in the previous section, we assume that $|\delta z_a|/t^2 \ll \Delta$ and the phase factor $e^{-i\e_n^a \delta z_a}$ is only needed when there is a convergence issue.
	In the case of  $\eta/(t^2+\eta^2) \gtrsim  \Delta$, 
	we can see from Eq.~\ref{eq:Ssp}  that the symmetry breaking $\eta$ determines the standard saddle point which minimizes the real part of the non-interacting action $\re S[\Lambda,0]$ among various saddle points. The standard saddle point is given by 
	\begin{align}\label{eqd:Lambda0}
	(\Lambda^{(0)})^{aa}_{nn}=\sgn (n+1/2),
	\end{align}  
	and yields the dominant contribution. By contrast, when $\eta/(t^2+\eta^2) \ll\Delta$ (including $\eta=0$), it is important to keep contributions from all saddle points which only differ by phases.

	\subsection{Quadratic fluctuations and the non-interacting SFF}
	
	We now consider the fluctuations $\delta Q$ of the matrix field $Q$ around a diagonal saddle point $\Lambda$, as well as the H.S. decoupling field $\phi$.  Up to quadratic order in $\delta Q=Q-\Lambda$ and $\phi$, $S[Q,\phi]$ can be expressed as 
	\begin{align}\label{eqd:S2}
	\begin{aligned}
	&
	\delta S[\delta Q,\phi;\Lambda^{(s)}]
	\equiv
	S[\Lambda^{(s)}+\delta Q,\phi]
	-
	S[\Lambda^{(s)},0]
	=
	\suml{\qb}
	\suml{a,b,m,n}
	M_{nm}^{ab}(\qb) \delta Q_{nm}^{ab}(\qb)\delta Q_{mn}^{ba}(-\qb)
	+
	\suml{a} M'^{a} \phi^{a}_0(\vex{0})
	\\
	&+
	\suml{\qb}
	\suml{a,m,n}
	\bar{M}^{a}_{nm}(\qb)
	\delta Q_{nm}^{aa}(\qb)\phi^{a}_{m-n}(-\qb)
	+
	\frac{1}{2}
	\suml{\qb}
	\suml{a,m} \tilde{M}^a_m
	\phi^{a}_{-m}(-\qb) \phi^{a}_{m}(\qb).
	\end{aligned}	
	\end{align}
	$M$, $M'$, $\bar{M}$ and $\bar{M}$ depend on the saddle point $\Lambda$ and acquire the forms 
	\begin{align}\label{eqd:Mab}
	\begin{aligned}
	M^{ab}_{nm}(\qb)
	=\, &
	\frac{\pi}{4\tel\Delta}
	\left( 1-\frac{\Delta}{2\tel \pi}
	\suml{\kb}
	G^{a}_{n}(\kb)
	G^{b}_{m}(\kb-\qb)
	\right) 
	\\
	\approx &
	\frac{\pi}{4\tel\Delta}
	\left\lbrace 1- \left(1-\delta_{s_m^b,s_n^a} \right)
	\left[ 
	1-	i \tel s_n^a \left( \zeta_b \e_m^b e^{-i\e_m^b \delta z_b}  - \zeta_a \e_n^a e^{-i\e_n^a \delta z_a} \right) -\tel D q^2
	\right] \right\rbrace,
	\\	
	M'^{a}
	=\, &
	-\suml{\kb}
	\suml{n}
	G^{a}_n(\kb)
	= 
	i\frac{\pi}{\Delta} \suml{n} s_n^a,
	\\
	\bar{M}^{a}_{nm}(\qb)
	=\, &
	i\frac{1}{2\tel}
	\suml{\kb}
	G^{a}_n(\kb)
	G^{a}_m(\kb-\qb)
	\approx 
	i\frac{\pi}{\Delta}\left(1-\delta_{s_m^a,s_n^a} \right),
	\\
	\tilde{M}^a_m
	=\, &
	\suml{\kb} \sum_{n}
	G^{a}_n(\kb)
	G^{a}_{n+m}(\kb+\qb)
	\approx
	\suml{\kb} \sum_{n} \left[ G^{a}(\kb,\e_n) \right]^2,
	\end{aligned}
	\end{align}
	where $D=\vf^2 \tel/2$ is the diffuson constant and $G$ is defined as
	\begin{align}
	G^{a}_n(\kb)
	=
	\dfrac{1}{\zeta_a \e_n^a e^{-i\e_n^a \delta z_a}-\xi_{\kb}+\frac{i}{2\tel} s_n^a }.
	\end{align}
	
	From Eq.~\ref{eqd:S2}, one can immediately see that the bare propagator of $\delta Q$ is determined by the matrix kernel  $M$.
	Depending on the values of $M^{ab}_{nm}(\qb)$, the fluctuations $\delta Q^{ab}_{nm}(\qb)$ and $\delta Q^{ba}_{mn}(-\qb)$ fall into two different categories: the massive mode and the soft mode.
	\begin{enumerate}
		\item 
		When $s_n^a = s_m^b$,  $M^{ab}_{nm}(\qb)$ is given approximately by a constant
		\begin{align}
		M^{ab}_{nm}(\qb)=\frac{\pi}{4\tel \Delta},
		\end{align}
		and the associated $\delta Q^{ab}_{nm}(\qb)$ and $\delta Q^{ba}_{mn}(-\qb)$ are massive mode fluctuations.
		\item
		By contrast, when $s_n^a = -s_m^b$,  we have
		\begin{align}\label{eqd:MS}
		M^{ab}_{nm}(\qb)=\frac{\pi}{4 \Delta}  \left[ i s_n^a \left( \zeta_b \e_m^b e^{-i\e_m^b \delta z_b}  - \zeta_a \e_n^a e^{-i\e_n^a \delta z_a} \right) +D q^2\right],
		\end{align}
		and the corresponding fluctuations belong to a special type of soft mode named diffuson.
		These soft modes can be generated by spatial dependent unitary rotation of the saddle point:
		\begin{align}\label{eqd:Qrot}
		Q(\rb)=R^{-1}(\rb) \Lambda R(\rb).
		\end{align} 
		In the special case when $a=-b$, $n=-m-1$ and $\qb=0$, Eq.~\ref{eqd:MS} becomes 
		\begin{align}
		M^{a,-a}_{n,-n-1}(0)=\frac{\pi}{\Delta} s_n^a  \frac{\eta}{t^2+\eta^2} \pi(n+1/2).
		\end{align}
		$M^{a,-a}_{n,-n-1}(0)$ vanishes when $\eta/(t^2+\eta^2) \ll 1/\Delta$, but assumes the value of $M^{a,-a}_{n,-n-1}(0) \gtrsim 1$ when $\eta/(t^2+\eta^2) \gtrsim 1/\Delta$.
		The corresponding soft modes are the zero modes generated by symmetry transformation, i.e., $R(\rb)$ applied to the $\Lambda$ in Eq.~\ref{eqd:Qrot} becomes the spatial uniform symmetry transformation $U$.
	\end{enumerate}
	
	
	Combing Eqs.~\ref{eqd:K} and~\ref{eqd:S2}, we find that the contribution to the SFF from a diagonal saddle point $\Lambda^{(s)}$ and small fluctuations around it $\delta Q^{(s)}=Q-\Lambda^{(s)}$ can be expressed as
	\begin{align}\label{eqd:K-2}
	\begin{aligned}
	&K'(\eta,t;\Lambda^{(s)})
	=\,
	\frac{-S[\Lambda^{(s)},0]}{Z_{\phi}Z_Q}
	\int \D \phi
	e^{ -S_{\phi 0}[\phi]}
	\int \D \delta Q^{(s)} \exp \left( -\delta S[\delta Q^{(s)},\phi;\Lambda^{(s)}]\right).
	\end{aligned}
	\end{align}
	In the non-interacting case, after setting $\phi=0$ in the equation above, we obtain
	\begin{align}\label{eqd:K0-1}
	\begin{aligned}
	&K_0'(\eta,t;\Lambda^{(s)})
	=\,
	e^{-S[\Lambda^{(s)},0]}
	\dfrac{\int \D \delta Q^{(s)} \exp 
		\left( -\suml{\qb}\suml{a,b,m,n}
		M_{nm}^{ab} (\qb) (\delta Q^{(s)})_{nm}^{ab}(\qb) (\delta Q^{(s)})_{mn}^{ba}(-\qb)
		\right)}
	{\int \D \delta Q^{(s)} \exp \left(-\frac{\pi}{4\tel \Delta} \suml{\qb}\suml{a,b,m,n} (\delta Q^{(s)})_{nm}^{ab} (\delta Q^{(s)})_{mn}^{ba} \right)  }.
	\end{aligned}
	\end{align} 
	The integration over each pair of $\delta Q^{ab}_{mn}(\qb)$ and $\delta Q^{ba}_{nm}(-\qb)$ gives rise to a factor of $4\tel \Delta/\pi M^{ab}_{nm}(\qb)$ after taking into account the corresponding contribution from the normalization constant $Z_Q$, as long as $M^{ab}_{nm}(\qb) \neq 0$. This leads to
	\begin{align}\label{eqd:K0-2}
	\begin{aligned}
	&K_0(\eta,t)
	=\,
	\suml{s}
	e^{-S[\Lambda^{(s)},0]}
	\exp 
	\left[  -	
	\suml{\qb, a,b,m,n}'
	\ln \left( \frac{4\tel \Delta}{\pi} M_{nm}^{ab}(\qb)\right) 
	\right] 
	\mathcal{Z}_0^{(s)},
	\end{aligned}
	\end{align} 
	where the summation $\sum_{\qb, a,b,m,n}'$ excludes the zero modes. The contribution from zero modes denoted by $\mathcal{Z}_0^{(s)}$ here might seem to be divergent in the limit of $\eta \rightarrow 0^+$ at the quadratic order.
	However, as mentioned earlier, taking into account higher order fluctuations, one finds
	\begin{align}\label{eqd:K0-2a}
	\begin{aligned}
	&\mathcal{Z}_0^{(s)}
	=\,
	\mathcal{V}_0^{(s)}
	\exp 
	\left[ 
	-	\frac{1}{2}
	\mathcal{N}_0^{(s)}
	\ln \left( 4\tel \Delta\right) 
	\right] .
	\end{aligned}
	\end{align} 
	$\mathcal{N}_0^{(s)}$ stands for the number of the zero modes, and $\mathcal{V}_0^{(s)}$ denotes the volume of the associated saddle point manifold generated by $Q_{sp}=U^{-1}\Lambda^{(s)}U$, where $U$ belongs to the symmetry group of the non-interacting theory~\cite{Kamenev}.
	
	We can see from Eq.~\ref{eqd:K0-2} that the massive modes give rise to a nonessential constant. By contrast, the soft mode fluctuations off-diagonal in the replica space ($\delta Q^{ab}$ with $a\neq b$) lead to the exponential ramp (see more details in the following sections).
	From the analysis of saddle point action, we believe that the SFF for $\eta/(t^2+\eta^2) \gtrsim \Delta$ can be approximated by the contribution associated with the standard saddle point $K(\eta,t) \approx K'(\eta,t;\Lambda^{(0)})$. In the opposite limit $\eta/(t^2+\eta^2) \gtrsim \Delta$, it is also necessary to take into account the contributions from fluctuations around nonstandard points.

	\subsection{Nonlinear $\sigma$-model for the standard saddle point}

	Let us now focus on the soft mode fluctuations around the standard saddle point $\Lambda^{(0)}$ and investigate their interactions with the H.S. decoupling field $\phi$. 
	These soft modes are generated by rotation of $\Lambda^{(0)}$, and can be expressed as 
	\begin{align}\label{eqd:Q0}
	Q(\rb)=R^{-1}(\rb)\Lambda^{(0)} R(\rb),
	\end{align}
	where $R(\rb) \in U(2N_{\e})/U(N_{\e}) \times U(N_{\e})$, with $N_{\e}$ being the total number of Matsubara frequencies smaller than the ultraviolet cutoff $|\e_n^{\pm}| \leq \tel^{-1}$. For zero modes, $Q$ can also be expressed as Eq.~\ref{eqd:Q0} with $R(\rb)=U\in\prod_{n=1}^{N_{\varepsilon}} U(2)/U(1)\times U(1)$ being a special spatial uniform unitary rotation which leaves the non-interacting action $S[Q,0]$ invariant when $\eta=0$.
	
	We employ the parameterization Eq.~\ref{eqd:Q0} and insert it into the action $S[Q,\phi]$ (Eq.~\ref{eqd:SQ}). After expanding the $\Tr \ln $ term in gradients of the rotation matrix $R$ and the decoupling field $\phi$,
	we arrive at the nonlinear $\sigma-$model describing soft mode fluctuations of $Q$ matrix around the standard saddle point $\Lambda^{(0)}$ :
	\begin{align}\label{eqd:NLSM}
	\begin{aligned}
	\delta S[Q,\phi;\Lambda^{(0)}]
	\equiv &
	S[Q,\phi]-S[\Lambda^{(0)},0]
	\\
	=&
	\frac{\pi }{4\Delta} D
	\suml{\qb}
	q^2
	\Tr \left[  Q(\qb)Q(-\qb) \right]
	+i\frac{\pi}{\Delta}
	\Tr
	\left[ \mathcal{E}  \sigma^3
	\left( Q(\vex{0})  -\Lambda^{(0)} \right)  \right]
	\\ 
	&
	+
	i\frac{\pi}{\Delta} \suml{\qb}\suml{a,n,m} Q^{aa}_{nm}(\qb)\phi^a_{m-n}(-\qb)
	+
	\frac{1}{2}  
	\suml{\qb} \suml{a,m}
	\tilde{M}^a_m
	\phi^{a}_m(-\qb)  \phi^a_m(\qb)
	.
	\end{aligned}
	\end{align}
	Here $\tilde{M}^a_m$ is defined in Eq.~\ref{eqd:Mab} and takes the value of $\tilde{M}^a_m=-i \zeta_a  z_a /\Delta$ for the standard saddle point $\Lambda^{(0)}$ (Eq.~\ref{eqd:Lambda0}).
	$Q$ is subject to the following constraints,
	\begin{align}\label{eqd:cons}
	\begin{aligned}
	\Tr Q (\rb)=0, 
	\qquad
	Q^2(\rb)=I,
	\qquad
	Q^{\dagger}(\rb)=Q(\rb).
	\end{aligned}
	\end{align}
	It can be reparameterized as 
	\begin{align}\label{eqd:Q}
	\begin{aligned}
	&
	\begin{array}{cc}
	\quad m \geq 0 & \qquad \qquad \qquad m<0
	\end{array}
	\\
	Q_{nm}(\rb)
	=&
	\begin{bmatrix}
	\sqrt{I- X (\rb)X^\dagger (\rb)} & X(\rb)
	\\
	X^\dagger(\rb)   & -\sqrt{I- X^\dagger(\rb) X(\rb) } 
	\end{bmatrix}
	\begin{array}{c}
	n \geq 0\\
	n < 0\\
	\end{array},
	\end{aligned}
	\end{align}		
	which resolves the constraints automatically. $X^{ab}_{nm}(\rb)$ ($X^{\dagger}\,^{ab}_{nm}(\rb)$) is an unconstrained $N_{\e} \times N_{\e}$ complex matrix labeled with nonnegative (negative) frequency row index $n$ and negative (nonnegative) frequency column index $m$ in addition to replica indices $a,b$.
	
	We then substitute Eq.~\ref{eqd:Q} into Eq.~\ref{eqd:NLSM} and expand in terms of matrix fields $X$ and $X^{\dagger}$. 
	One can see, after a rescaling similar to Eq.~\ref{eq:rescale} (see also Ref.~\cite{AP}),  that terms of higher order in $X$ and $X^{\dagger}$ are also of higher order in disorder strength.
	We keep the terms up to quadratic order in $X$ in the total action $\delta S[Q,\phi;\Lambda^{(0)}]$
	as well as the quartic terms in the $\phi$-independent part of the action,
	\begin{subequations}\label{eqd:SX}
		\begin{align}
		&\delta S'[Q,\phi;\Lambda^{(0)}]
		=\,
		S_{\phi}+S_X^{(2)} +S_X^{(4)},
		\\
		\label{eqd:Sphi}
		&S_{\phi}
		=\,
		\frac{1}{2}  
		\suml{\qb} \suml{a,m}
		\tilde{M}^a_m
		\phi^{a}_{-m}(-\qb)  \phi^{a}_m(\qb),
		\\
		&\begin{aligned}\label{eqd:SX2}
		S_X^{(2)} 
		=\, &
		\suml{\kb_1,\kb_2} 
		\suml{abcd}
		\suml{m,m'\geq 0>n,n'}
		X^{\dagger}\,^{ab}_{nm}(\kb_1)
		D^{ba;dc}_{mn;n'm'}(\kb_1,\kb_2)
		X^{cd}_{m'n'}(\kb_2)
		\\
		&
		+
		\suml{\kb}\suml{ab}\suml{m>0>n}
		\left[ 
		\bar{J}\,^{ab}_{nm}(\kb)
		X\,^{ba}_{mn}(\kb)
		+
		X^{\dagger}\,^{ab}_{nm}(\kb)
		J\,^{ba}_{mn}(\kb)
		\right] ,
		\end{aligned}
		\\
		&\begin{aligned}\label{eqd:SX4}
		&S_X^{(4)}
		= 
		\frac{\pi}{16 \Delta}
		\suml{\kb_1,\kb_2,\kb_3,\kb_4} \suml{abcd}
		\suml{m,v\geq 0>n,u}
		\delta_{\kb_1+\kb_3,\kb_2+\kb_4}
		X^{\dagger}\,^{ab} _{nm} (\kb_1) X^{bc}_{mu}(\kb_2)
		X^{\dagger}\,^{cd} _{uv} (\kb_3) X^{da}_{vn}(\kb_4)
		\\
		& \times 
		\left[ 
		-2D(\kb_1 \cdot \kb_3 +\kb_2 \cdot \kb_4) 
		+D(\kb_1 + \kb_3)\cdot (\kb_2 + \kb_4)  
		+i\left( 
		\zeta_a \e_n^a e^{-i\e_n^a \delta z_a}
		-\zeta_b \e_m^b e^{-i\e_m^b \delta z_b}
		+\zeta_c \e_u^c e^{-i\e_u^c \delta t_c}
		-\zeta_d \e_v^d e^{-i\e_v^d \delta t_d}
		\right) 
		\right]  .
		\end{aligned}
		\end{align}	
	\end{subequations}
	Here $\mathcal{K}$, $\bar{J}$ and $J$ are defined as
	\begin{align}\label{eq:M}
	\begin{aligned}
	&\mathcal{K}^{ba;dc}_{mn;n'm'}(\kb_1,\kb_2)
	=
	\delta_{ad}\delta_{bc}
	\frac{\pi}{2\Delta}
	\left\lbrace 
	\left[ D k_1^2 + i  \left( \zeta_a\e_n^a e^{-i\e_n^a \delta z_a}-\zeta_b\e_m^b e^{-i\e_m^b \delta z_b} \right)  \right] 
	\delta_{mm'} \delta_{nn'} \delta_{\kb_1,\kb_2}
	\right. 
	\\
	&\left. 
	\qquad\qquad\qquad\qquad
	-i 
	\delta_{nn'} \phi^{b}_{m-m'}(\kb_1-\kb_2)
	+i 
	\delta_{mm'} \phi^{a}_{n'-n}(\kb_1-\kb_2)
	\right\rbrace 
	,
	\\
	&\bar{J}^{ab}_{nm}(\kb)
	=
	\delta_{ab}
	i \frac{\pi}{\Delta}
	\phi^{a}_{n-m}(-\kb)
	,
	\qquad
	J\,^{ba}_{mn}(\kb)
	=
	\delta_{ab}
	i \frac{\pi}{\Delta} 
	\phi^{a}_{m-n}(\kb)
	.
	\end{aligned}
	\end{align}	
	If one equates $\delta Q_{nm}^{ab}(\qb)$ with $X^{ab}_{nm}(\qb)$ for $n \geq 0 >m$ and $X^{\dagger}\,^{ab}_{nm}(-\qb)$ for $m \geq 0 >n$,
	the action derived earlier by considering only the quadratic order fluctuations, i.e., Eq.~\ref{eqd:S2}, is consistent with the first few terms in Eq.~\ref{eqd:SX} as expected.

	\subsubsection{Feynman rules}

	From the quadratic action $S_X^{(2)}$ (Eq.~\ref{eqd:SX2}), we find the bare $X$ propagator, i.e., the bare diffuson propagator
	\begin{align}\label{eqd:GX0}
	\begin{aligned}
	(\GG^{(0)}_{X}\,)^{ab;b'a'}_{nm;m'n'}(\kb,\kb')
	\equiv 
	\braket{X^{ab}_{nm}(\kb)X^{\dagger}\,^{b'a'}_{m'n'}(\kb')}_0
	=
	\frac{2\Delta}{\pi} \D_0(\kb,\zeta_a \e_n^a e^{-i\e_n^a \delta z_a}- \zeta_b\e_m^b e^{-i\e_m^b \delta z_b})\delta_{nn'}\delta_{mm'}\delta_{a,a'}\delta_{b,b'} \delta_{\kb,\kb'},
	\end{aligned}
	\end{align}
	where we have defined
	\begin{align}\label{eqd:D0}
	\begin{aligned}
	\D_0(\kb;\ww)
	=
	\frac{1}{Dk^2-i \ww }.
	\end{aligned}
	\end{align}
	
	The bare propagator of the H.S. decoupling field $\phi$ arises from the part of the action involving only the bosonic field $\phi$, including $S_{\phi0}$ (Eq.~\ref{eqd:Sphi0}) and $S_{\phi}$ (Eq.~\ref{eqd:Sphi}),
	and takes the form of
	\begin{align}
	\begin{aligned}
	i\left[ \GG_{\phi}^{(0)}(\qb,\ww_m^a)\right]^{ab} 
	=
	\braket{\phi^{a}_m(\qb)\phi^{b}_{-m}(-\qb)}_0
	=
	i \delta_{ab}   \frac{\zeta_a}{z_a } \left[ L^2 V^{-1}+i\frac{\zeta_a}{z_a} \tilde{M}_m^a \right]^{-1}
	=
	i \delta_{ab}   \frac{\zeta_a}{z_a } \gamma  \Delta,
	\end{aligned}
	\end{align}
	where
	\begin{align}
	\begin{aligned}
	\gamma 
	\equiv&
	\dfrac{\Delta^{-1}}{L^2 V^{-1}+\Delta^{-1}}
	=
	\dfrac{\nu_0}{V^{-1}+\nu_0}.
	\end{aligned}
	\end{align}
	
	The diagrammatic representations of the bare $X$ and $\phi$ propagators are depicted in Figs.~\ref{fig:p1}~(a) and~(g), respectively. As for the previous RMT model, the double solid black lines with oppositely directed
	arrows indicate the matrix field $X$, while the red wavy line denotes the bosonic $\phi$. In Figs.~\ref{fig:p1}~(b)-(f), we also show the 4-point vertex for matrix field $X$ as well as the interaction vertices that couple $X$ and $\phi$. The corresponding analytical expressions are as follows:
	\begin{align}
	\begin{aligned}
	&(b)	=
	-\frac{\pi}{8 \Delta}
	\delta_{\kb_1+\kb_3,\kb_2+\kb_4}
	X^{\dagger}\,^{ab} _{nm} (\kb_1) X^{bc}_{mu}(\kb_2)
	X^{\dagger}\,^{cd} _{uv} (\kb_3) X^{da}_{vn}(\kb_4)
	\\
	& \times 
	\left[ 
	-2D(\kb_1 \cdot \kb_3 +\kb_2 \cdot \kb_4) 
	+D(\kb_1 + \kb_3)\cdot (\kb_2 + \kb_4)  
	+i\left( 
	\zeta_a \e_n^a e^{-i\e_n^a \delta z_a}
	-\zeta_b \e_m^b e^{-i\e_m^b \delta z_b}
	+\zeta_c \e_u^c e^{-i\e_u^c \delta t_c}
	-\zeta_d \e_v^d e^{-i\e_v^d \delta t_d}
	\right) 
	\right],
	\\
	&(c)=+i\frac{\pi}{2\Delta}\phi^{a}_{n-n'}(\kb_1-\kb_2) 
	X^{\dagger}\,^{ba}_{mn}(\kb_1)
	X^{ab}_{n'm}(\kb_2),
	\qquad
	(d)=-i\frac{\pi}{2\Delta}
	\phi^{b}_{m'-m}(\kb_1-\kb_2)
	X^{\dagger}\,^{ba}_{mn}(\kb_1)
	X^{ab}_{nm'}(\kb_2),
	\\
	&
	(f)=
	-
	i \frac{\pi}{\Delta}
	\phi^{a}_{m-n}(-\kb)
	X\,^{aa}_{nm}(\kb),
	\qquad \qquad \qquad\qquad\quad\,\,
	(e)=
	-i \frac{\pi}{\Delta} 
	\phi^{a}_{n-m}(\kb)
	X^{\dagger}\,^{aa}_{mn}(\kb).
	\end{aligned}
	\end{align}

	\subsubsection{Interaction dressed $\phi$ propagator}
	
	The leading order self-energy diagram for the bosonic field $\phi$ is shown in Fig.~\ref{fig:p1}(i) and its analytical expression takes the form of
	\begin{align}\label{eqd:Sigphi}
	\begin{aligned}
	-i \Sigma_{\phi}^{ab}(\qb,\ww_m^a)
	=&
	-\delta_{ab} \frac{2\pi}{\Delta} 
	\suml{0>n\geq -|m|}
	\frac{1}{ D q^2 - i\zeta_a \left(  \e_{n+|m|}^ae^{-i\e_{n+|m|}^a \delta z_a}-\e_n^a e^{-i\e_n^a \delta z_a} \right)  }
	\\
	=&
	-\delta_{ab}
	\frac{z_a}{\Delta} \frac{\ww_{|m|}^a}{ D q^2 - i\zeta_a  \ww_{|m|}^a}.
	\end{aligned}
	\end{align}
	Here we have used
	\begin{align}
	\begin{aligned}
	\suml{0>n\geq -|m|}
	=|m|= z_a\frac{\ww_{|m|}^a}{2\pi}.
	\end{aligned}
	\end{align}
	
	Inserting this result into the Dyson equation for bosonic field $\phi$, we find the interaction dressed $\phi$ propagator
	\begin{align}\label{eqd:Gphi}
	\begin{aligned}
	i\GG_{\phi}^{ab}(\qb,\ww_m^a)
	=
	\left[ \left( 
	-i\GG^{(0)}_{\phi}\,^{-1} (\qb,\ww_m^a)+i\Sigma_{\phi}(\qb,\ww_m^a)
	\right)^{-1}\right]^{ab} 
	=
	i\delta_{ab} \zeta_a \frac{\gamma \Delta }{z_a} 
	\dfrac{\D_{\gamma}(\qb, \zeta_a \ww_{|m|}^a)}
	{\D_{0}(\qb, \zeta_a \ww_{|m|}^a)},
	\end{aligned}
	\end{align} 
	where $D_{\gamma}(\kb,\ww)$ is defined as
	\begin{align}\label{eqd:Dgamma}
	\begin{aligned}
	\D_{\gamma}(\kb,\ww)
	=
	\frac{1}{Dk^2-i(1-\gamma) \ww}.
	\end{aligned}
	\end{align}
	We use a red wavy line with a solid dot to represent diagrammatically the dressed $\phi$ propagator $\GG_{\phi}$ (Fig.~\ref{fig:p1}(h)),  and for the bare $\phi$ propagator $\GG_{\phi}^{(0)}$ the solid dot is replaced with an open one (Fig.~\ref{fig:p1}(g)).
	
	\subsubsection{Interaction dressed diffuson propagator}
	
	For matrix field $X$, the leading order self-energy diagrams are shown in Fig.~\ref{fig:p2}, and their total contribution can be expressed as a summation of two parts:
	\begin{align}\label{eqd:Sigma}
	(\Sigma_{X})\,^{ab;b'a'}_{nm;m'n'} (\kb,\kb')
	=
	\left[ 
	(\Sigma_{\msf{diag}})\,^{ab;ba}_{nm;mn}(\kb;\kb)\delta_{n,n'}\delta_{m,m'} 
	+
	(\Sigma_{\msf{off}})\,^{aa;aa}_{nm;m'n'}(\kb;\kb)\delta_{n-m,n'-m'}
	\delta_{ab}
	\right]
	\delta_{aa'}\delta_{bb'} \delta_{\kb,\kb'}.
	\end{align} 
	$\Sigma_{\msf{diag}}$ is diagonal in the Matsubara frequency space and represents the contribution from diagrams (a)-(d) in Fig.~\ref{fig:p2}. The remaining contribution from diagrams (e)-(f) in Fig.~\ref{fig:p2} is denoted as $\Sigma_{\msf{off}}$ with nonvanishing off-diagonal components in the Matsubara frequency space (i.e. $(\Sigma_{\msf{off}})_{nm;m'n'}$ is non-zero for $m' \neq m$ and $n'\neq n$). 
	We emphasize that $\Sigma_{\msf{off}}^{ab;ba}$ vanishes when $a \neq b$ and therefore is irrelevant to the inter-replica $X$ propagator $\GG_{X}^{a,-a;-a,a}$.
	Self-energy diagrams (g) and (h) in Fig.~\ref{fig:p2}  are not considered here as this kind of diagrams usually cancel with the Jacobian from the parameterization Eq.~\ref{eqd:Q}.
	
	The analytical expressions for diagrams (a)-(f) in Fig.~\ref{fig:p2} are given by, in respective order,
	\begin{align}\label{eqd:Sigma1-4}
	\begin{aligned}
	(\Sigma^{(a)}_{X})^{ab;ba}_{nm;mn}(\kb,\kb)
	=&
	-i
	\frac{\pi}{2\Delta}
	\suml{\qb}\suml{u \leq n}
	\D_{0} \left( \kb-\qb, 
	\zeta_a \e_{n-u}^a e^{-i\e_{n-u}^a \delta z_a}-\zeta_b \e_m^b e^{-i\e_m^b \delta z_b} \right) 
	\GG_{\phi}^{aa}(\qb,\ww_u^a),
	\\
	(\Sigma^{(b)}_{X})^{ab;ba}_{nm;mn}(\kb,\kb)
	=&
	-i
	\frac{\pi}{2\Delta}
	\suml{\qb}\suml{u<-m}
	\D_{0} \left( \kb-\qb, \zeta_a \e_n^a e^{-i\e_n^a \delta z_a}-\zeta_b \e_{m+u}^b e^{-i\e_{m+u}^b \delta z_b} \right) 
	\GG_{\phi}^{bb}(\qb,\ww_u^b),
	\\
	(\Sigma^{(c)}_{X})^{ab;ba}_{nm;mn}(\kb,\kb)
	=&
	i
	\frac{\pi}{2\Delta} 
	\suml{\qb}\suml{u<-n}
	\left[ 
	\D_{0}^{-1} \left( \qb, \zeta_a \e_n^a e^{-i\e_n^a \delta z_a}-\zeta_a \e_{n+u}^a  e^{-i\e_{n+u}^a \delta z_a} \right) 
	+\D_{0}^{-1}\left( \kb, \zeta_a \e_n^a e^{-i\e_n^a \delta z_a}-\zeta_b \e_m^b e^{-i\e_m^b \delta z_b} \right)
	\right] 
	\\
	& \times
	\D_{0}^2 \left( \qb, \zeta_a \e_n^a e^{-i\e_n^a \delta z_a}-\zeta_a \e_{n+u}^a  e^{-i\e_{n+u}^a \delta z_a}\right) 
	\GG_{\phi}^{aa}(\qb,\ww_u^a),
	\\
	(\Sigma^{(d)}_{X})^{ab;ba}_{nm;mn}(\kb,\kb)
	=&
	i
	\frac{\pi}{2\Delta} 
	\suml{\qb}\suml{u\leq m}
	\left[ 
	\D_{0}^{-1} \left( \qb, \zeta_b \e_{m-u} e^{-i\e_{m-u}^b \delta z_b}-\zeta_b \e_m^b e^{-i\e_m^b \delta z_b}  \right) 
	+\D_{0}^{-1}\left( \kb, \zeta_a \e_n^a e^{-i\e_n^a \delta z_a}-\zeta_b \e_m^b e^{-i\e_m^b \delta z_b} \right) 
	\right] 
	\\
	& \times
	\D_{0}^2 \left( \qb, \zeta_b \e_{m-u}e^{-i\e_{m-u}^b \delta z_b}-\zeta_b \e_m^b e^{-i\e_m^b \delta z_b}  \right) 
	\GG_{\phi}^{bb}(\qb,\ww_u^b),
	\\
	(\Sigma^{(e)}_{X})^{aa;aa}_{nm;m'n'}(\kb,\kb)
	=&
	i\frac{\pi}{2\Delta}
	\delta_{n-m,n'-m'}\suml{\qb}
	\D_0 \left( \kb-\qb, \zeta_a \e_{n'}e^{-i\e_{n'}^a \delta z_a} -\zeta_a\e_{m}^a e^{-i\e_m^a \delta z_a}\right) 
	\GG_{\phi}^{aa}(\qb,\e_{n-n'}^a),
	\\
	(\Sigma^{(f)}_{X})^{aa;aa}_{nm;m'n'}(\kb,\kb)
	=&
	i\frac{\pi}{2\Delta}
	\delta_{n-m,n'-m'}\suml{\qb}
	\D_0 \left( \kb-\qb, \zeta_a \e_{n}e^{-i\e_{n}^a \delta z_a} -\zeta_a\e_{m'}^a e^{-i\e_{m'}^a \delta z_a} \right) 
	\GG_{\phi}^{aa}(\qb,\e_{n-n'}^a),
	\end{aligned}
	\end{align}
	where the external Matsubara frequency indices satisfy $n,n'\geq 0>m,m'$.  
	
	Using the Dyson equation for $X$ propagator:
	\begin{align}\label{eqd:GX1}
	\begin{aligned}
	\suml{\kb'}\suml{a',b'}\suml{n'\geq >m'}
	\left( 
	(\GG^{(0)}_{X} )^{-1}-\Sigma_{X}
	\right)^{ab;b'a'}_{nm;m'n'}(\kb,\kb')
	\left( \GG_{X}\right)^{a'b';b''a''}_{n'm';m''n''}(\kb',\kb'')
	=
	\delta_{aa''}\delta_{bb''}\delta_{nn''}\delta_{mm''}\delta_{\kb,\kb''},
	\end{aligned}
	\end{align}
	as well as the structure of the self-energy (Eq.~\ref{eqd:Sigma}), we find that the inter-replica $X$ propagator $\GG_{X}^{ab;b''a''}$ is determined entirely by the bare propagator $\GG^{(0)}_{X}\,^{ab;ba}$ and the inter-replica self-energy $\Sigma_{\msf{dia}}^{ab;ba}$:
	\begin{align}
	&\begin{aligned}\label{eqd:GX2a}
	\GG_{X}\,^{ab;b''a''}_{nm;m''n''}(\kb,\kb'')
	=
	\dfrac{1}{	
		(\GG^{(0)}_{X})^{-1}\,^{ab;ba}_{nm;mn}(\kb,\kb)-(\Sigma_{\msf{dia}})^{ab;ba}_{nm;mn}(\kb,\kb)
	}
	\delta_{aa''}\delta_{bb''}\delta_{nn''}\delta_{mm''},
	\qquad \qquad\qquad \qquad\quad\,\,
	a \neq b.
	\end{aligned}
	\end{align}
	We're interested in the connected SFF defined in Eq.~\ref{eq:conSFF}, and therefore will focus on the inter-replica components of $X$ propagator and self-energy.
	
	The summation of contributions from diagrams (a)-(d) in Fig.~\ref{fig:p2} gives rise to the inter-replica self energy $(\Sigma_{\msf{diag}}^{ab;ba})$ which can be expressed as
	\begin{align}\label{eqd:Sigdiag}
	(\Sigma_{\msf{diag}})\,^{ab;ba}_{nm;mn}(\kb,\kb)
	=
	-\frac{\pi}{2\Delta}
	\left[ 
	\delta D k^2 -i \delta z (\zeta_a \e_n^a e^{-i\e_n^a \delta z_a}-\zeta_b \e_m^b e^{-i\e_m^b \delta z_b})
	+
	\lambda^{ab}_{nm}
	\right].
	\end{align} 
	Here $\delta D$ and $\delta z$ are related to the renormalization effect and will be ignored in the following. We concentrate on the influence of the mass term $\lambda^{ab}_{nm}$ given by
	\begin{align}\label{eqd:lambda}
	\begin{aligned}
	\lambda^{ab}_{nm}
	=&\,
	+i
	\suml{\qb}\suml{-n\leq u \leq n}
	\D_{0} \left( \qb, -\zeta_a \ww_u^a \right) 
	\GG_{\phi}^{aa}(\qb,\ww_u^a)
	+i
	\suml{\qb}\suml{m<u<-m}
	\D_{0} \left(\qb, -\zeta_b \ww_u^b \right) 
	\GG_{\phi}^{bb}(\qb,\ww_u^b)
	\\
	=
	&
	+
	\suml{\qb}\suml{-n\leq u\leq n}
	i \zeta_a \frac{\gamma \Delta }{z_a} 
	\D_{0} \left( \qb, -\zeta_a \ww_u^a \right) 
	\dfrac{\D_{\gamma}(\qb, \zeta_a  \ww_{|u|}^a)}
	{\D_{0}(\qb, \zeta_a \ww_{|u|}^a)}
	+
	\suml{\qb}\suml{m<u<-m}
	i \zeta_b \frac{\gamma \Delta }{z_b} 
	\D_{0} \left( \qb, -\zeta_b \ww_{u}^b\right) 
	\dfrac{\D_{\gamma}(\qb, \zeta_b \ww_{|u|}^b)}
	{\D_{0}(\qb, \zeta_b \ww_{|u|}^b)},
	\end{aligned}
	\end{align}
	where in the second equality, we have substituted the expression for $\phi$ propagator $\GG_{\phi}$ (Eq.~\ref{eqd:Gphi}).
	The inter-replica $X$ propagator can be obtained by
	inserting Eqs.~\ref{eqd:Sigdiag} and~\ref{eqd:GX0} into Eq.~\ref{eqd:GX2a}, and takes the form
	\begin{align}\label{eq:GXoff}
	\begin{aligned}
	(\GG_{X}\,)^{ab;b'a'}_{nm;m'n'}(\kb,\kb')
	=
	\frac{2\Delta}{\pi} 
	\dfrac{1}{
		Dk^2-i(\zeta_a \e_n^a e^{-i\e_n^a \delta z_a}- \zeta_b\e_m^b e^{-i\e_m^b \delta z_b})+\lambda^{ab}_{nm}
	}
	\delta_{nn'}\delta_{mm'}\delta_{a,a'}\delta_{b,b'} \delta_{\kb,\kb'},
	\qquad
	a\neq b.
	\end{aligned}
	\end{align}
	Here we have ignored the renormalization effect, and set $\delta D$ and $\delta z$ to zero.
	
	\subsubsection{Inter-replica mass}
	
	We will now evaluate the inter-replica mass $\lambda^{ab}_{nm}$ with $a \neq b$ (Eq.~\ref{eqd:lambda}) in the limit of $\eta\rightarrow 0^+$.
	To proceed, we rewrite it as a summation of three terms:
	\begin{align}
	\begin{aligned}
	&\lambda^{ab}_{nm}=
	\lambda_0
	+
	\lambda_1
	+
	\lambda_2,
	\\
	&\lambda_0
	=
	i  \left( \frac{ \zeta_a}{z_a} +\frac{ \zeta_b}{z_b} \right) 
	\gamma\Delta
	\suml{\qb} \D_{0} \left( \qb,0 \right),
	\\
	&\lambda_1
	=
	i\gamma \Delta
	\suml{1 \leq u\leq n}
	\frac{ \zeta_a }{z_a} 
	I_1(\ww_u^a)
	+
	i\gamma \Delta 
	\suml{1 \leq u\leq -m-1}
	\frac{ \zeta_b}{z_b} 
	I_1(\ww_u^b),
	\\
	&\lambda_2
	=
	i\gamma \Delta\suml{1 \leq u\leq n}
	\frac{\zeta_a }{z_a} 
	I_2(\ww_u^a)
	+
	i\gamma \Delta
	\suml{1 \leq u\leq -m-1}
	\frac{ \zeta_b }{z_b} 
	I_2(\ww_u^b),
	\end{aligned}
	\end{align}
	where
	\begin{align}
	\begin{aligned}
	&
	I_1(x)
	\equiv
	\frac{1}{2}
	\suml{\qb}
	\left[ \D_{0} \left( \qb, -x \right) +\D_{0} \left( \qb, x \right) \right] 
	\left[ 
	\dfrac{\D_{\gamma}(\qb, x)}
	{\D_{0}(\qb, x)}
	-	
	\dfrac{\D_{\gamma}(\qb, -x)}
	{\D_{0}(\qb, -x)}
	\right] ,
	\\
	&
	I_2(x)
	\equiv
	\frac{1}{2}
	\suml{\qb}
	\left[ \D_{0} \left( \qb, -x \right) +\D_{0} \left( \qb, x \right) \right] 
	\left[ 
	\dfrac{\D_{\gamma}(\qb, x)}
	{\D_{0}(\qb, x)}
	+
	\dfrac{\D_{\gamma}(\qb, -x)}
	{\D_{0}(\qb, -x)}
	\right] .
	\end{aligned}
	\end{align}
	
		The first term $\lambda_0$ contains a momentum integral that is infrared divergent. However, it vanishes in the limit of $\eta \rightarrow 0^+$  due to the overall factor $(\zeta_a/z_a+\zeta_b/z_b)\rightarrow 0$, and therefore can be omitted.
	In  the case where $n=-m-1$ and $a=-b$, the last term $\lambda_2$ vanishes in the $\eta \rightarrow 0^+$ limit since $I_2(x)$ is an even function of $x$. This means that $\lambda_2$ contributes to the renormalization effect and can also be neglected.
	As a result, we're left with $\lambda_{1}$, which after the momentum integration reduces to
	\begin{align}\label{eqd:lambda}
	\begin{aligned}
	&
	\lambda^{ab}_{nm}
	\approx
	\lambda_1
	=
	\frac{1}{4}\frac{L^2 \Delta }{D} \frac{\gamma^2}{2-\gamma} 
	\left( 
	\frac{n }{z_a} 
	+
	\frac{-m-1 }{z_b} 
	\right)
	=
	\frac{1}{8\pi}\frac{1 }{\nu_0D} \frac{\gamma^2}{2-\gamma} 
	\left( 
	\e_n^a
	-
	\e_m^b
	\right)
	-
	\frac{1}{4}\frac{1 }{\nu_0D} \frac{\gamma^2}{2-\gamma} \frac{1}{t},
	\qquad
	a\neq b.
	\end{aligned}
	\end{align}

	\subsection{ Contribution to the SFF from the diffusons around the standard saddle point}
	
	As mentioned earlier, for the case of $\eta/(t^2+\eta^2) \gtrsim \Delta$,  we can consider only the contribution from the soft mode fluctuations around the dominant saddle point $\Lambda^{(0)}$ and ignore that from all nonstandard ones. In this case, to the leading order, the SFF can be expressed as the following equation up to an unessential constant,
	\begin{align}\label{eqd:K2-1}
	\begin{aligned}
	&K'(\eta,t;\Lambda^{(0)})
	\propto
	e^{-S[\Lambda^{(0)},0]}
	\frac{
		\int \D \phi
		\exp \left[ \frac{i}{2} \suml{\qb}\phi(-\qb)  \GG_{\phi}^{-1}(\qb) \phi(\qb)  \right] 
		\int \D (X^{\dagger},X) 
		\exp \left( 
		- \suml{\kb} X^{\dagger}(\kb) \GG_X^{-1} (\kb,\kb) X(\kb)
		\right) 
	}
	{
		\int \D \phi
		e^{ -S_{\phi 0}[\phi] }
		\int \D (X^{\dagger},X) 
		e^{-\frac{\pi}{2\tel\Delta} \suml{\qb} \Tr X^{\dagger}(\qb) X(\qb)}
	}.
	\end{aligned}
	\end{align}
	Here $\GG_X$ (Eq.~\ref{eqd:GX1})  and $\GG_{\phi}$ (Eq.~\ref{eqd:Gphi}) denote, respectively, the interaction dressed $X$ and $\phi$ propagators, and we have neglected the massive modes' contribution.
	In the opposite case of $\eta/(t^2+\eta^2) \ll \Delta$, additional contribution from the nonstandard saddle points is needed as well.

	Performing the integration in Eq.~\ref{eqd:K2-1} leads to
	\begin{align}
	&\begin{aligned}\label{eqd:K2-2a}
	&K'(\eta,t;\Lambda^{(0)})
	\propto
	e^{-S[\Lambda^{(0)},0]}
	\exp
	\left\lbrace 
	-\suml{\kb}
	\sum_{a b} \suml{n\geq 0>m} 
	\ln  \left( 
	\frac{2\tel \Delta }{\pi }
	\GG_{X}^{-1}\,^{ab;ba}_{nm;mn}(\kb,\kb)\right) 
	+\frac{1}{2} \suml{\qb}  \suml{a,m}  
	\ln  
	\left( 
	(1-\gamma)
	\dfrac{\D_{\gamma}(\qb, \zeta_a \ww_{|m|}^a)}
	{\D_{0}(\qb, \zeta_a \ww_{|m|}^a)} 
	\right) 
	\right\rbrace.
	\end{aligned}
	\end{align}
	Similarly, for the non-interacting case, one can show that
	\begin{align}
	\begin{aligned}\label{eqd:K2-2b}
	&K'_0(\eta,t;\Lambda^{(0)})
	\propto
	e^{-S[\Lambda^{(0)},0]}
	\exp
	\left\lbrace 
	-\suml{\kb}\sum_{a b} \suml{n\geq 0>m}  \ln  \left( 
	\frac{2\tel \Delta }{\pi }
	(\GG_{X}^{(0)})^{-1}\,^{ab;ba}_{nm;mn} (\kb,\kb)\right) 
	\right\rbrace,
	\end{aligned}
	\end{align}
	consistent with Eq.~\ref{eqd:K0-2} as expected.
	
	To calculate the connected SFF $K^{\msf{con}}(\eta,t)$ ($K_{0}^{\msf{con}}(\eta,t)$), we can now focus on the terms involving only the inter-replica $X$ propagator $\GG_{X}^{a,-a;-a,a}$ ($\GG_{X}^{(0)}\,^{a,-a;-a,a}$)  in Eq.~\ref{eqd:K2-2a} (Eq.~\ref{eqd:K2-2b}). All remaining terms, including the one from saddle point action $S[\Lambda^{(0)},0]$, the intra-replica $X$ propagator $\GG_{X}^{aa;aa}$ ($\GG_{X}^{(0)}\,^{aa;aa}$) and the $\phi$ propagator $\GG_{\phi}^{aa}$, contribute to the disconnected SFF.
	Substituting the approximated expression for the inter-replica $X$ propagator Eq.~\ref{eq:GXoff} which ignores the renormalization effect, we find the contribution from soft mode fluctuations around $\Lambda^{(0)}$ to $\ln K'^{\msf{con}}$:
	\begin{align}\label{eqd:K-3}
	\begin{aligned}
	&\ln K'^{\msf{con}}(\eta,t;\Lambda^{(0)})
	=
	-\suml{\kb}\sum_{a\neq b} \suml{n\geq 0>m}  \ln  \left( 
	\frac{2\tel \Delta }{\pi }
	(\GG_{X}^{-1})^{ab;ba}_{nm;mn} (\kb,\kb)\right) 
	\\
	=	&
	-\suml{\kb} \suml{a\neq b} \suml{n\geq 0>m} 
	\ln \left\lbrace  \tel  \left[ D  k^2-i (\zeta_a \e_n^a e^{-i\e_n^a \delta z_a}-\zeta_b \e_m^b e^{-i\e_m^b \delta z_b})+\lambda^{ab}_{nm} \right] \right\rbrace 
	\\
	=&
	\suml{a\neq b} 
	\int_{-\infty}^{\infty} \frac{d\e}{2\pi}
	\int_{-\infty}^{\infty} \frac{d\e'}{2\pi}
	\ln \left( 1+e^{-i \zeta_a z_a\e} \right) 
	\ln \left( 1+e^{-i \zeta_b z_b \e'} \right) 
	\suml{\kb} 
	\dfrac{1}{\left( Dk^2-i(\e-\e')+\lambda^{ab}(\zeta_a\e,\zeta_b \e')\right)^2}
		\left( \frac{\partial \lambda^{ab}}{\partial \e}-i \right) 
		\left( \frac{\partial \lambda^{ab}}{\partial \e'}+i\right).
	\end{aligned}
	\end{align}
	Here in the last equality, we have converted the Matsubara frequency summations into a continuous integral.
	$\lambda^{ab}(\zeta_a \e,\zeta_b \e')$ represents the mass $\lambda^{ab}_{nm}$ (Eq.~\ref{eqd:lambda}) after the analytic continuation $\e_n^a \rightarrow \zeta_a\e$, $\e_m^b \rightarrow \zeta_b\e'$. It assumes the following form in the limit of $\eta\rightarrow 0^+$,
	\begin{align}\label{eqd:lambdaoff}
	\begin{aligned}
	&
	\lambda^{ab}(\zeta_a\e,\zeta_b\e')
	\approx
	\frac{1}{8\pi}\frac{1 }{\nu_0D} \frac{\gamma^2}{2-\gamma} 
	\left( 
	\zeta_a\e
	-
	\zeta_b\e'
	\right)
	-
	\frac{1}{4}\frac{1 }{\nu_0D} \frac{\gamma^2}{2-\gamma} \frac{1}{t},
	\qquad
	a\neq b.
	\end{aligned}
	\end{align}
	In the non-interacting case, the corresponding contribution to $\ln K_{0}^{\msf{con}}$ can be simply obtained from Eq.~\ref{eqd:K-3} by setting $\lambda^{ab}$ to 0:
	\begin{align}\label{eqd:K0-3}
	\begin{aligned}
	\ln K_{0}'^{\msf{con}}(\eta,t;\Lambda^{(0)})
	=&
	\suml{a\neq b} 
	\int_{-\infty}^{\infty} \frac{d\e}{2\pi}
	\int_{-\infty}^{\infty} \frac{d\e'}{2\pi}
	\ln \left( 1+e^{-i \zeta_a z_a\e} \right) 
	\ln \left( 1+e^{-i \zeta_b z_b \e'} \right) 
	\suml{\kb} 
	\dfrac{1}{\left( Dk^2-i(\e-\e')\right)^2}.
	\end{aligned}
	\end{align}
	If one keeps only the $\kb=0$ terms in the summations, Eqs.~\ref{eqd:K-3} and~\ref{eqd:K0-3} become equivalent to their zero-dimensional versions Eqs.~\ref{eq:K-2} and~\ref{eq:K0-2} for the RMT model.

	As in the previous section, we replace $	\ln \left( 1+e^{-i \zeta_a z_a\e} \right) 
		\ln \left( 1+e^{-i \zeta_b z_b \e'} \right) $ with $e^{-i \zeta_a z_a\e} e^{-i \zeta_b z_b \e'} $ in Eqs.~\ref{eqd:K-3} and~\ref{eqd:K0-3} assuming that the neglected terms from this replacement will cancel with part of higher order corrections.
	After the replacement and a change of variables as in Eq.~\ref{eq:cov},  Eq.~\ref{eqd:K-3} becomes
	\begin{align}\label{eqd:K2-4}
	\begin{aligned}
	\ln K''^{\msf{con}}(\eta,t;\Lambda^{(0)})
	=&
	\suml{a\neq b} 
	\int_{-\infty}^{\infty} \frac{d\e}{2\pi}
	\int_{-\infty}^{\infty} \frac{d\e'}{2\pi}
	e^{-i \zeta_a z_a\e-i \zeta_b z_b \e'} 
	\suml{\kb} 
	\dfrac{1}{\left( Dk^2-i(\e-\e')+\lambda^{ab}(\zeta_a \e,\zeta_b \e')\right)^2}
		\left( \frac{\partial \lambda^{ab}}{\partial \e}-i \right) 
		\left( \frac{\partial \lambda^{ab}}{\partial \e'}+i\right)
	\\
	=&
	\suml{a} 	\suml{\kb} 
	\int_{-E_{\msf{UV}}}^{E_{\msf{UV}}} \frac{dE}{2\pi}
	\int_{-\infty}^{\infty} \frac{d\e}{2\pi}
	e^{-i \zeta_a t \ww-2 \eta E} 
	\left( \dfrac{1}{Dk^2-i \ww+\lambda^{a,-a}(E)} \right)^2
		\left[ 	\left(\frac{1}{2} \frac{\partial \lambda^{a,-a}(E)}{\partial E}\right)^2+1\right] 
	\\
	=&
	\suml{a} \suml{\kb} 
	\int_{-E_{\msf{UV}}}^{E_{\msf{UV}}} \frac{dE}{2\pi}
	t e^{-\zeta_a t(\lambda^{a,-a}(E) +Dk^2)} e^{-2 \eta E}
	\Theta \left( \zeta_a(\re\lambda^{a,-a}(E)+Dk^2) \right) 
		\left[ 	\left( \frac{1}{2}\frac{\partial \lambda^{a,-a}(E)}{\partial E}\right)^2+1\right] .
	\end{aligned}
	\end{align}
	Here we have used the fact that $\lambda^{a,-a}$ depends on $E=(\e+\e')/2$ but not $\ww=\e-\e'$.
	We have also imposed the ultraviolet cutoff $E_{\msf{UV}}=\tel^{-1}$ and ignored the variation of single-density of states near Fermi surface.
	
	For the non-interacting case, following an analogous procedure, we instead have
	\begin{align}\label{eqd:K2-40}
	\begin{aligned}
	\ln K_{0}''^{\msf{con}}(\eta,t)
	=&
	\suml{a\neq b} 
	\int_{-\infty}^{\infty} \frac{d\e}{2\pi}
	\int_{-\infty}^{\infty} \frac{d\e'}{2\pi}
	e^{-i \zeta_a z_a\e-i \zeta_b z_b \e'} 
	\suml{\kb} 
	\dfrac{1}{\left( Dk^2-i(\e-\e')\right)^2}
	\\
	=&
	\suml{a} 	\suml{\kb} 
	\int_{-E_{\msf{UV}}}^{E_{\msf{UV}}} \frac{dE}{2\pi}
	\int_{-\infty}^{\infty} \frac{d\e}{2\pi}
	e^{-i \zeta_a t \ww} e^{-2\eta E}
	\left( \dfrac{1}{Dk^2-i \ww} \right)^2
	=
	\int_{-E_{\msf{UV}}}^{E_{\msf{UV}}} \frac{dE}{2\pi}
	e^{-2\eta E}
	\suml{\kb} 
	t e^{-tDk^2}  .
	\end{aligned}
	\end{align}
	In the regime where $t\gg E_{\msf{Th}}^{-1}$, with $E_{\msf{Th}}^{-1}=L^2/D$ being the Thouless time (i.e., the time it takes to diffuse the whole system),
		the dominate contribution to the integral in Eq.~\ref{eqd:K2-40} comes from the region $|\ww|\sim t^{-1} \ll E_{\msf{Th}}$ and $\kb=0$.
		The contribution from $\kb \neq 0$ is suppressed since $Dk^2 \gtrsim E_{\msf{Th}} \gg |\ww|$.
		As a result, only the $\kb=0$ term in the momentum summation needs to be retained, and the equation above become equivalent to Eq.~\ref{eq:K2-30}.
	If one compares just the $\kb=0$ terms in Eq.~\ref{eqd:K2-4} and Eq.~\ref{eqd:K2-40}, we find the presence of mass suppresses the growth of the connected SFF as in the zero-dimensional case discussed in the previous section. 
	
	In the diffusive regime where $E_{\msf{TH}}^{-1} \gg t\gg\tel$, we need to include contribution from nonzero momentum $\kb$ as well.
	Taking the $\eta\rightarrow 0^+$ limit and performing the momentum integrations in Eqs.~\ref{eqd:K2-4} and~\ref{eqd:K2-40}, we obtain
	\begin{align}\label{eq:K2-5}
	\begin{aligned}
	\ln K''^{\msf{con}}(\eta \rightarrow 0^+,t)
	=&
	\frac{L^2}{8\pi^2D}
	\frac{1}{C_1} (\frac{C_1^2}{4}+1)
	\left[ 2C_1E_{\msf{UV}}+\frac{1}{t} \left( e^{-(1+C_1)E_{\msf{UV}} t} -e^{-(1-C_1)E_{\msf{UV}} t}\right) \right], 
	\\
	\ln K_{0}''^{\msf{con}}(\eta\rightarrow 0^+,t)
	=&
	\frac{L^2}{4\pi^2D}
	E_{\msf{UV}} \left( 1-e^{-E_{\msf{UV}} t} \right),
	\end{aligned}
	\end{align}
	where 
	\begin{align}
	C_1\equiv \frac{1}{8\pi}\frac{1 }{\nu_0D} \frac{\gamma^2}{2-\gamma} .
	\end{align}
	We have employed the fact that $C_1<1$ for weak enough disorder strength and $C_1E_{\msf{UV}} \gg \frac{1}{4}\frac{1 }{\nu_0D} \frac{\gamma^2}{2-\gamma}\frac{1}{t}$ in the regime of interest $t \gg E_{\msf{UV}}^{-1}=\tel^{-1}$. 
	From Eq.~\ref{eq:K2-5}, we can see that in the current case, the leading order contribution is $t$ independent, and therefore one needs to consider higher order fluctuations to see the influence of the inter-replica mass on the time dependence of the connected SFF.
	
	\subsection{Cumulant expansion and higher order fluctuations}
	
	For a two-dimensional non-interacting disordered system, an expression similar to the cumulant expansion Eq.~\ref{eq:CumSum} for the many-body SFF of the RMT model can be derived.
	We start from the following equation 
	\begin{align}\label{eqd:Kave}
	\begin{aligned}
	K_0(\eta,t)
	=
	\left\langle
	\exp
	\left\lbrace 
	L^2
	\suml{a}
	\int_{-\infty}^{\infty} d\e
	\ln \left[ 1+e^{-i \zeta_a z_a \e } \right] 
	\nu(\e)
	\right\rbrace 
	\right\rangle ,
	\end{aligned}
	\end{align}
	which expresses the non-interacting SFF in terms of the single-particle density of states $\nu(\e)$:
	\begin{align}\label{eqd:nu}
	\nu(\e)=\frac{1}{L^2} \Tr \delta (\e -h),
	\end{align}
	with $h$ being the single-particle Hamiltonian.
	A cumulant expansion leads to
	\begin{align}\label{eqd:CumSum}
	\begin{aligned}
	\ln K_0(\eta,t)  
	=\,&
	L^2
	\int_{-\infty}^{\infty} d \e
	\suml{a}\ln \left( 1+e^{-i \zeta_a z_a \e } \right) 
	\R_1(\e)
	\\
	&
	+
	\frac{L^4}{2} 
	\int_{-\infty}^{\infty} d \e
	\int_{-\infty}^{\infty} d \e'
	\left[ \suml{a}\ln \left(  1+e^{-i \zeta_a z_a \e } \right) \right] 
	\left[ \suml{b}\ln \left( 1+e^{-i \zeta_{b} z_b \e' } \right) \right] 
	\R_2^{\msf{con}}(\e,\e')
	\\
	&+...+
	\frac{L^{2n}}{n!} 
	\int_{-\infty}^{\infty} ...\int_{-\infty}^{\infty} 
	\left\lbrace 
	\prod_{k=1}^{n}
	d \e_k
	\left[ \suml{a_k}
	\ln \left( 1+e^{-i \zeta_{a_k} z_{a_k} \e_k }\right) \right] 
	\right\rbrace 
	\R_n^{\msf{con}}(\e_1,...,\e_n)
	+... \,\,\,,
	\end{aligned}
	\end{align}
	where
	$\R_n^{\msf{con}}(\e_1,...,\e_n)$ is the connected part of the bare $n$-point single-particle energy level correlation function
	\begin{align}\label{eqd:Rn}
	\begin{aligned}
	\R_n(\e_1,...,\e_n)
	=
	\left\langle 
	\nu(\e_1)...\nu(\e_n)
	\right\rangle.
	\end{aligned}
	\end{align}
	
	One can immediately see that the second order term in the cumulant expansion in Eq.~\ref{eqd:CumSum} is equivalent to the leading order fluctuation correction Eq.~\ref{eqd:K2-40} (after restoring the contribution from intra-replica $X$ propagator),
	\begin{align}\label{eqd:CUM-2}
	\begin{aligned}
	&-\suml{\kb} \sum_{a, b} \suml{n>0>m} \Tr \ln  (\GG_{X}^{(0)}\,^{-1})^{ab;ba}_{nm;mn}(\kb,\kb)
	=
	\frac{L^{4}}{2}  \suml{a,b}
	\int_{-\infty}^{\infty} d \e
	\int_{-\infty}^{\infty} d \e'
	\ln \left(  1+e^{-i \zeta_a z_a \e } \right) 
	\ln \left( 1+e^{-i \zeta_{b} z_b \e' } \right)
	\R_2^{\msf{con}}(\e,\e'),
	\end{aligned}
	\end{align}
	by using the explicit form of the non-interacting single-particle level correlation~\cite{altshuler1986repulsion,Kamenev} 
	\begin{align}\label{eqd:R2}
	\begin{aligned}
	L^{4}\R_2^{\msf{con}}(\e,\e')=\frac{1}{2\pi^2} \re \suml{\kb}
	\frac{1}{\left( Dk^2-i(\e-\e')\right)^2},
	\qquad
	|\e-\e'| \gg \Delta.
	\end{aligned}
	\end{align}
	Here an oscillatory part in the single-particle energy level correlation $\R_2^{\msf{con}}$ has been ignored for $|\e-\e'| \gg \Delta$, and contribution from nonstandard saddle points~\cite{AndreevAltshuer,Kamenev} is needed to recover this part.
	In the ergodic regime where $|\e-\e'| \ll E_{\msf{TH}}$, only the $\kb=0$ term in the momentum summation in Eq.~\ref{eqd:R2} needs to be retained, and the two level correlation function $\R_2^{\msf{con}}(\e,\e')$ becomes equivalent to Eq.~\ref{eq:R2}, meaning that the single-particle energy levels follow the distribution of GUE.

	Setting $\phi=0$ and $Q=\Lambda^{(0)}$ in the action $S[Q,\phi]$ in Eq.~\ref{eqd:SQ}, we find that the action of the standard saddle point yields the first term in the cumulant expansion in Eq.~\ref{eqd:CumSum}:
	\begin{align}\label{eqd:CUM-1}
	\begin{aligned}
	S[\Lambda^{(0)},0]
	=
	-L^2
	\int_{-\infty}^{\infty} d \e
	\suml{a}\ln \left( 1+e^{-i \zeta_a z_a \e } \right) 
	\R_1(\e),
	\end{aligned}
	\end{align}
	where the average single-particle level density is given by
	\begin{align}
	\R_1(\e)=-\frac{1}{L^2} \frac{1}{\pi}\im \suml{\kb}  \frac{1}{\e-\xi_{\kb}+i\frac{1}{2\tel}} 
	\approx \nu_0.
	\end{align}
	
	In summary, for the non-interacting SFF, the leading two terms in the cumulant expansion Eq.~\ref{eqd:CumSum} can be recovered by considering the standard saddle point action and the leading order correction from fluctuations around the standard saddle point (see Eqs.~\ref{eqd:K2-2b} and~\ref{eqd:K0-3}).
	To recover the higher order terms in the cumulant expansion which are also essential for the SFF, one has to consider the higher order fluctuation correction.
	
	In the presence of interactions, the calculation in the previous section shows that the leading order fluctuation correction associated with $\Lambda^{(0)}$ to the connected SFF $K^{\msf{con}}(\eta,t)$  can also by expressed as the second term in the cumulant expansion Eq.~\ref{eq:CumSum}  (with $a\neq b$), after effecting the replacement, 
	\begin{align}
	\begin{aligned}
	2\pi^2 L^4 \R_2^{\msf{con}}(\e,\e')= \suml{\kb} \frac{1 }{\left( Dk^2-i(\e-\e')^2 \right) } \rightarrow \suml{\kb}\frac{1 }{\left( Dk^2-i(\e-\e')^2+\lambda \right) },
	\qquad
	|\e-\e'| \gg \Delta.
	\end{aligned}
	\end{align} 
	We expect to see similar interaction effect on higher order fluctuation corrections, but this is beyond the scope of the current study.

\bibliography{main}